\documentclass[aps, prd,twocolumn,superscriptaddress,showpcs,amsmath,amssymb]{revtex4-1} 

\usepackage{mathrsfs}
\usepackage{graphicx,overpic,rotating}
\usepackage{hyperref}
\usepackage{dcolumn}
\usepackage{bm}
\usepackage{xspace}
\usepackage{multirow}
\usepackage{booktabs}
\usepackage{longtable}
\usepackage{lineno}

\newcommand{\Acp}{\ensuremath{A_\textit{CP}}}

\newcommand{\stat}{\ensuremath{\mathrm{~(stat)}}}
\newcommand{\syst}{\ensuremath{\mathrm{~(syst)}}}
\newcommand{\Dbar}{\ensuremath{\overline{D}{}}}
\newcommand{\Dhh}{\ensuremath{D^0\to h^+h'^-}}

\def \rightdownarrow
 {\kern.3em
 \rule[.5ex]{.15mm}{2ex}
 {\mbox{$\kern-0.1em{\longrightarrow}$}}      
 }

\def\lessim{\mathrel {\vcenter {\baselineskip 0pt \kern 0pt  
\hbox{$<$} \kern 0pt \hbox{$\sim$} }}}

\def\gessim{\mathrel {\vcenter {\baselineskip 0pt \kern 0pt   
\hbox{$>$} \kern 0pt \hbox{$\sim$} }}}

\newcommand{\pipi}{\ensuremath{\pi^+\pi^-}}
\newcommand{\KK}{\ensuremath{K^+K^-}}
\newcommand{\Kpi}{\ensuremath{K^+\pi^-}}
\newcommand{\piK}{\ensuremath{K^-\pi^+}}

\newcommand{\mum}{\mbox{$\mu$m}}				

\newcommand{\tev}{\ensuremath{\mathrm{Te\kern -0.1em V}}}
\newcommand{\gev}{\ensuremath{\mathrm{Ge\kern -0.1em V}}}	
\newcommand{\mev}{\ensuremath{\mathrm{Me\kern -0.1em V}}}	
\newcommand{\kev}{\ensuremath{\mathrm{ke\kern -0.1em V}}}	
\newcommand{\massgev}{\mbox{\gev/$c^2$}}			
\newcommand{\massmev}{\mbox{\mev/$c^2$}}			

\newcommand{\CP}{{\it CP}}                                            

\newcommand{\Fig}[1]{Figure \ref{fig:#1}}

\newcommand{\refcita}[1]{Ref.~\cite{#1}}

\newcommand{\gauss}{\ensuremath{\mathscr{G}}}
\newcommand{\tail}{\ensuremath{\mathscr{T}}}

\newcommand{\pdf}{\ensuremath{\wp}}

\newcommand{\babar}{{\mbox{\sl B\hspace{-0.4em} {\small\sl A}\hspace{-0.37em}\sl B\hspace{-0.4em}}} {\small\sl A\hspace{-0.02em}R}}

\def\beq{\begin{equation}}
\def\eeq{\end{equation}}
\def\bea{\begin{eqnarray}}
\def\eea{\end{eqnarray}}

\def\sss{\scriptscriptstyle}

\def\barp{{\raise.35ex\hbox
{${\sss (}$}}---{\raise.35ex\hbox{${\sss )}$}}}
\def\bdbarp{\hbox{$B_d$\kern-1.4em\raise1.4ex\hbox{\barp}}}
\def\bsbarp{\hbox{$B_s$\kern-1.4em\raise1.4ex\hbox{\barp}}}

\def\roughly#1{\mathrel{\raise.3ex\hbox
{$#1$\kern-.75em\lower1ex\hbox{$\sim$}}}}

\newcommand{\bear}{\begin{array}}
\newcommand{\ear}{\end{array}}
\newcommand{\bet}{\begin{tabular}}
\newcommand{\eet}{\end{tabular}}
\newcommand{\beqn}{\begin{eqnarray}}
\newcommand{\eeqn}{\end{eqnarray}}

\newcommand{\Dpipi}{\ensuremath{D^{0} \rightarrow \pi^+ \pi^-}}
\newcommand{\Dkpi}{\ensuremath{D^{0} \rightarrow K^- \pi^+}}

\newcommand{\Dkk}{\ensuremath{D^{0} \rightarrow K^+ K^-}}

\newcommand{\DKpi}{\ensuremath{D^{0} \rightarrow K^- \pi^+}}
\newcommand{\aDKpi}{\ensuremath{\overline{D}^{0} \rightarrow K^+ \pi^-}}

\begin{document}
\graphicspath{{plots_for_paper/}}
\title{Measurement of CP--violating asymmetries in $D^0\to\pi^+\pi^-$ and $D^0\to K^+K^-$ decays at CDF}
\affiliation{Institute of Physics, Academia Sinica, Taipei, Taiwan 11529, Republic of China}
\affiliation{Argonne National Laboratory, Argonne, Illinois 60439, USA}
\affiliation{University of Athens, 157 71 Athens, Greece}
\affiliation{Institut de Fisica d'Altes Energies, ICREA, Universitat Autonoma de Barcelona, E-08193, Bellaterra (Barcelona), Spain}
\affiliation{Baylor University, Waco, Texas 76798, USA}
\affiliation{Istituto Nazionale di Fisica Nucleare Bologna, $^{ee}$University of Bologna, I-40127 Bologna, Italy}
\affiliation{University of California, Davis, Davis, California 95616, USA}
\affiliation{University of California, Los Angeles, Los Angeles, California 90024, USA}
\affiliation{Instituto de Fisica de Cantabria, CSIC-University of Cantabria, 39005 Santander, Spain}
\affiliation{Carnegie Mellon University, Pittsburgh, Pennsylvania 15213, USA}
\affiliation{Enrico Fermi Institute, University of Chicago, Chicago, Illinois 60637, USA}
\affiliation{Comenius University, 842 48 Bratislava, Slovakia; Institute of Experimental Physics, 040 01 Kosice, Slovakia}
\affiliation{Joint Institute for Nuclear Research, RU-141980 Dubna, Russia}
\affiliation{Duke University, Durham, North Carolina 27708, USA}
\affiliation{Fermi National Accelerator Laboratory, Batavia, Illinois 60510, USA}
\affiliation{University of Florida, Gainesville, Florida 32611, USA}
\affiliation{Laboratori Nazionali di Frascati, Istituto Nazionale di Fisica Nucleare, I-00044 Frascati, Italy}
\affiliation{University of Geneva, CH-1211 Geneva 4, Switzerland}
\affiliation{Glasgow University, Glasgow G12 8QQ, United Kingdom}
\affiliation{Harvard University, Cambridge, Massachusetts 02138, USA}
\affiliation{Division of High Energy Physics, Department of Physics, University of Helsinki and Helsinki Institute of Physics, FIN-00014, Helsinki, Finland}
\affiliation{University of Illinois, Urbana, Illinois 61801, USA}
\affiliation{The Johns Hopkins University, Baltimore, Maryland 21218, USA}
\affiliation{Institut f\"{u}r Experimentelle Kernphysik, Karlsruhe Institute of Technology, D-76131 Karlsruhe, Germany}
\affiliation{Center for High Energy Physics: Kyungpook National University, Daegu 702-701, Korea; Seoul National University, Seoul 151-742, Korea; Sungkyunkwan University, Suwon 440-746, Korea; Korea Institute of Science and Technology Information, Daejeon 305-806, Korea; Chonnam National University, Gwangju 500-757, Korea; Chonbuk National University, Jeonju 561-756, Korea}
\affiliation{Ernest Orlando Lawrence Berkeley National Laboratory, Berkeley, California 94720, USA}
\affiliation{University of Liverpool, Liverpool L69 7ZE, United Kingdom}
\affiliation{University College London, London WC1E 6BT, United Kingdom}
\affiliation{Centro de Investigaciones Energeticas Medioambientales y Tecnologicas, E-28040 Madrid, Spain}
\affiliation{Massachusetts Institute of Technology, Cambridge, Massachusetts 02139, USA}
\affiliation{Institute of Particle Physics: McGill University, Montr\'{e}al, Qu\'{e}bec, Canada H3A~2T8; Simon Fraser University, Burnaby, British Columbia, Canada V5A~1S6; University of Toronto, Toronto, Ontario, Canada M5S~1A7; and TRIUMF, Vancouver, British Columbia, Canada V6T~2A3}
\affiliation{University of Michigan, Ann Arbor, Michigan 48109, USA}
\affiliation{Michigan State University, East Lansing, Michigan 48824, USA}
\affiliation{Institution for Theoretical and Experimental Physics, ITEP, Moscow 117259, Russia}
\affiliation{University of New Mexico, Albuquerque, New Mexico 87131, USA}
\affiliation{The Ohio State University, Columbus, Ohio 43210, USA}
\affiliation{Okayama University, Okayama 700-8530, Japan}
\affiliation{Osaka City University, Osaka 588, Japan}
\affiliation{University of Oxford, Oxford OX1 3RH, United Kingdom}
\affiliation{Istituto Nazionale di Fisica Nucleare, Sezione di Padova-Trento, $^{ff}$University of Padova, I-35131 Padova, Italy}
\affiliation{University of Pennsylvania, Philadelphia, Pennsylvania 19104, USA}
\affiliation{Istituto Nazionale di Fisica Nucleare Pisa, $^{gg}$University of Pisa, $^{hh}$University of Siena and $^{ii}$Scuola Normale Superiore, I-56127 Pisa, Italy}
\affiliation{University of Pittsburgh, Pittsburgh, Pennsylvania 15260, USA}
\affiliation{Purdue University, West Lafayette, Indiana 47907, USA}
\affiliation{University of Rochester, Rochester, New York 14627, USA}
\affiliation{The Rockefeller University, New York, New York 10065, USA}
\affiliation{Istituto Nazionale di Fisica Nucleare, Sezione di Roma 1, $^{jj}$Sapienza Universit\`{a} di Roma, I-00185 Roma, Italy}
\affiliation{Rutgers University, Piscataway, New Jersey 08855, USA}
\affiliation{Texas A\&M University, College Station, Texas 77843, USA}
\affiliation{Istituto Nazionale di Fisica Nucleare Trieste/Udine, I-34100 Trieste, $^{kk}$University of Udine, I-33100 Udine, Italy}
\affiliation{University of Tsukuba, Tsukuba, Ibaraki 305, Japan}
\affiliation{Tufts University, Medford, Massachusetts 02155, USA}
\affiliation{University of Virginia, Charlottesville, Virginia 22906, USA}
\affiliation{Waseda University, Tokyo 169, Japan}
\affiliation{Wayne State University, Detroit, Michigan 48201, USA}
\affiliation{University of Wisconsin, Madison, Wisconsin 53706, USA}
\affiliation{Yale University, New Haven, Connecticut 06520, USA}

\author{T.~Aaltonen}
\affiliation{Division of High Energy Physics, Department of Physics, University of Helsinki and Helsinki Institute of Physics, FIN-00014, Helsinki, Finland}
\author{B.~\'{A}lvarez~Gonz\'{a}lez$^z$}
\affiliation{Instituto de Fisica de Cantabria, CSIC-University of Cantabria, 39005 Santander, Spain}
\author{S.~Amerio}
\affiliation{Istituto Nazionale di Fisica Nucleare, Sezione di Padova-Trento, $^{ff}$University of Padova, I-35131 Padova, Italy}
\author{D.~Amidei}
\affiliation{University of Michigan, Ann Arbor, Michigan 48109, USA}
\author{A.~Anastassov$^x$}
\affiliation{Fermi National Accelerator Laboratory, Batavia, Illinois 60510, USA}
\author{A.~Annovi}
\affiliation{Laboratori Nazionali di Frascati, Istituto Nazionale di Fisica Nucleare, I-00044 Frascati, Italy}
\author{J.~Antos}
\affiliation{Comenius University, 842 48 Bratislava, Slovakia; Institute of Experimental Physics, 040 01 Kosice, Slovakia}
\author{G.~Apollinari}
\affiliation{Fermi National Accelerator Laboratory, Batavia, Illinois 60510, USA}
\author{J.A.~Appel}
\affiliation{Fermi National Accelerator Laboratory, Batavia, Illinois 60510, USA}
\author{T.~Arisawa}
\affiliation{Waseda University, Tokyo 169, Japan}
% ...

\collaboration{CDF Collaboration\footnote{With visitors from
$^a$Istituto Nazionale di Fisica Nucleare, Sezione di Cagliari, 09042 Monserrato (Cagliari), Italy,
$^b$University of CA Irvine, Irvine, CA 92697, USA,
$^c$University of CA Santa Barbara, Santa Barbara, CA 93106, USA,
$^d$University of CA Santa Cruz, Santa Cruz, CA 95064, USA,
$^e$Institute of Physics, Academy of Sciences of the Czech Republic, Czech Republic,
$^f$CERN, CH-1211 Geneva, Switzerland,
$^g$Cornell University, Ithaca, NY 14853, USA,
$^h$University of Cyprus, Nicosia CY-1678, Cyprus,
$^i$Office of Science, U.S. Department of Energy, Washington, DC 20585, USA,
$^j$University College Dublin, Dublin 4, Ireland,
$^k$ETH, 8092 Zurich, Switzerland,
$^l$University of Fukui, Fukui City, Fukui Prefecture, Japan 910-0017,
$^m$Universidad Iberoamericana, Mexico D.F., Mexico,
$^n$University of Iowa, Iowa City, IA 52242, USA,
$^o$Kinki University, Higashi-Osaka City, Japan 577-8502,
$^p$Kansas State University, Manhattan, KS 66506, USA,
$^q$Korea University, Seoul, 136-713, Korea,
$^r$University of Manchester, Manchester M13 9PL, United Kingdom,
$^s$Queen Mary, University of London, London, E1 4NS, United Kingdom,
$^t$University of Melbourne, Victoria 3010, Australia,
$^u$Muons, Inc., Batavia, IL 60510, USA,
$^v$Nagasaki Institute of Applied Science, Nagasaki, Japan,
$^w$National Research Nuclear University, Moscow, Russia,
$^x$Northwestern University, Evanston, IL 60208, USA,
$^y$University of Notre Dame, Notre Dame, IN 46556, USA,
$^z$Universidad de Oviedo, E-33007 Oviedo, Spain,
$^{aa}$CNRS-IN2P3, Paris, F-75205 France,
$^{bb}$Texas Tech University, Lubbock, TX 79609, USA,
$^{cc}$Universidad Tecnica Federico Santa Maria, 110v Valparaiso, Chile,
$^{dd}$Yarmouk University, Irbid 211-63, Jordan,
}}
\noaffiliation
\date{\today}

\begin{abstract}
We report on a measurement of \CP--violating asymmetries (\Acp) in the Cabibbo-suppressed \mbox{$D^0\to\pi^+\pi^-$} and \mbox{$D^0\to K^+K^-$} decays reconstructed in a data sample corresponding to  $5.9$~fb$^{-1}$ of integrated luminosity collected by the upgraded Collider Detector at Fermilab. We use the strong decay $D^{*+}\to D^0\pi^+$  to identify the flavor of the charmed meson at production and exploit \CP--conserving strong $c\bar{c}$ pair-production in $p\bar{p}$ collisions. High-statistics samples of Cabibbo-favored $D^0\to K^-\pi^+$ decays with and without a $D^{*\pm}$ tag are used to correct for instrumental effects and significantly reduce systematic uncertainties. We measure $\Acp(D^0\to\pi^+\pi^-) = \bigl(+0.22\pm0.24\stat\pm0.11\syst\bigr)\%$  and \mbox{$\Acp(D^0\to K^+K^-) = \bigl(-0.24\pm0.22\stat\pm0.09\syst\bigr)\%$}, in agreement with \CP\ conservation.
These are the most precise determinations from a single experiment to date. Under the assumption of negligible direct \CP\ violation in  \mbox{$D^0\to\pi^+\pi^-$} and \mbox{$D^0\to K^+K^-$} decays, the results provide an upper limit to the \CP--violating asymmetry in $D^0$ mixing, $|\Acp^{\rm{ind}}(D^0)|< 0.13\%$ 
at the 90\% confidence level. 
\end{abstract}



\maketitle

\section{Introduction\label{sec:intro}}
The rich phenomenology of neutral flavored mesons provides many experimentally accessible observables sensitive to virtual contributions of non-standard model (SM) particles or couplings. Presence of non-SM physics may alter the expected decay or flavor-mixing rates, or introduce additional sources of \CP\ violation besides the Cabibbo-Kobayashi-Maskawa (CKM) phase. The physics of neutral kaons and bottom mesons has been mostly explored in dedicated experiments using kaon beams and $e^+e^-$ collisions~\cite{Antonelli:2009ws}. The physics of bottom-strange mesons is currently being studied in detail  in hadron collisions~\cite{Antonelli:2009ws}. 
In spite of the success of several dedicated experiments in the 1980's and 1990's, experimental sensitivities to parameters related to mixing and \CP\ violation in the charm sector were still orders of magnitude from most SM and non-SM expectations~\cite{Bianco:2003vb}. Improvements from early measurements at dedicated $e^+e^-$ colliders at the $\Upsilon(4S)$ resonance ($B$-factories) and the Tevatron were still insufficient for discriminating among SM and non-SM scenarios~\cite{pdg,hfag,Artuso:2008vf,Shipsey:2006zz,Burdman:2003rs}. 
Since charm transitions are described by physics of the first two quark generations, \CP--violating effects are expected to be smaller than $\mathcal{O}(10^{-2})$. Thus, relevant measurements require large event samples and careful control of systematic uncertainties to reach the needed sensitivity. Also, \CP--violating effects for charm have significantly more uncertain predictions compared to the bottom and strange sectors because of the intermediate value of the charm quark mass (too light for factorization of hadronic amplitudes and too heavy for applying chiral symmetry). All these things taken together have made the advances in the charm sector slower.
\par Studies of \CP\ violation in charm decays provide a unique probe for new physics.  The neutral $D$ system is the only one where up-sector quarks are involved in the initial state. Thus it probes scenarios where up-type quarks  play a special role, such as supersymmetric models where the down quark and the squark mass matrices are aligned~\cite{Nir:1993mx,Ciuchini:2007cw} and, more generally, models in which CKM mixing is generated in the up-quark sector. 
The interest in charm dynamics has increased recently with the observation of charm oscillations~\cite{Aubert:2007wf,Staric:2007dt,:2007uc}. The current measurements~\cite{hfag} indicate $\mathcal{O}(10^{-2})$ magnitudes for the parameters governing their phenomenology. Such values are on the upper end of most theory predictions~\cite{Petrov:2006nc}. Charm oscillations could be enhanced by a broad class of non-SM physics processes~\cite{Golowich:2007ka}. Any generic non-SM contribution to the mixing would naturally carry additional \CP--violating phases, which could enhance the observed \CP--violating asymmetries relative to SM predictions. Time integrated \CP--violating asymmetries of singly-Cabibbo-suppressed decays  into \CP\ eigenstates such as $D^0\to\pi^+\pi^-$ and $D^0\to K^+ K^-$ are powerful probes of non-SM physics contributions in the ``mixing" transition amplitudes. They also probe the magnitude of ``penguin"  contributions, which are negligible in the SM, but could be greatly enhanced by the exchange of additional non-SM particles. Both phenomena would,  in general, increase the size of the observed \CP\ violation with respect to the SM expectation. Any significant \CP--violating asymmetry above 
the   10$^{-2}$ level expected in the CKM hierarchy would  indicate non-SM physics.
The current experimental status is summarized in Table~\ref{tab:today}. No \CP\ violation has been found within the precision of about 0.5\% attained by the Belle and \babar\ experiments. The previous CDF result dates from 2005 and was obtained using data from only 123 pb$^{-1}$ of integrated luminosity. Currently, CDF has the world's largest samples of exclusive charm meson decays in charged final states, with competitive signal purities, owing to the good performance of the trigger for displaced tracks. With the current sample CDF can achieve a sensitivity that allows probing more extensive portions of the space of non-SM physics parameters.\par
We present measurements of time-integrated \CP--violating asymmetries in the Cabibbo-suppressed $D^0\to\pi^+\pi^-$ and $D^0\to K^+K^-$ decays (collectively referred to as $D^0\to h^+h^-$ in this article) using 1.96 TeV proton-antiproton collision data collected by the upgraded Collider Detector at Fermilab (CDF~II) and corresponding to
5.9 fb$^{-1}$ of integrated luminosity. Because the final states are common to charm and anti-charm meson decays, the time-dependent asymmetry between decays of states identified as $D^0$ and $\Dbar^0$ at the time of production ($t=0$)  defined as
\begin{equation}\label{eq:acp}
\Acp(h^+h^-, t) = \frac{N(D^0\to h^+h^-;t)-N(\Dbar^0\to h^+h^-;t)}{N(D^0\to h^+h^-;t)+N(\Dbar^0\to h^+h^-;t)}, \nonumber
\end{equation}
receives contributions from any difference in decay widths between $D^0$ and $\Dbar^0$ mesons in the chosen final state (direct \CP\ violation), any difference in mixing probabilities between $D^0$ and $\Dbar^0$ mesons, and the interference between direct decays and decays preceded by flavor oscillations (both indirect \CP\ violation). Due to the slow mixing rate of charm mesons, the time-dependent asymmetry is approximated at first order as the sum of two terms,
\begin{equation}\label{eq:acp2}
\Acp(h^+h^-;t) \approx  \Acp^{\rm{dir}}(h^+h^-)+\frac{t}{\tau}\ \Acp^{\rm{ind}}(h^+h^-),
\end{equation}
where $t/\tau$ is the proper decay time in units of $D^0$ lifetime ($\tau \approx 0.4$ ps), and the asymmetries are related to the decay amplitude $\mathcal{A}$ and the usual parameters used to describe flavored-meson mixing $x,\ y,\ p$,  and $q$  \cite{pdg} by 
\begin{widetext}
\begin{align}
\Acp^{\rm{dir}}(h^+h^-) &\equiv\Acp(t=0)=\frac{\left|\mathcal{A}(D^0\to h^+h^-)\right|^2-\left|\mathcal{A}(\Dbar^0\to h^+h^-)\right|^2}{\left|\mathcal{A}(D^0\to h^+h^-)\right|^2+\left|\mathcal{A}(\Dbar^0\to h^+h^-)\right|^2},\\*
\Acp^{\rm{ind}}(h^+h^-)  &= \frac{\eta_{\CP}}{2}\left[y \left(\left|\frac{q}{p}\right|-\left|\frac{p}{q}\right|\right)\cos\varphi-x \left(\left|\frac{q}{p}\right|+\left|\frac{p}{q}\right|\right)\sin\varphi\right],
\end{align}
\end{widetext}
where $\eta_{\CP} = +1$ is the \CP-parity of the decay final state and $\varphi$ is the \CP--violating phase.
The time-integrated asymmetry is then the time integral of Eq.\ (\ref{eq:acp2}) over the observed distribution of proper decay time ($D(t)$),
\begin{align}\label{eq:acp3}
\Acp(h^+h^-) 	&= \Acp^{\rm{dir}}(h^+h^-)+\Acp^{\rm{ind}}(h^+h^-)\int_0^\infty \frac{t}{\tau}\ D(t)dt  \nonumber \\ 
			&=  \Acp^{\rm{dir}}(h^+h^-) + \frac{\langle t \rangle}{\tau}\ \Acp^{\rm{ind}}(h^+h^-).
\end{align}
The first term arises from direct and the second one from indirect \CP\ violation. Since the value of $\langle t \rangle$ depends on $D(t)$, different values of time-integrated asymmetry could be observed in different experiments, depending on the detector acceptances as a function of decay time. Thus, each experiment may provide different sensitivity to   $\Acp^{\rm{dir}}$ and $\Acp^{\rm{ind}}$. 
Since the data used in this analysis were collected with an online event selection (trigger) that imposes requirements on the displacement of the $D^0$-meson decay point from the production point, our sample is enriched in higher-valued decay time candidates with respect to experiments at the $B$-factories. This makes the present measurement more sensitive to mixing-induced \CP\ violation. In addition, combination of our results with those from Belle and \babar\  provides some discrimination between the two contributions to the asymmetry.


\begin{table}[t]
\centering
\caption{Summary of recent experimental measurements of \CP\--violating asymmetries. The first quoted uncertainty is statistical, the second uncertainty is systematic.}\label{tab:today}
\begin{tabular}{lcc}
\hline\hline
Experiment      					& 		$\Acp(\pi^+\pi^-)\ (\%)$ 			&		$\Acp(K^+K^-)\ (\%)$			\\
\hline
\babar\ 2008 \cite{Aubert:2007if}	& 		$-0.24\pm0.52 \pm0.22$	& 		$+0.00\pm0.34 \pm0.13 $ 	\\
Belle 2008 \cite{:2008rx} 			& 		$-0.43\pm0.52 \pm0.12$ 	& 		$-0.43\pm0.30 \pm0.11 $ 	\\
CDF 2005   \cite{Acosta:2004ts} 	& 		$+1.0\pm1.3 \pm0.6 $ 		& 		$+2.0\pm1.2 \pm0.6$		\\
\hline\hline
\end{tabular}
\end{table}

\section{Overview\label{sec:overview}}
In the present work we measure the \CP--violating asymmetry in decays of $D^0$ and $\overline{D}^0$ mesons into $\pi^+\pi^-$ and $K^+K^-$ final states. Because the final states are charge-symmetric, to know whether they originate from a $D^0$ or a $\Dbar^0$ decay, we need the neutral charm candidate to be produced in the decay of an identified $D^{* +}$ or  $D^{* -}$ meson. Flavor conservation in the  strong-interaction decay of the $D^{*\pm}$ meson allows identification of the initial charm flavor through the sign of the charge of the $\pi$ meson:
 $D^{* +} \to D^0~\pi^+$ and 
$ D^{* -} \to \Dbar^0~\pi^-.$
We refer to $D$ mesons coming from identified $D^{*\pm}$ decays as the {\em tagged} sample and to the tagging pion as the {\em soft} pion, $\pi_s$.\par
In the data collected by CDF between February 2002 and January 2010, corresponding to an integrated luminosity of about 5.9~fb$^{-1}$,  we reconstruct approximately 215~000 $D^*$--tagged $D^0\to\pi^+\pi^-$ decays and 476~000 $D^*$--tagged $D^0\to K^+K^-$ decays. To measure the asymmetry, we determine the number of detected decays of opposite flavor and use the fact that primary charm and anti-charm mesons are produced in equal numbers by the \CP--conserving strong interaction. The observed asymmetry is the combination of the contributions from \CP\ violation and from charge asymmetries in the detection efficiency between positive and negative soft pions from the $D^{*\pm}$ decay. To correct for such instrumental asymmetries,  expected to be of the order of a few $10^{-2}$, we use two additional event samples: 5 million tagged, and 29 million untagged Cabibbo--favored $D^0\to K^-\pi^+$ decays. We achieve cancellation of instrumental asymmetries with high accuracy and measure the \CP--violating asymmetries of  $D^0\to\pi^+\pi^-$ and $D^0\to K^+K^-$ with a systematic uncertainty of about $10^{-3}$. \par
The paper is structured as follows.  In Sec.\ \ref{sec:detector} we briefly describe the components of the CDF detector
relevant for this analysis. 
 In Sec.\  \ref{sec:trigger} we summarize how the CDF trigger system was used to collect the event sample. We describe the strategy of the analysis and how we correct for detector-induced asymmetries in Sec.\  \ref{sec:method}. The event selection and the kinematic requirements applied to isolate the event samples are presented 
 in Sec.\  \ref{sec:sel}; the reweighting of kinematic distributions is discussed in Sec.\ \ref{sec:kin}.  The determination of observed asymmetries from data is described in 
 Sec.\  \ref{sec:fit}. 
 In Sec.\  \ref{sec:syst} we discuss possible sources of systematic uncertainties  and finally, in Sec.\  \ref{sec:final}, we present the results and compare with measurements performed by other experiments. We also show that by combining the present measurement with results from other experiments, we can partially disentangle the contribution of direct and indirect \CP\ violation. A brief summary is presented in Sec.\ \ref{sec:theend}. A mathematical derivation of the method employed to correct for instrumental  asymmetries is discussed in Appendix \ref{sec:method_math} and its validation on simulated samples is summarized in Appendix \ref{sec:mcvalidation}. 

\section{The CDF II detector\label{sec:detector}}
The CDF II detector has a cylindrical geometry with forward-backward symmetry and a tracking system in a 1.4 T magnetic field, coaxial with the beam. The tracking system is surrounded by calorimeters~\cite{calorimetro} and muon-detection chambers~\cite{muoni}. A cylindrical coordinate system, $(r,\phi,z)$, is used with origin at the geometric center of the detector, where $r$ is the perpendicular distance from the beam, $\phi$ is the azimuthal angle, and the $\hat{z}$ vector is in the direction of the proton beam. The polar angle $\theta$ with respect to the proton beam defines the pseudorapidity $\eta = -\ln\tan(\theta/2)$.

The CDF II detector tracking system  determines the trajectories of charged particles (tracks) and consists of an open cell argon-ethane gas drift chamber called the central outer tracker (COT)~\cite{COT} and a silicon vertex microstrip detector (SVX~II)~\cite{SVX}. The COT active volume covers $|z|<155$ cm from a radius of 40 to 140 cm and consists of 96 sense wire layers grouped into eight alternating axial and 2$^{\circ}$ stereo superlayers. To improve the resolution on their parameters, tracks found in the COT are extrapolated inward and matched to hits in the silicon detector. The SVX~II has five layers of silicon strips at radial distances ranging from 2.5 cm to 10.6 cm from the beamline. Three of the five layers are double-sided planes with $r-z$ strips oriented at 90$^\circ$ relative to $r-\phi$ strips, and the remaining two layers are double-sided planes with strips oriented at $\pm1.2^\circ$ angles relative to the $r-\phi$ strips. The SVX~II detector consists of three longitudinal barrels, each 29 cm in length, and covers approximately 90\% of the $p\overline{p}$ interaction region. The SVX~II provides precise information on the trajectories of long-lived particles (decay length),  which is  used for the identification of displaced,  secondary track vertices of $B$ and $D$ hadron decays. An innermost single-sided silicon layer (L00), installed at 1.5 cm from the beam,  further improves the resolution for vertex reconstruction \cite{L00}. Outside of the SVX~II, two additional layers of silicon assist pattern recognition and extend the sensitive region of the tracking detector to $|\eta|\approx 2$~\cite{ISL}. These intermediate silicon layers (ISL) are located between the SVX~II and the COT and consist of one layer at a radius of 23 cm in the central region, $|\eta|\leq 1$, and two layers in the forward region $1\leq|\eta|\leq 2$, at radii of 20 and 29 cm.  The component of a charged particle's momentum transverse to the beam ($p_T$) is determined with a resolution of \mbox{$\sigma_{p_T}/p_T \approx 0.07\% \ p_T$~($p_T$ in GeV/c)} for tracks with $p_T>2$ GeV/$c$. The excellent momentum resolution yields precise mass resolution  for fully reconstructed $B$ and $D$ decays, which provides good signal-to-background.  
 The typical resolution on the reconstructed position of decay vertices is approximately 30 $\mu$m in the transverse direction, effective to identify vertices from charmed meson decays, which are typically displaced by 250~$\mu$m from the beam. In the longitudinal direction, the resolution is approximately 70 $\mu$m, allowing suppression of backgrounds from charged particles originating from decays of distinct heavy hadrons in the event.

\section{Online sample selection\label{sec:trigger}}
The CDF~II trigger system  plays an important role in this measurement. Identification of hadronic decays of heavy-flavored mesons is challenging in the Tevatron collider environment due to the large inelastic $p\overline{p}$ cross section and high particle multiplicities at 1.96 TeV. In order to collect these events, the trigger system must reject more than 99.99\% of the collisions while retaining good efficiency for signal. In this Section, we describe the CDF~II trigger system and the algorithms used in collecting the samples of hadronic $D$ decays in this analysis.

The CDF~II trigger system has a three-level architecture: the first two levels, level 1 (L1) and level 2 (L2), are implemented in hardware and the third, level 3 (L3), is implemented in software on a cluster of computers using reconstruction algorithms that are similar to those used off line.

Using information from the COT, at L1,  the extremely fast tracker (XFT)~\cite{XFT} reconstructs trajectories of charged particles in the $r-\phi$ plane for each proton-antiproton bunch crossing. Events are selected for further processing when two tracks that satisfy trigger criteria on basic variables are found. The variables include the product of any combination of two particles' charges (opposite or same sign), the opening angle of the two tracks in the transverse plane ($\Delta\phi$), the two particles' transverse momenta, and their scalar sum.

At L2 the silicon vertex trigger (SVT)~\cite{SVT} incorporates information from the SVX~II detector into the trigger track reconstruction. The SVT identifies tracks displaced from the $p\bar{p}$ interaction point, such as those that arise from weak decays of heavy hadrons and have sufficient transverse momentum. Displaced tracks are those that have a distance of closest approach to the beamline (impact parameter $d_0$) inconsistent with having originated from the  $p\bar{p}$ interaction point (primary vertex). The impact parameter resolution of the SVT is approximately 50 $\mu$m, which includes a contribution of 35 $\mu$m from the width of the $p\overline{p}$ interaction region. The trigger selections used in this analysis require two tracks, each with impact parameter typically greater than 120~$\mu$m and smaller than 1 mm. In addition, the L2 trigger requires the transverse decay length ($L_{xy}$) to exceed $200~\mu$m,  where $L_{xy}$ is calculated as the projection of the vector from the primary vertex to the two-track vertex in the transverse plane along the vectorial sum of the transverse momenta of the tracks. The trigger based on the SVT collects large quantities of long-lived $D$ hadrons,  rejecting most of the prompt background. However,  through its impact-parameter-based selection, the SVT trigger also biases the observed proper decay time distribution. This has important consequences for the results of this analysis, which will be discussed in Sec.\ \ref{sec:final}.

The L3 trigger uses a full reconstruction of the event with all detector information, but uses a simpler tracking algorithm and preliminary calibrations relative to the ones used off line. The L3 trigger retests the criteria imposed by the L2 trigger. In addition, the difference in $z$ of the two tracks at the point of minimum distance from the primary vertex, $\Delta z_0$, is required not to exceed 5 cm,  removing events where the pair of tracks originate from different collisions within the same crossing of $p$ and $\bar{p}$ bunches.

\begin{table}[t]
\centering
\caption{Typical selection criteria for the three versions of the displaced-tracks trigger used in this analysis. The criteria refer to track pairs. The $p_T$, $d_0$, and $\eta$ requirements are applied to both tracks. The $\sum p_T$ refers to the scalar sum of the $p_T$ of the two tracks. The  $\sum p_T$ threshold in each of the three vertical portions of the table identifies the high-$p_T$ (top), medium-$p_T$ (middle), and low-$p_T$ (bottom) trigger selections.}\label{tab:mulbodtrig}
\begin{tabular}{lll}
\hline\hline
Level-1 & Level-2 & Level-3 \\
\hline
 $p_T > 2.5$ GeV/$c$ & $p_T > 2.5$ GeV/$c$ & $p_T> 2.5$ GeV/$c$\\
  $\sum p_T > 6.5$ GeV/$c$ & $\sum p_T > 6.5$ GeV/$c$ & $\sum p_T > 6.5$ GeV/$c$ \\
 Opposite charge & Opposite charge & Opposite charge \\
 $\Delta\phi < 90^\circ$ & $2^\circ < \Delta\phi < 90^\circ$ & $2^\circ < \Delta\phi < 90^\circ$ \\
 &  $0.12 < d_0 <1.0$ mm & $0.1 < d_0 <1.0$ mm\\
 & $L_{xy} > 200~\mu$m &  $L_{xy} >200~\mu$m \\
 & & $|\Delta z_0|<5$ cm \\
 & & $|\eta|<1.2$ \\ 
\hline
 $p_T > 2$ GeV/$c$ & $p_T > 2$ GeV/$c$ & $p_T>2$ GeV/$c$\\
  $\sum p_T > 5.5$ GeV/$c$ & $\sum p_T > 5.5$ GeV/$c$ & $\sum p_T > 5.5$ GeV/$c$ \\
 Opposite charge & Opposite charge & Opposite charge \\
 $\Delta\phi < 90^\circ$ & $2^\circ < \Delta\phi < 90^\circ$ & $2^\circ < \Delta\phi < 90^\circ$ \\
 & $0.12 < d_0 <1.0$ mm & $0.1 < d_0 <1.0$ mm \\
 & $L_{xy} > 200~\mu$m &  $L_{xy} >200~\mu$m \\
 & & $|\Delta z_0|<5$ cm \\
 & & $|\eta|<1.2$ \\ 
\hline
$p_T > 2$ GeV/$c$ & $p_T > 2$ GeV/$c$ & $p_T>2$ GeV/$c$\\
  $\sum p_T > 4$ GeV/$c$ & $\sum p_T > 4$ GeV/$c$ & $\sum p_T > 4$ GeV/$c$ \\
 $\Delta\phi < 90^\circ$ & $2^\circ < \Delta\phi < 90^\circ$ & $2^\circ < \Delta\phi < 90^\circ$ \\
 & $0.1 < d_0 <1.0$ mm & $0.1 < d_0 <1.0$ mm \\
 & $L_{xy} > 200~\mu$m &  $L_{xy} >200~\mu$m \\
 & & $|\Delta z_0|<5$ cm \\
 & & $|\eta|<1.2$ \\ 
\hline
\hline
\end{tabular}
\end{table}

Over the course of a single continuous period of Tevatron collisions (a store), the available trigger bandwidth varies because trigger rates fall as instantaneous luminosity falls. Higher trigger rates at high luminosity arise from both a larger rate for real physics processes as well as multiplicity-dependent backgrounds in multiple $p\overline{p}$ interactions.
To fully exploit the available trigger bandwidth, we employ three main variants of the displaced-tracks trigger. The three selections are summarized  in Table~\ref{tab:mulbodtrig} and are referred to as the low-$p_T$, medium-$p_T$, and high-$p_T$ selections according to their requirements on minimum transverse momentum. At high luminosity, the higher purity but less efficient high-$p_T$ selection is employed. As the luminosity decreases over the course of a store, trigger bandwidth becomes available and the other selections are utilized to fill the available trigger bandwidth and maximize the charm yield. The rates are controlled by the application of a prescale, which rejects a predefined fraction of events accepted by each trigger selection,  depending on the instantaneous luminosity.

\section{Suppressing detector-induced charge asymmetries\label{sec:method}}

The procedure used to cancel detector-induced asymmetries is briefly outlined here, while a detailed mathematical treatment is given in Appendix \ref{sec:method_math}.\par
We directly measure the observed ``raw'' asymmetry:
\begin{equation}
A(D^0) = \frac{N_{\text{obs}}(D^0)-N_{\text{obs}}(\Dbar^0)}{N_{\text{obs}}(D^0)+N_{\text{obs}}(\Dbar^0)}, \nonumber
\end{equation}
that is, the number of observed $D^0$ decays into the selected final state ($\pi^+\pi^-$ or $K^+K^-$) minus the number of $\Dbar^0$ decays, 
divided by the sum.

\begin{figure}[ht]
\centering
\includegraphics[width=8.6cm]{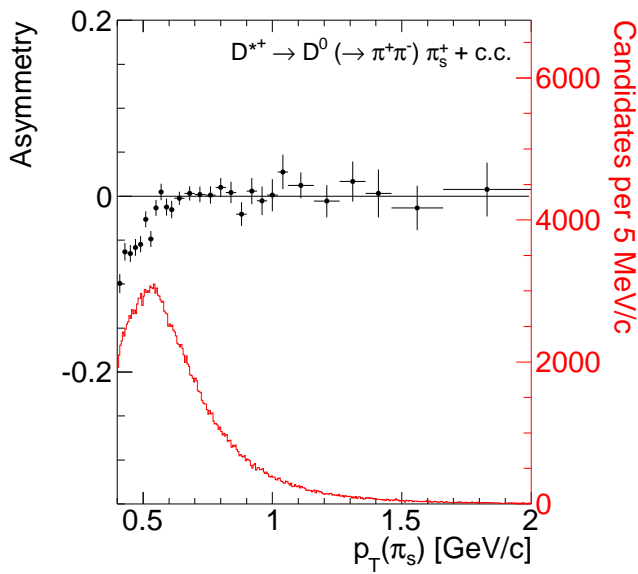}
\caption{Observed asymmetry between the number of reconstructed $D^{*+}$ and $D^{*-}$ mesons as a function of the soft pion's transverse momentum for pure samples of $D^{*+}\to D^0(\to\pi^+\pi^-)\pi_s^+$ and  $D^{*-}\to \overline{D}^0(\to\pi^+\pi^-)\pi_s^-$ decays. The soft pion transverse momentum spectrum is also shown.}\label{fig:soft}
\end{figure}

The main experimental difficulty of this measurement comes from the small differences in the detection efficiencies of tracks of opposite charge which may lead, if not properly taken into account, to spuriously-measured charge asymmetries. Relevant instrumental effects include differences in interaction cross sections with matter between positive and negative low-momentum hadrons and the geometry of the main tracking system. The drift chamber layout is intrinsically charge asymmetric because of  a $\approx 35^\circ$ tilt angle between the cell orientation and the radial direction,  designed to partially correct for the Lorentz angle in the charge drift direction caused by crossed electric and magnetic fields.  In the COT, different detection efficiencies are expected for positive and negative low-momentum tracks (especially, in our case,  for soft pions), which induce an instrumental asymmetry in the number of reconstructed $D^{*}$--tagged $D^0$ and $\Dbar^0$ mesons. Other possible asymmetries may originate in slightly different performance between positive and negative tracks in pattern-reconstruction and track-fitting algorithms. The combined effect of these is a net asymmetry in the range of a few percent, as shown in Fig.~\ref{fig:soft}. This must be corrected to better than one per mil to match the expected statistical precision of the present measurement.
In order to cancel detector effects, we extract the value of $\Acp(D^0\to h^+h^-)$ using a fully data-driven method, based on an appropriate combination of charge-asymmetries observed in three different event samples: $D^*$-tagged $D^0\to h^+h^-$ decays (or simply $hh^*$), $D^*$-tagged $D^0\to K^-\pi^+$ decays ($K\pi^*$), and untagged $D^0\to K^-\pi^+$ decays ($K\pi$). We assume the involved physical and instrumental asymmetries to be small, as indicated by previous measurements. Neglecting terms of order $\Acp\delta$ and $\delta^2$, the observed asymmetries in the three samples are
\begin{equation}\label{eq:acpraw}
\begin{aligned}
A(hh^*) &= \Acp(hh) + \delta(\pi_s)^{hh^*},\\
A(K\pi^*) &= \Acp(K\pi) + \delta(\pi_s)^{K\pi^*} + \delta(K\pi)^{K\pi^*},\\
A(K\pi) &= \Acp(K\pi) + \delta(K\pi)^{K\pi},
\end{aligned}
\end{equation}
where
$\delta(\pi_s)^{hh^*}$ is the instrumental asymmetry for reconstructing a positive or negative soft pion associated with a $h^+h^-$ charm decay induced by charge-asymmetric interaction cross section and reconstruction efficiency for low transverse momentum pions;
$\delta(\pi_s)^{K\pi^*}$ is the same as above for tagged $K^+\pi^-$ and $K^-\pi^+$ decays; and 
$\delta(K\pi)^{K\pi}$ and $\delta(K\pi)^{K\pi^*}$ are the instrumental asymmetries for reconstructing a $K^+\pi^-$ or a $K^-\pi^+$ decay for the untagged and the tagged case, respectively. 
All the above effects can vary as functions of a number of kinematic variables or environmental conditions in the detector.
If the kinematic distributions of soft pions are consistent in $K\pi^*$  and $hh^*$ samples,  and if the distributions of $D^0$ decay products are consistent in $K\pi^*$ and $K\pi$ samples,  then $\delta(\pi_s)^{hh^*} \approx \delta(\pi_s)^{K\pi^*}$ and $\delta(K\pi)^{K\pi^*}\approx \delta(K\pi)^{K\pi}$. The \CP--violating asymmetries then become accessible as
\begin{equation}\label{eq:formula}
\Acp(hh) = A(hh^*) - A(K\pi^*) + A(K\pi).
\end{equation}
This formula relies on cancellations based on two assumptions. At the Tevatron,  charm and anticharm mesons are expected to be created in almost equal numbers. Since the overwhelming majority of them are produced by \CP--conserving strong interactions, and the $p\bar{p}$ initial state is \CP\ symmetric, any small difference between the abundance of charm and anti-charm flavor is constrained to be antisymmetric in pseudorapidity.  As a consequence, we assume that the net effect of any possible charge asymmetry in the production cancels out, as long as the distribution of the decays in the sample used for this analysis is symmetric in pseudorapidity. An upper limit to any possible residual effect is evaluated as part of the study of systematic uncertainties  (Sec.\ \ref{sec:syst}). The second assumption is that the detection efficiency for the $D^*$ can be expressed as the product of the efficiency for the soft pion times the efficiency for the $D^0$ final state. This assumption has been tested  (Sec.\ \ref{sec:syst}), and any residual effect included in the systematic uncertainties.\par
Before applying this technique to data,  we show that our approach achieves the goal of suppressing detector induced asymmetries down to the per mil level using the full Monte Carlo simulation (Appendix~\ref{sec:mcvalidation}).  The simulation contains only charmed signal decays. The effects of the underlying event and multiple interactions are not simulated. We apply the method to samples simulated with a wide range of physical and detector asymmetries to verify that the cancellation works. The simulation is used here only to test the validity of the technique; all final results are derived from data only,  with no direct input from simulation.

\section{Analysis event selection\label{sec:sel}}
The offline selection is designed to retain the maximum number of \Dhh\ decays with accurately measured momenta and decay vertices. Any requirements that may induce asymmetries between the number of selected $D^0$ and $\Dbar^0$ mesons are avoided. The reconstruction is based solely on tracking, disregarding any information on particle identification. Candidate decays are reconstructed using only track pairs compatible with having fired the trigger.  Standard quality criteria on the minimum number of associated silicon-detector and drift-chamber hits are applied to each track to ensure precisely measured momenta and decay vertices in three-dimensions \cite{tesi-angelo}. Each final-state particle is required to have $p_T>2.2$ GeV/$c$, $|\eta|<1$, and impact parameter between 0.1 and 1 mm. 
The reconstruction of $D^0$ candidates considers all pairs of oppositely-charged particles in an event, which are arbitrarily assigned the charged pion mass.  The two tracks are constrained to originate from a common vertex by a kinematic fit subject to standard quality requirements. 
The $\pi^+\pi^-$ mass of candidates is required to be in the range 1.8 to 2.4~GeV$/c^2$, to retain all signals of  interest  and sideband regions sufficiently wide to study backgrounds. The two tracks are required to have an azimuthal separation $2^{\circ} < \Delta\phi < 90^{\circ}$, and correspond to a scalar sum of the two particles' transverse momenta greater than 4.5 GeV/$c$. We require $L_{xy}$ to exceed 200~$\mu$m to reduce background from decays of hadrons that don't contain heavy quarks. We also require the impact parameter of the $D^0$ candidate with respect to the beam, $d_0(D^0)$,  to be smaller than $100\ \mu$m to reduce the contribution from charmed mesons produced in long-lived $B$ decays (secondary charm). In the rare (0.04\%) occurrence that multiple \Dhh\  decays sharing the same tracks are reconstructed in the event,  we retain the one having the best vertex fit quality.
\begin{figure*}[t]
\centering
\begin{overpic}[width=8.6cm]{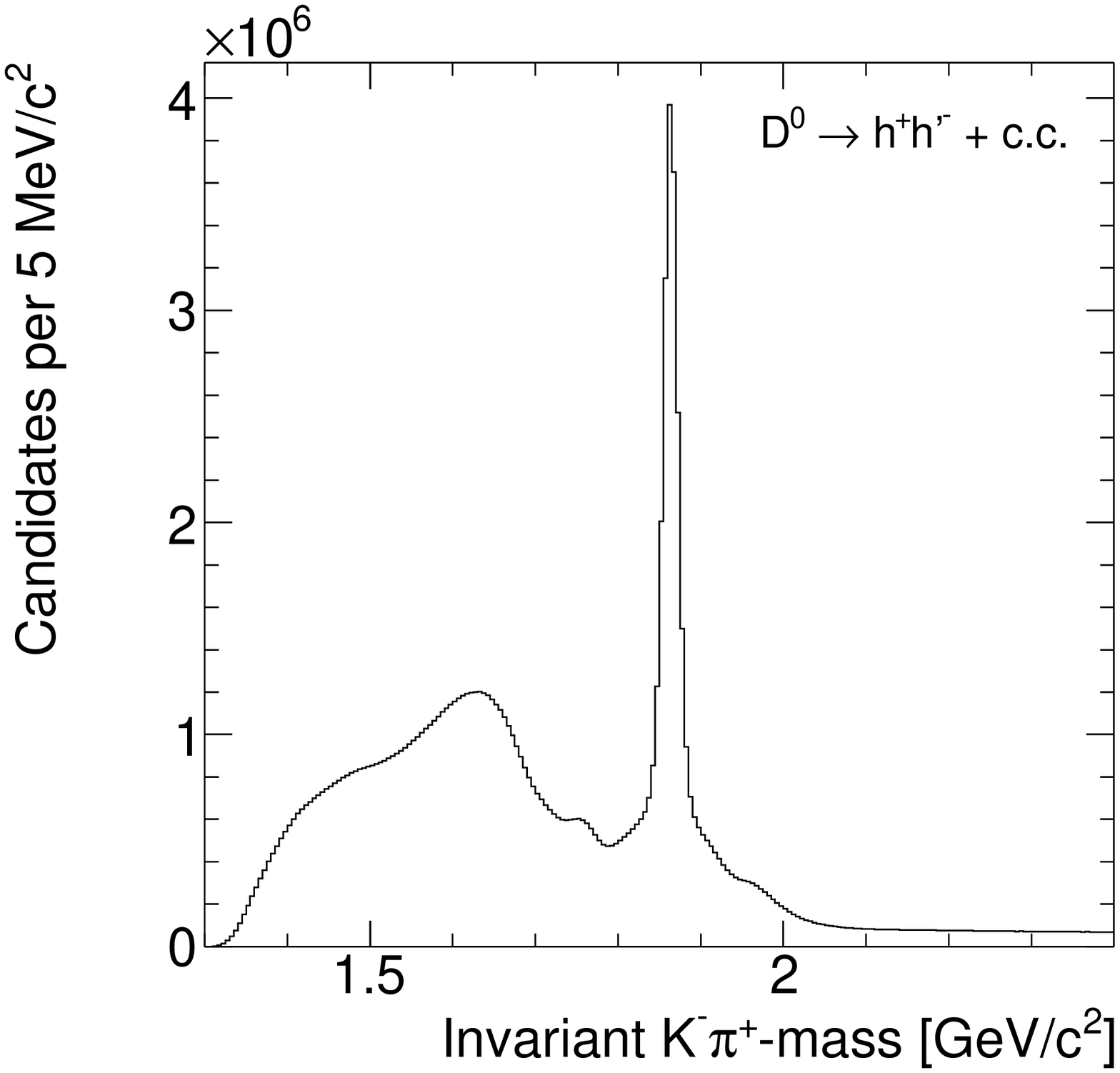}
\put(22,78){(a)}
\end{overpic}\hfil
\begin{overpic}[width=8.6cm]{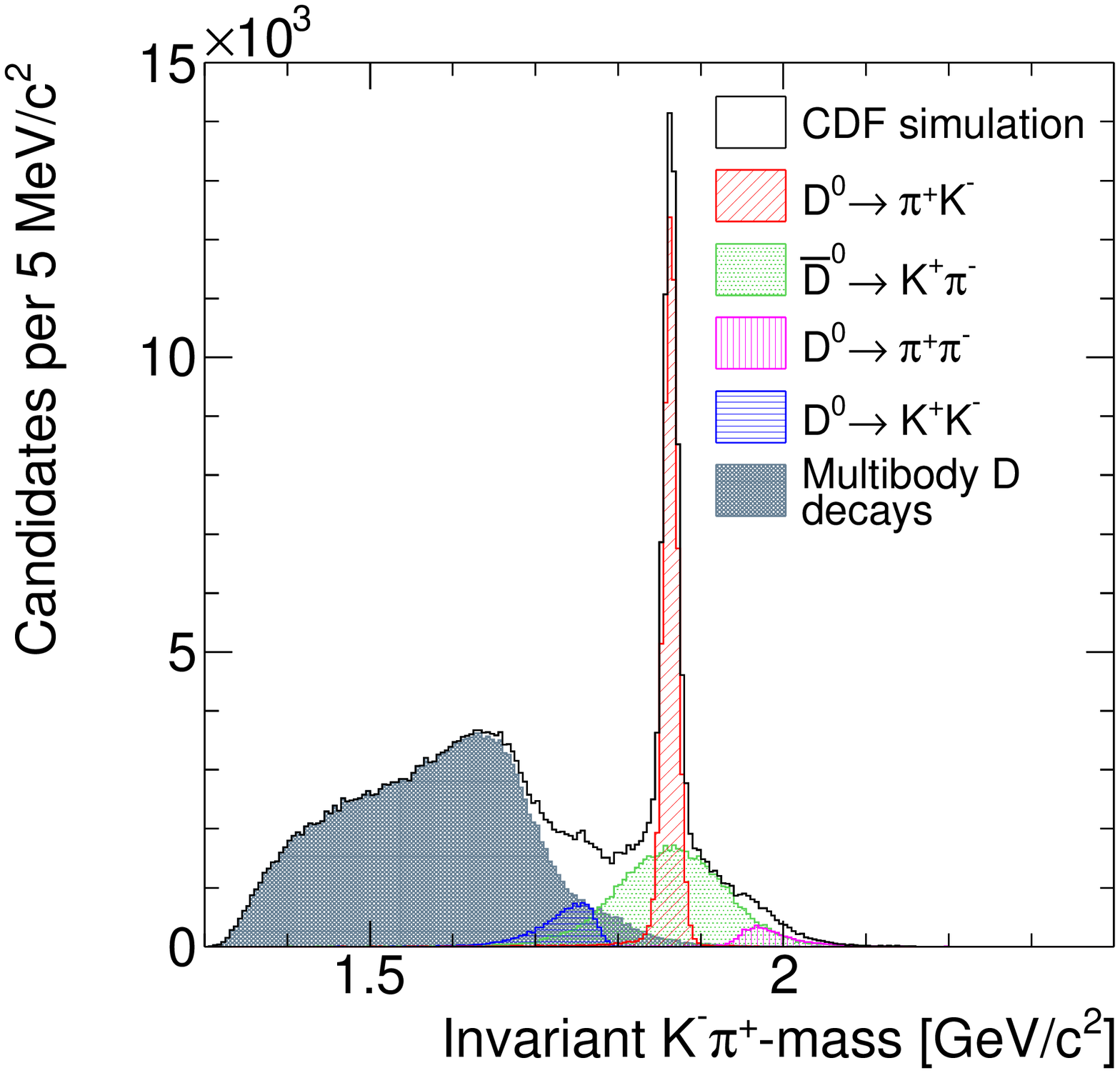}
\put(22,78){(b)}
\end{overpic}
\caption{Comparison between the  $K^-\pi^+$--mass distributions of (a) the untagged sample and of (b) a simulated sample of inclusive charm decays. See text for explanation of
contributions.}\label{fig:mass_1d}
\end{figure*}

Figure~\ref{fig:mass_1d} shows the $K^-\pi^+$ mass distribution for the resulting sample, which is referred to as ``untagged" in the following since no $D^*$ decay reconstruction has been imposed at this stage. The distribution of a sample of simulated inclusive charmed decays is also shown for comparison.  Only a single charmed meson decay per event is simulated without the underlying event.  In both distributions the kaon (pion) mass is arbitrarily assigned to the negative (positive) particle. The prominent narrow signal is dominated by $D^0\to K^-\pi^+$ decays.  A broader structure, also centered on the known $D^0$ mass,  are $\Dbar^0\to K^+\pi^-$ candidates reconstructed with swapped $K$ and $\pi$ mass assignments to the decay products. Approximately 29 million $D^0$ and $\Dbar^0$ mesons decaying into $K^{\pm}\pi^{\mp}$ final states are reconstructed. The two smaller enhancements at lower and higher masses than the $D^0$ signal are due to mis-reconstructed $D^0\to K^+K^-$ and $D^0\to\pi^+\pi^-$ decays, respectively. Two sources of background contribute. A component of random track pairs that accidentally meet the selection requirements (combinatorial background) is most visible at masses higher than 2 GeV/$c^2$, but populates almost uniformly the whole mass range. A large shoulder due to mis-reconstructed multi-body charm decays peaks at a mass of approximately 1.6 GeV/$c^2$. 

In the ``tagged"-samples reconstruction, we form $D^{*+}\to D^0\pi_s^{+}$ candidates by associating with each $D^0$ candidate all tracks present in the same event. The additional particle is required to satisfy basic quality requirements for the numbers of associated silicon and drift chamber hits, to be central ($|\eta|<1$), and to have transverse momentum greater that 400~MeV/$c$.  
We assume this particle to be a pion (``soft pion") and we match its trajectory to the $D^0$ vertex with simple requirements on relative separation:  impact parameter smaller than  600~$\mu$m and longitudinal distance from the primary vertex smaller than 1.5~cm. 
Since the impact parameter of the low-energy pion has degraded resolution with respect to those of the $D^0$ tracks, no real benefit is provided by a full three--track vertex fit for the $D^*$ candidate. We retain $D^*$ candidates with $D^0\pi_s$ mass smaller than 2.02~GeV/$c^2$. In the 2\% of cases in which multiple $D^*$ candidates are associated with a single $D^0$ candidate, we randomly choose only one $D^*$ candidate  for further analysis. \par The  $D^0\pi_s$ mass is calculated using the vector sum of the momenta of the three particles as $D^*$ momentum, and the known $D^0$ mass in the determination of the $D^*$ energy. This quantity has the same resolution advantages of  the more customary $M(h^+ h^{(')-} \pi_s)- M(h^+ h^{(')-})$ mass difference, and has the additional advantage that it is independent of the mass assigned to the $D^0$ decay products. Therefore all $D^{*+}\to D^0(\to h^+ h^{(')-})\pi_s^{+}$ modes have the same $D^0\pi_s$ mass distribution, which is not true for the mass difference distribution.

\begin{figure*}[t]
\centering
\begin{overpic}[width=5.9cm,grid=false]{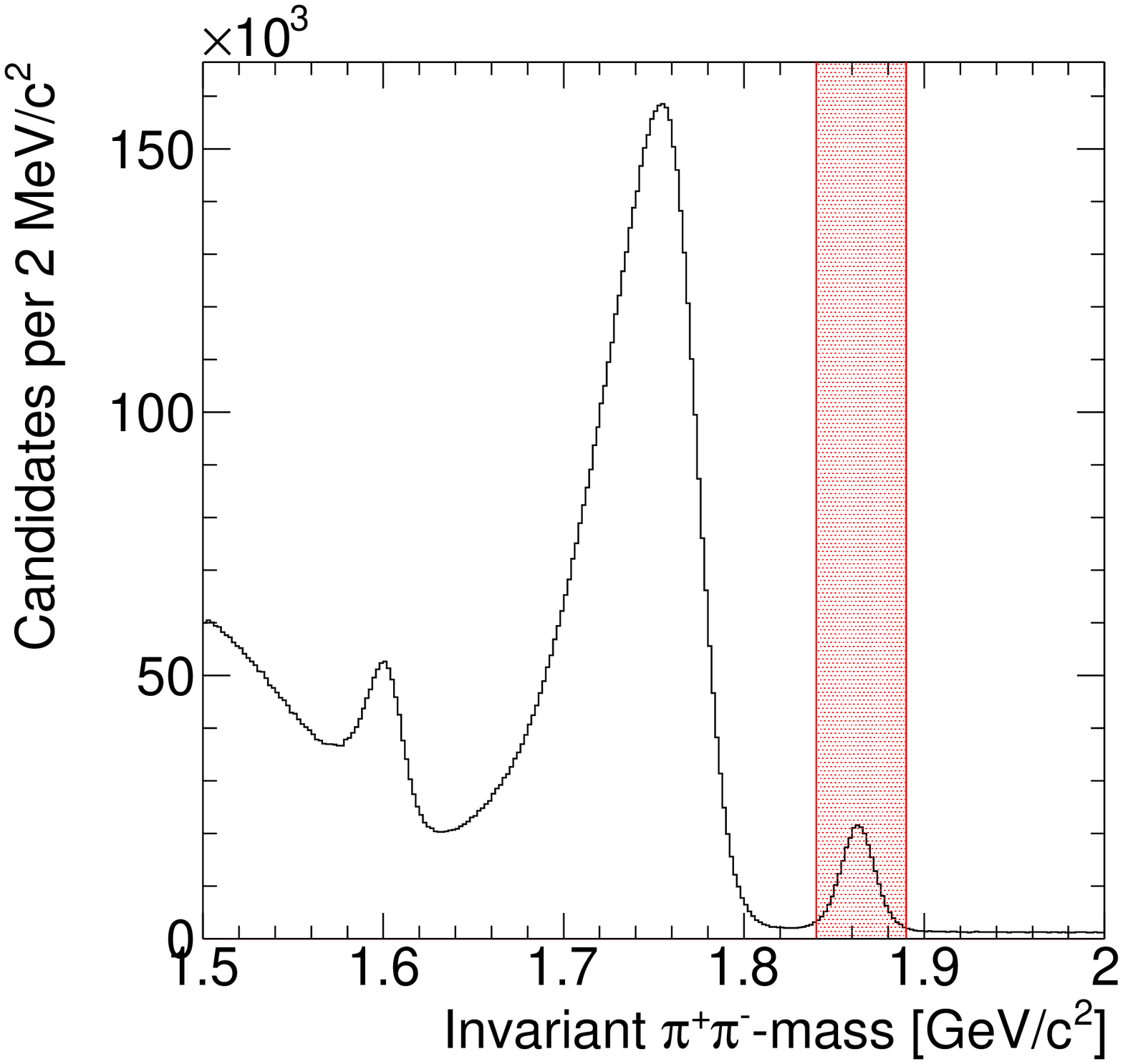}
\put(22,78){(a)}
\end{overpic}\hfil
\begin{overpic}[width=5.9cm,grid=false]{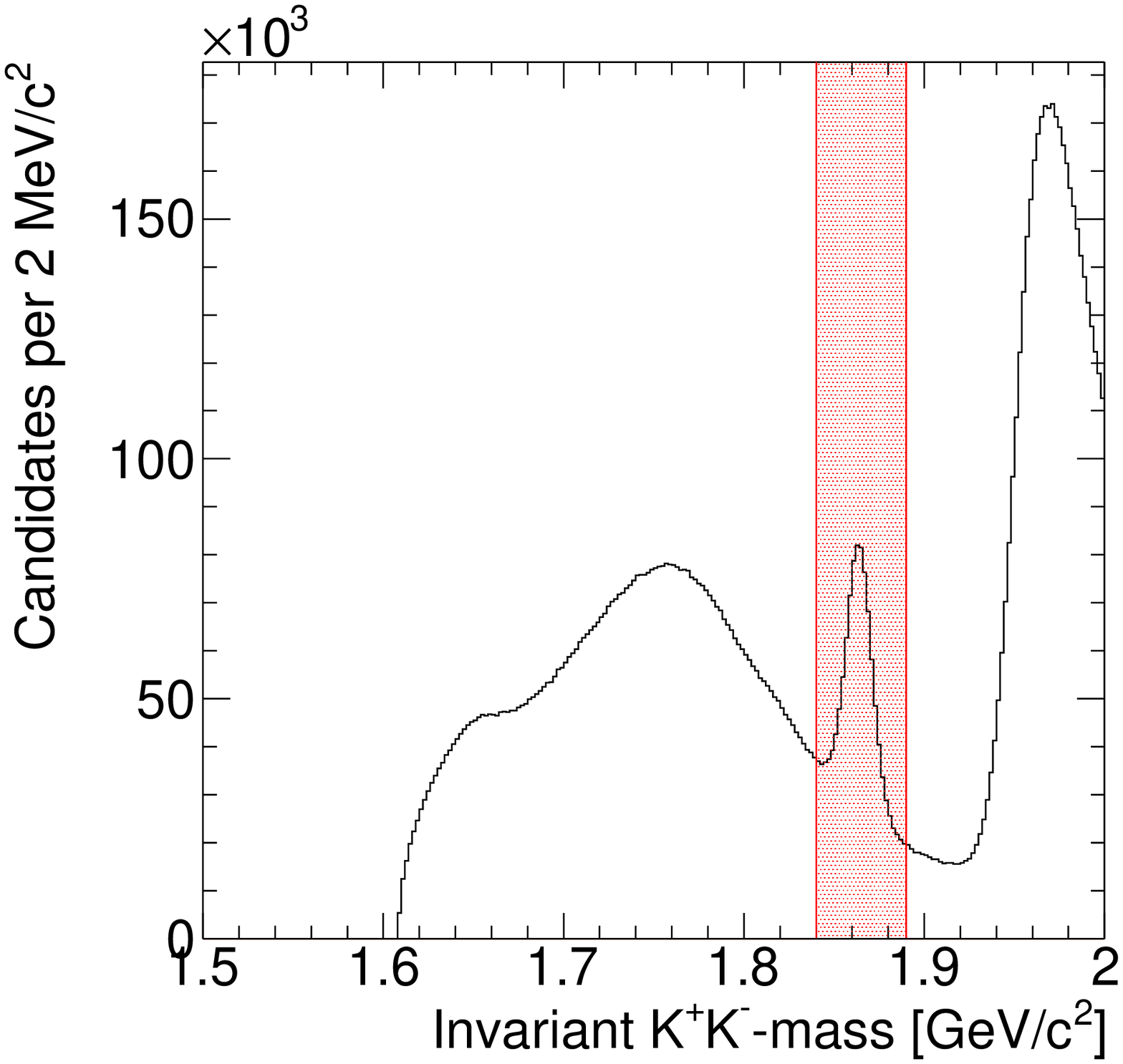}
\put(22,78){(b)}
\end{overpic}\hfil
\begin{overpic}[width=5.9cm,grid=false]{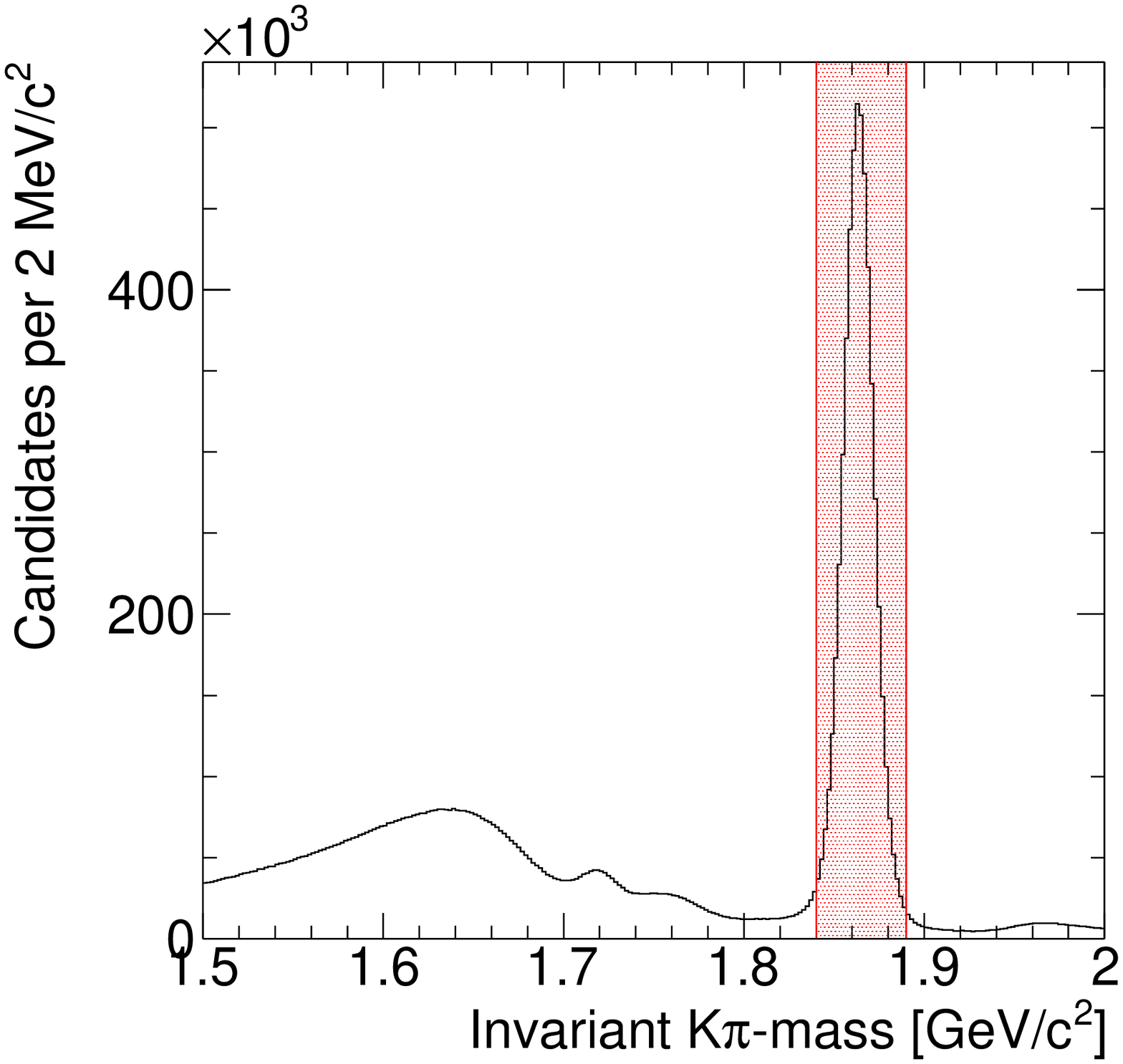}
\put(22,78){(c)}
\end{overpic}\\
\begin{overpic}[width=5.9cm,grid=false]{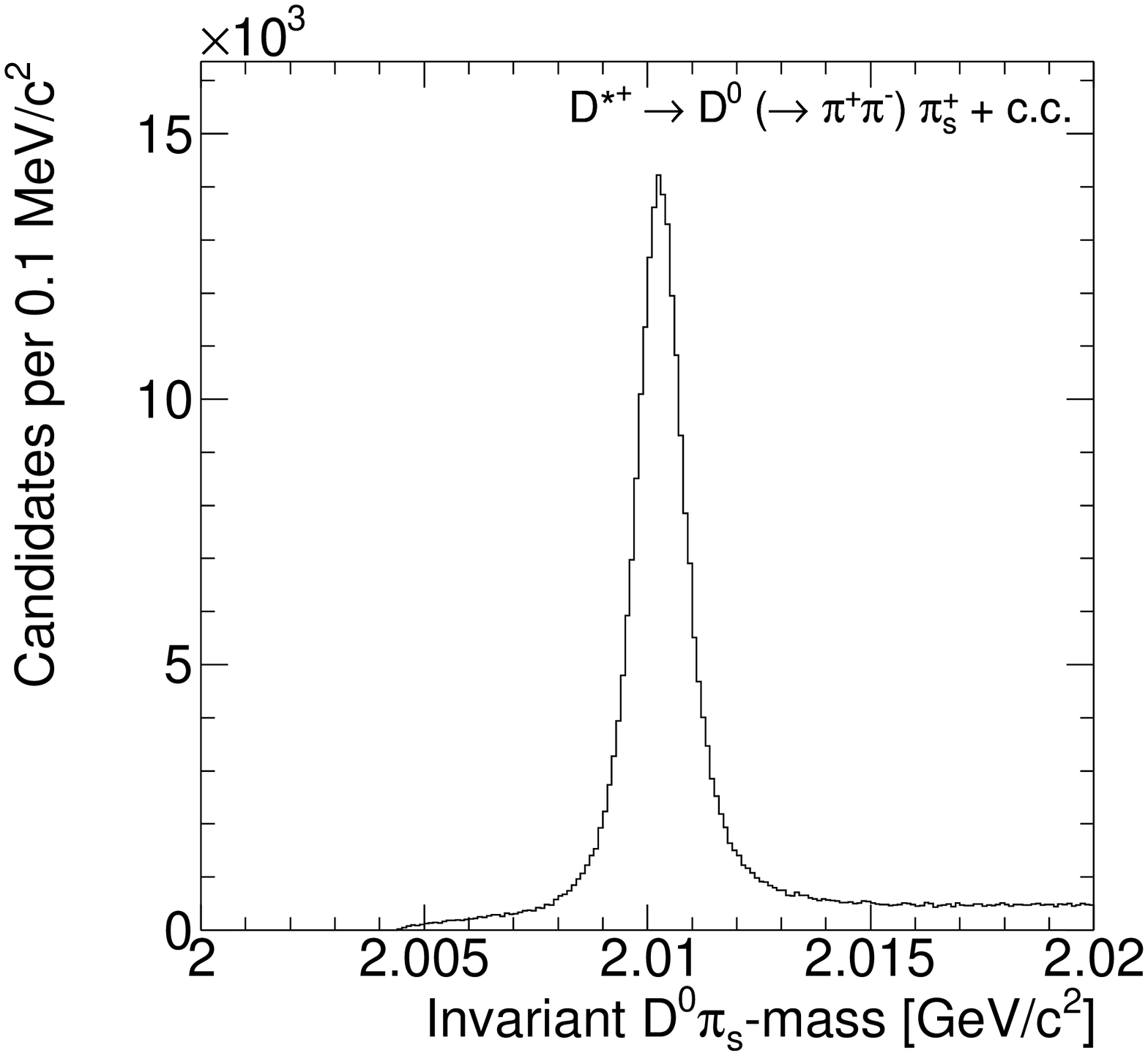}
\put(22,78){(d)}
\end{overpic}\hfil
\begin{overpic}[width=5.9cm,grid=false]{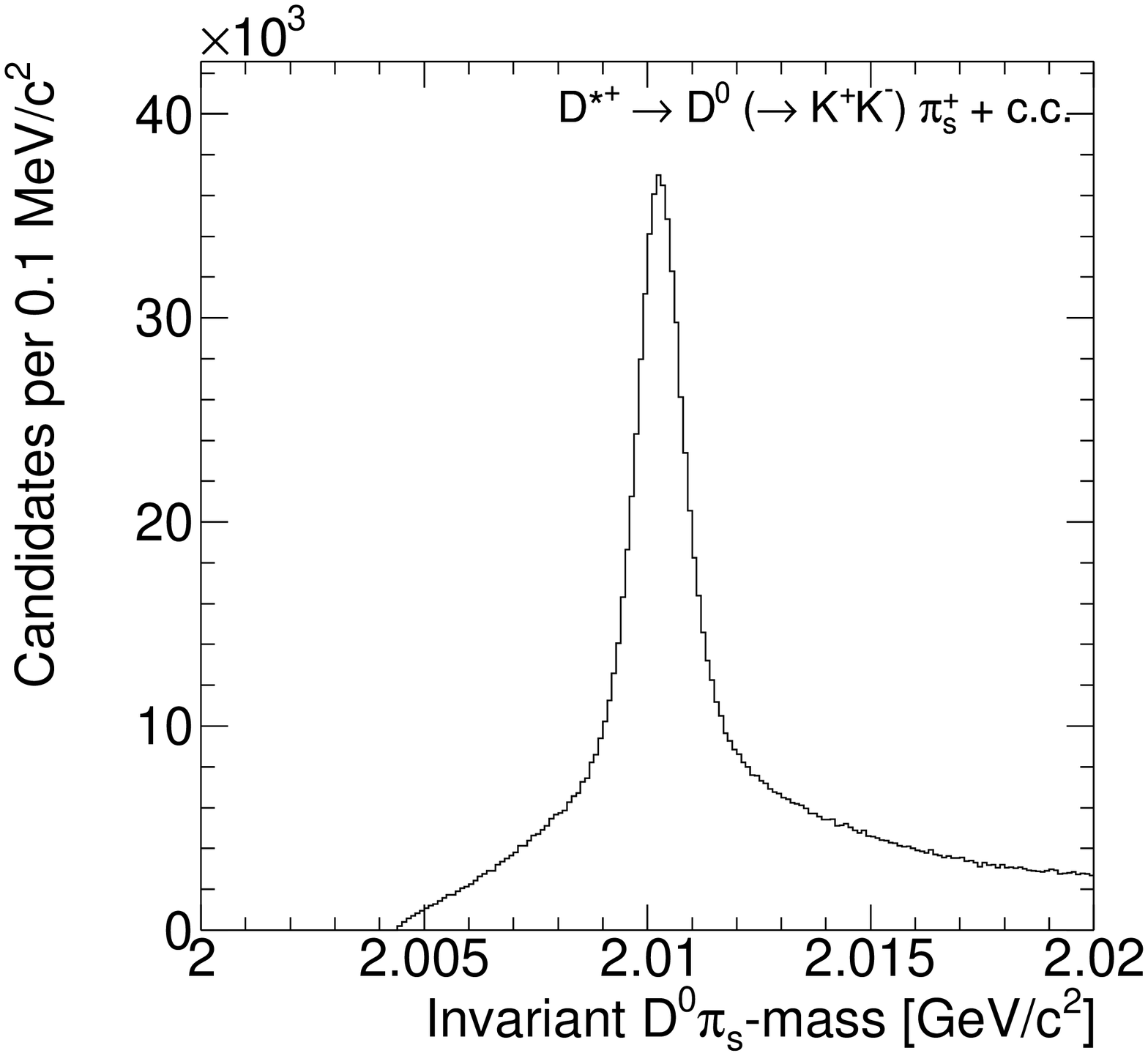}
\put(22,78){(e)}
\end{overpic}\hfil
\begin{overpic}[width=5.9cm,grid=false]{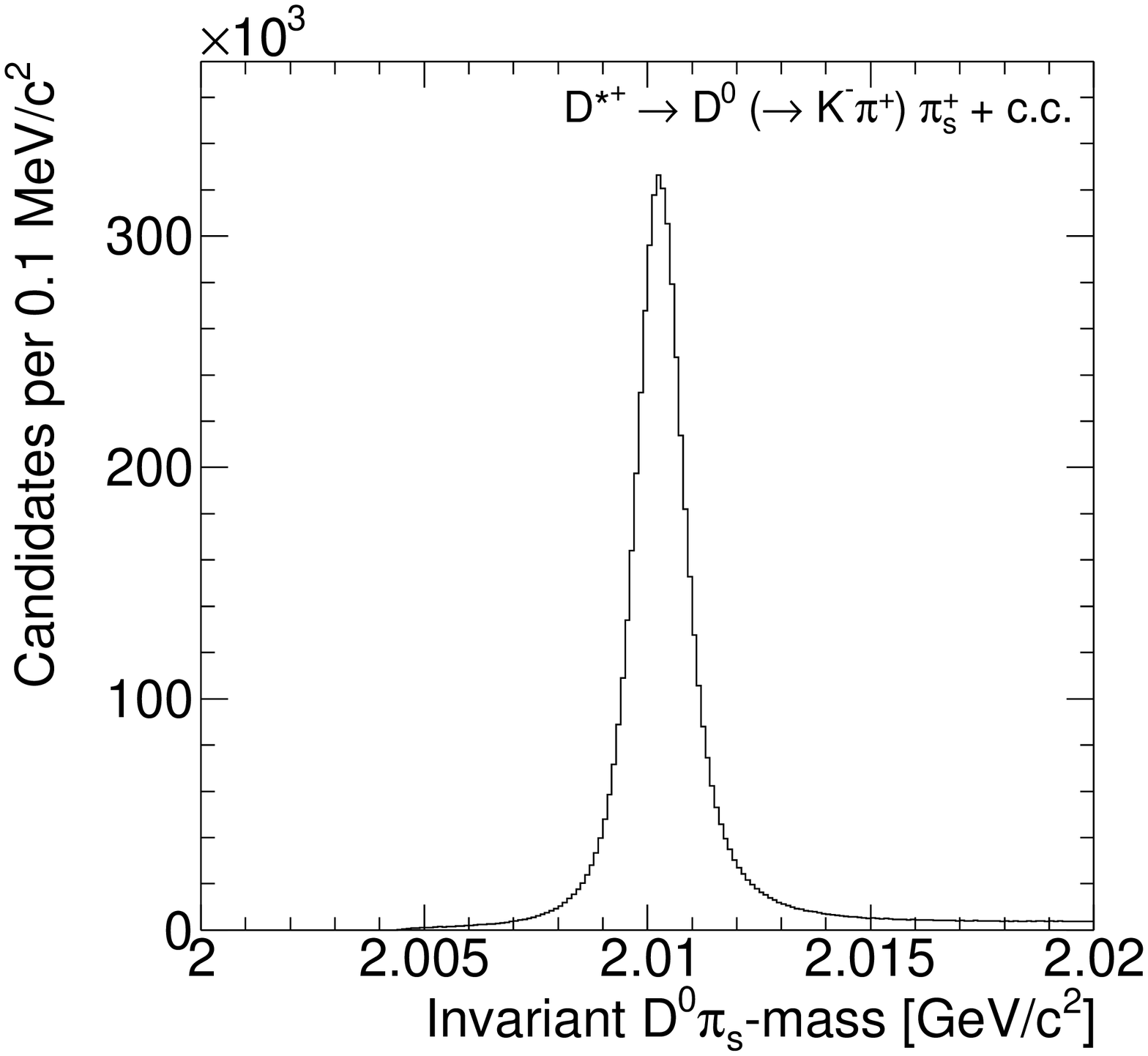}
\put(22,78){(f)}
\end{overpic}
\caption{Distributions of  (a) \pipi, (b) \KK, and (c) $K\pi$ mass. Regions used to define the tagged samples are shaded. Distribution of $D^0\pi_s$ mass for  tagged (d) \Dpipi, (e) \Dkk,  and (f) \Dkpi\  samples selected in the shaded regions.}\label{fig:mass_distr}
\end{figure*}

In each tagged sample ($D^0\to\pi^+\pi^-$ , $D^0\to K^+K^-$ and $D^0\to K^-\pi^+$) we require the corresponding two-body mass to lie within 24 MeV/$c^2$ of the known $D^0$ mass \cite{pdg}, as shown in Figs.~\ref{fig:mass_distr} (a)--(c).
Figures~\ref{fig:mass_distr}~(d)--(f) show the resulting $D^0\pi_s$ mass distribution. A clean $D^*$ signal is visible superimposed on background components that are different in each $D^0$ channel. As will be shown in Sec.\ \ref{sec:fit},   the backgrounds in the $D^0 \pi_s$ distributions for $D^0 \to \pi^+ \pi^-$ and $D^0 \to K^+ K^-$
decays are mainly due to associations of random pions with real $D^0$ candidates. In the $D^0\to K^+K^-$ case,  there is also a substantial contribution from mis-reconstructed multi-body charged and neutral charmed decays (mainly $D^{*+}\to D^0(\to K^-\pi^+\pi^0)\pi_s^+$ where the neutral pion is not reconstructed) that yield a broader enhancement underneath the signal peak.
We reconstruct approximately 215~000 $D^*$--tagged $D^0\to\pi^+\pi^-$ decays, 476~000 $D^*$--tagged $D^0\to K^+K^-$ decays,  and 5 million $D^*$--tagged $D^0\to\pi^+K^-$ decays.

\section{Kinematic distributions equalization}\label{sec:kin}
Because detector--induced asymmetries depend on kinematic properties, the asymmetry cancellation is realized accurately only if the kinematic distributions across the three samples are the same. Although the samples have been selected using the same requirements, small kinematic differences between decay channels may persist due to the different masses involved. We extensively search for any such residual effect across several kinematic distributions and reweight the tagged $D^0\to h^+h^-$  and untagged $D^0\to K^-\pi^+$ distributions to reproduce the tagged $D^0\to K^-\pi^+$ distributions when necessary. For each channel, identical reweighting functions are used for charm and anti-charm decays.

\begin{figure*}[t]
\centering
\begin{overpic}[width=5.9cm,grid=false]{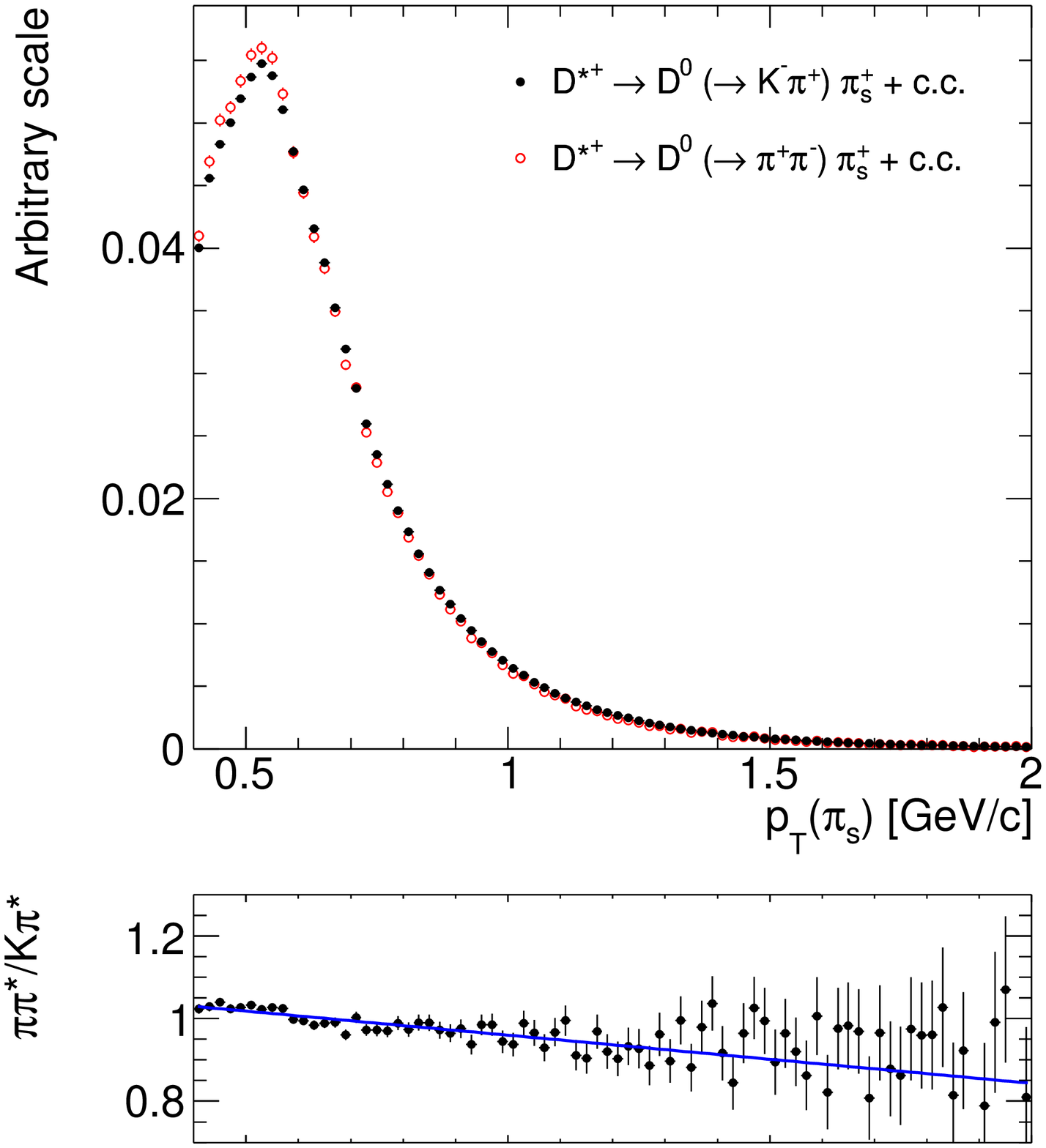}
\put(30,70){(a)} 
\end{overpic}\hfil
\begin{overpic}[width=5.9cm,grid=false]{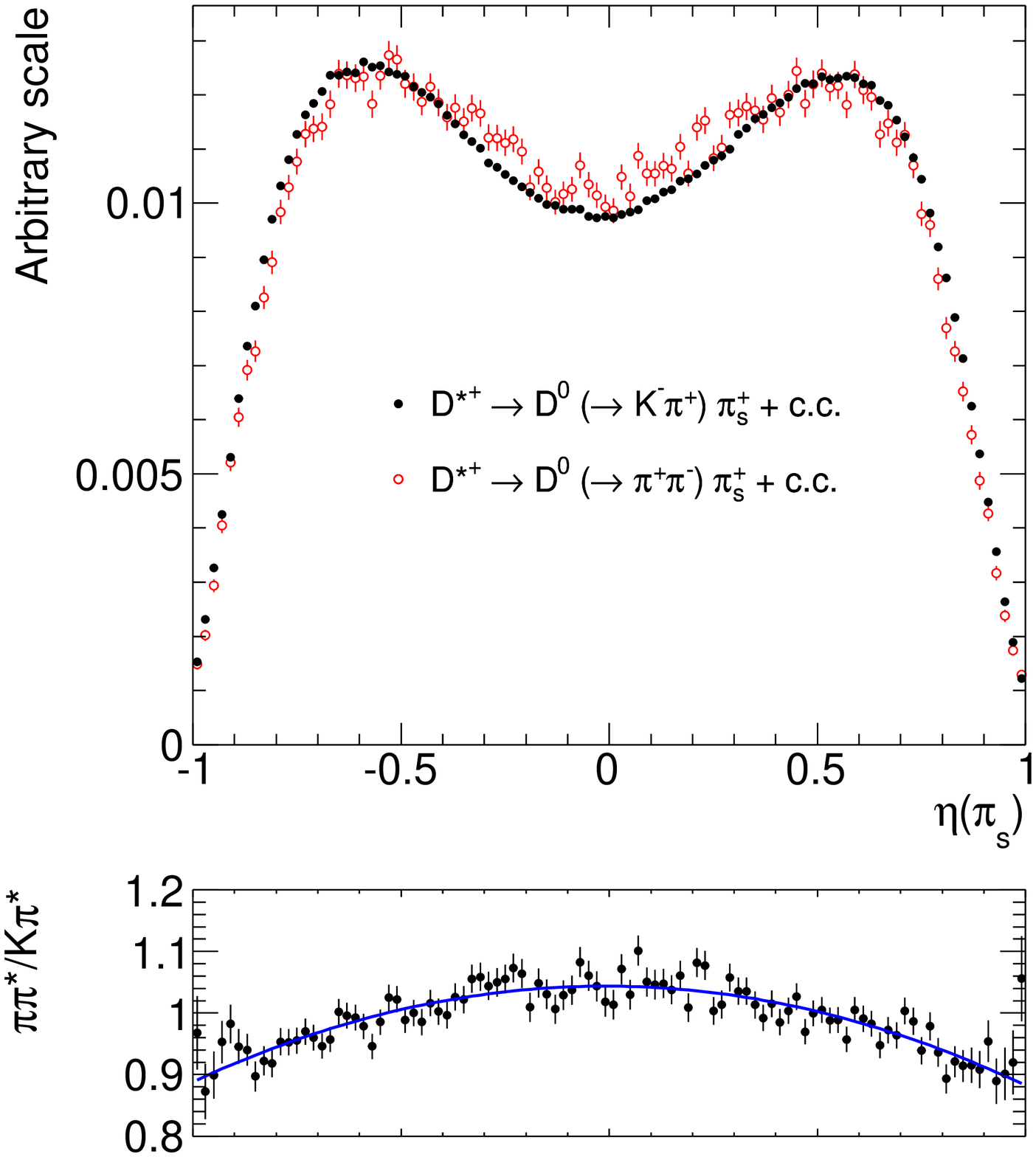}
\put(30,70){(b)} 
\end{overpic}
\begin{overpic}[width=5.9cm,grid=false]{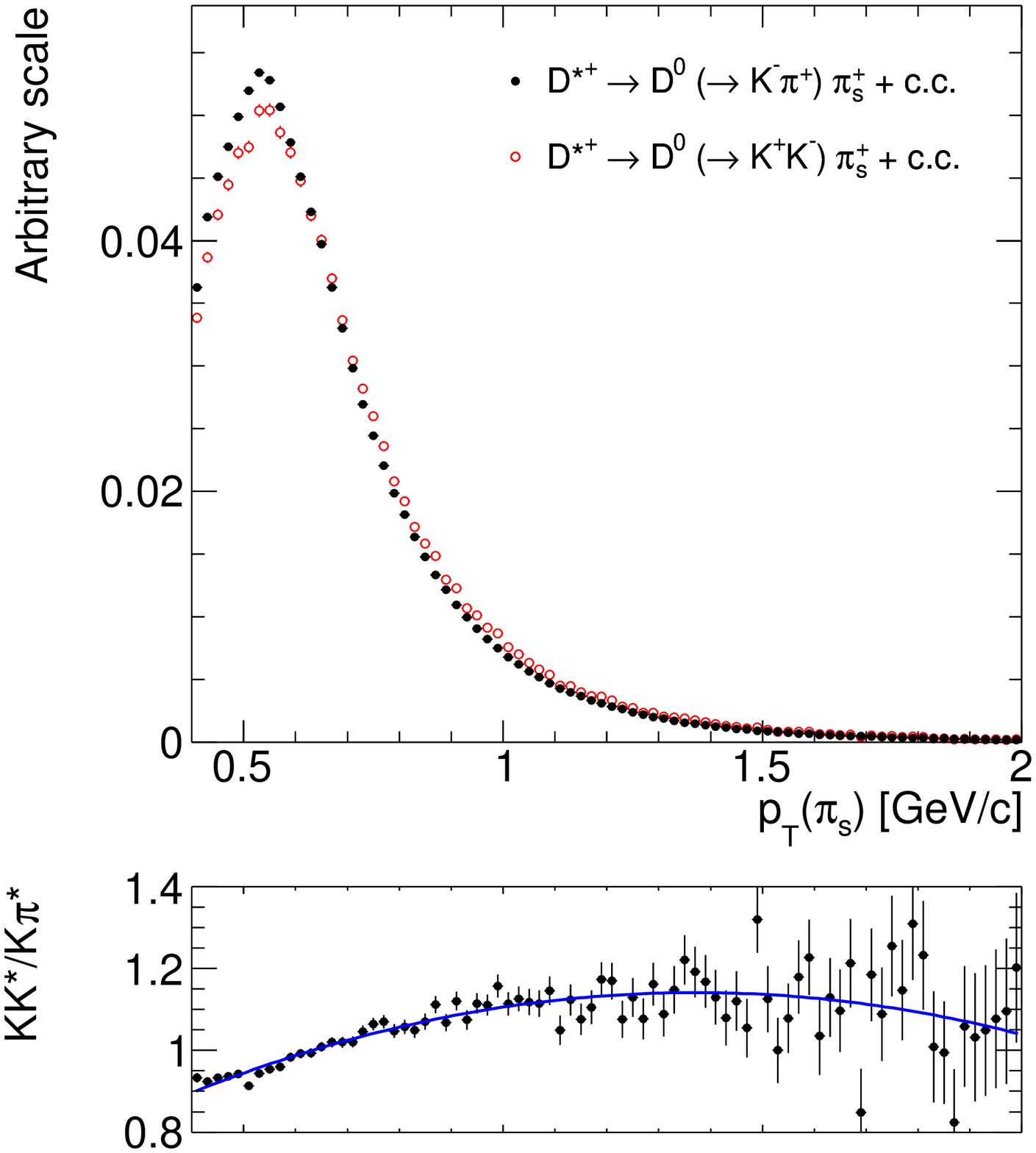}
\put(30,70){(c)} 
\end{overpic}\hfil
\begin{overpic}[width=5.9cm,grid=false]{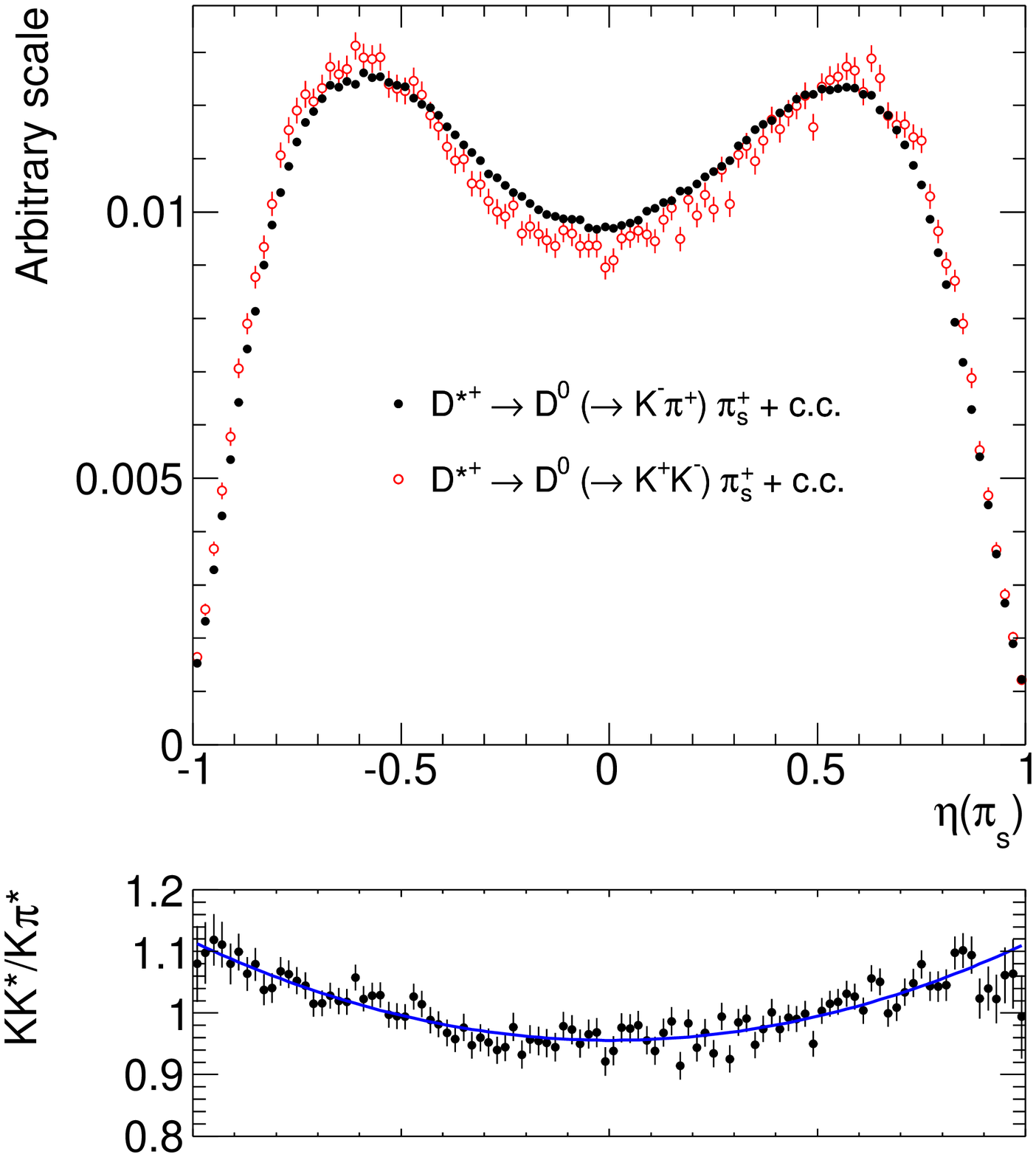}
\put(30,70){(d)} 
\end{overpic}
\begin{overpic}[width=5.9cm,grid=false]{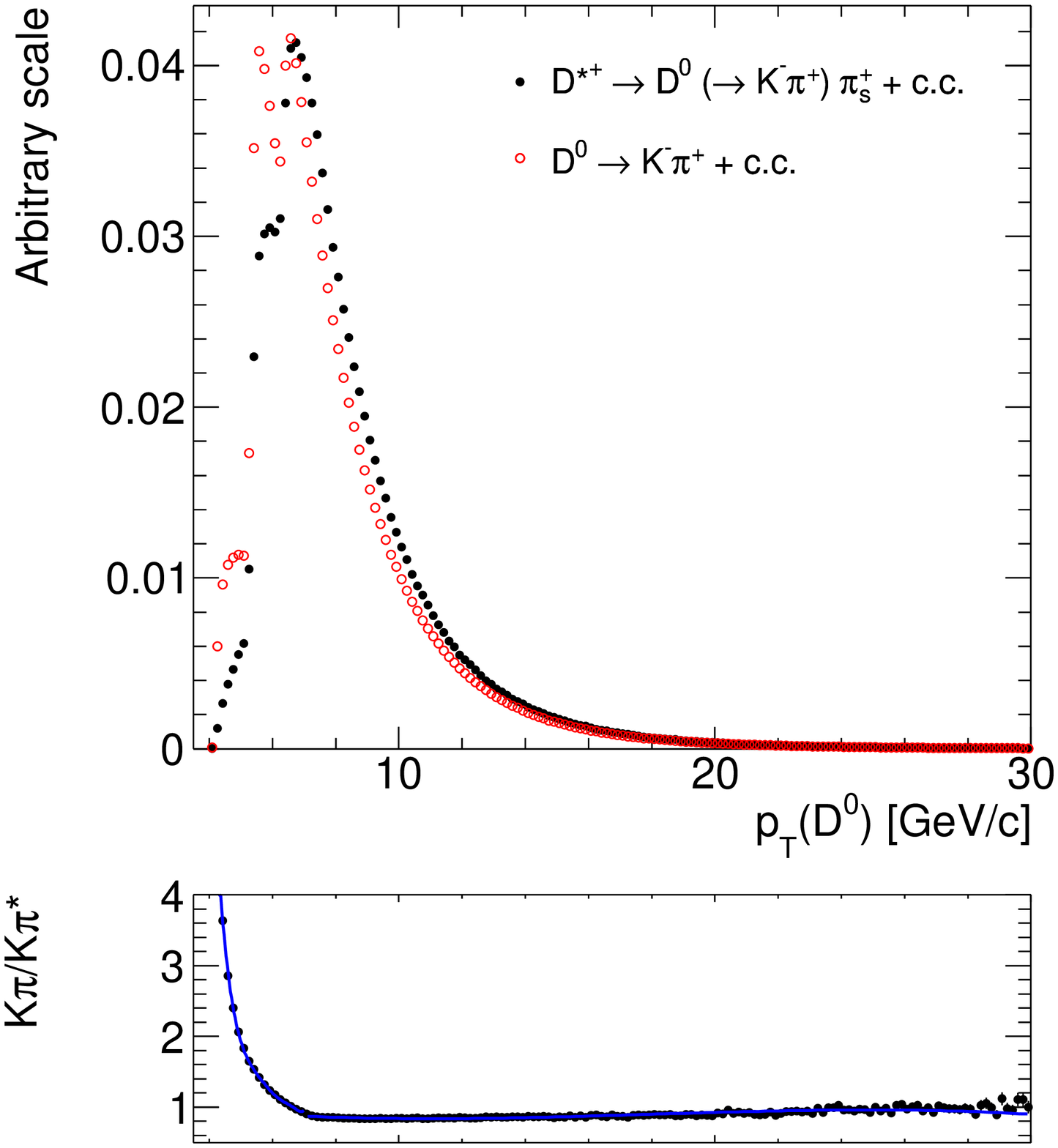}
\put(30,70){(e)} 
\end{overpic}\hfil
\begin{overpic}[width=5.9cm,grid=false]{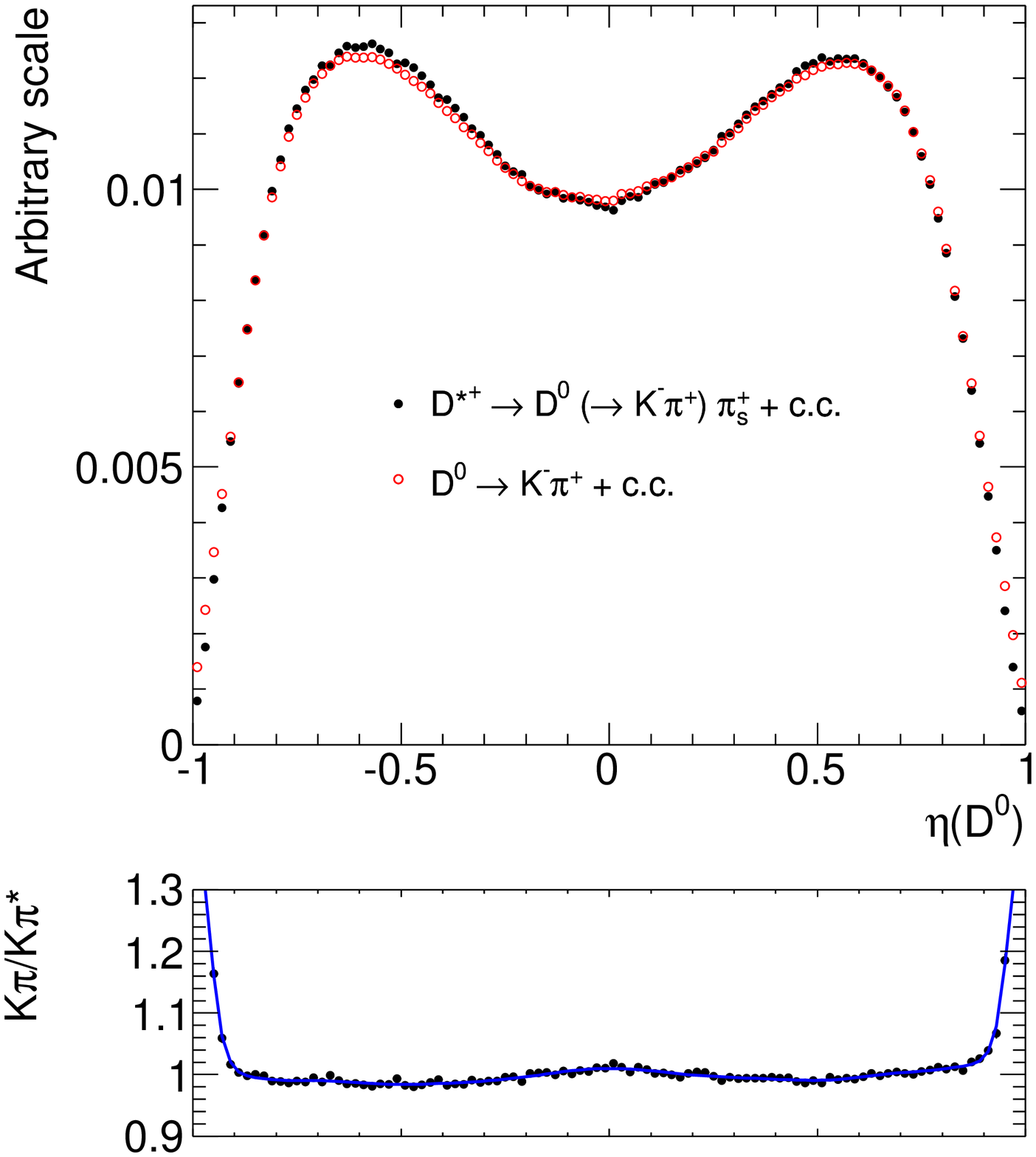}
\put(30,70){(f)} 
\end{overpic}
\caption{Comparison between normalized kinematic distributions of the various tagged and untagged samples used in the analysis: (a), (c) soft pion transverse momentum, and (b),(d) pseudorapidity of $hh^*$ and $K\pi^*$ events; (e) $D^0$ transverse momentum and (f) pseudorapidity of $K\pi$ and $K\pi^*$ events. Tagged distributions are background-subtracted.}\label{fig:rew}
\end{figure*}

We define appropriate sideband regions according to the specific features of each tagged sample (Fig.\ \ref{fig:mass_distr} (a)--(c)). Then we compare background-subtracted distributions for tagged $h^+h^{(')-}$  decays, studying a large set of $\pi_s$ kinematic variables ($p_T$, $\eta$, $\phi$, $d_0$, and $z_0$) \cite{tesi-angelo}.  We observe small discrepancies only in the transverse momentum and pseudorapidity distributions as shown in Fig.~\ref{fig:rew}~(a)--(d). The ratio between the two distributions is used to extract a smooth curve used as a candidate-specific weight. A similar study of $D^0$ distributions for tagged and untagged \DKpi\ decays shows discrepancies only in the distributions of transverse momentum and pseudorapidity  (Fig.~\ref{fig:rew}) which are reweighted accordingly.

\begin{figure}[t]
\centering
\begin{overpic}[width=8.6cm,grid=false]{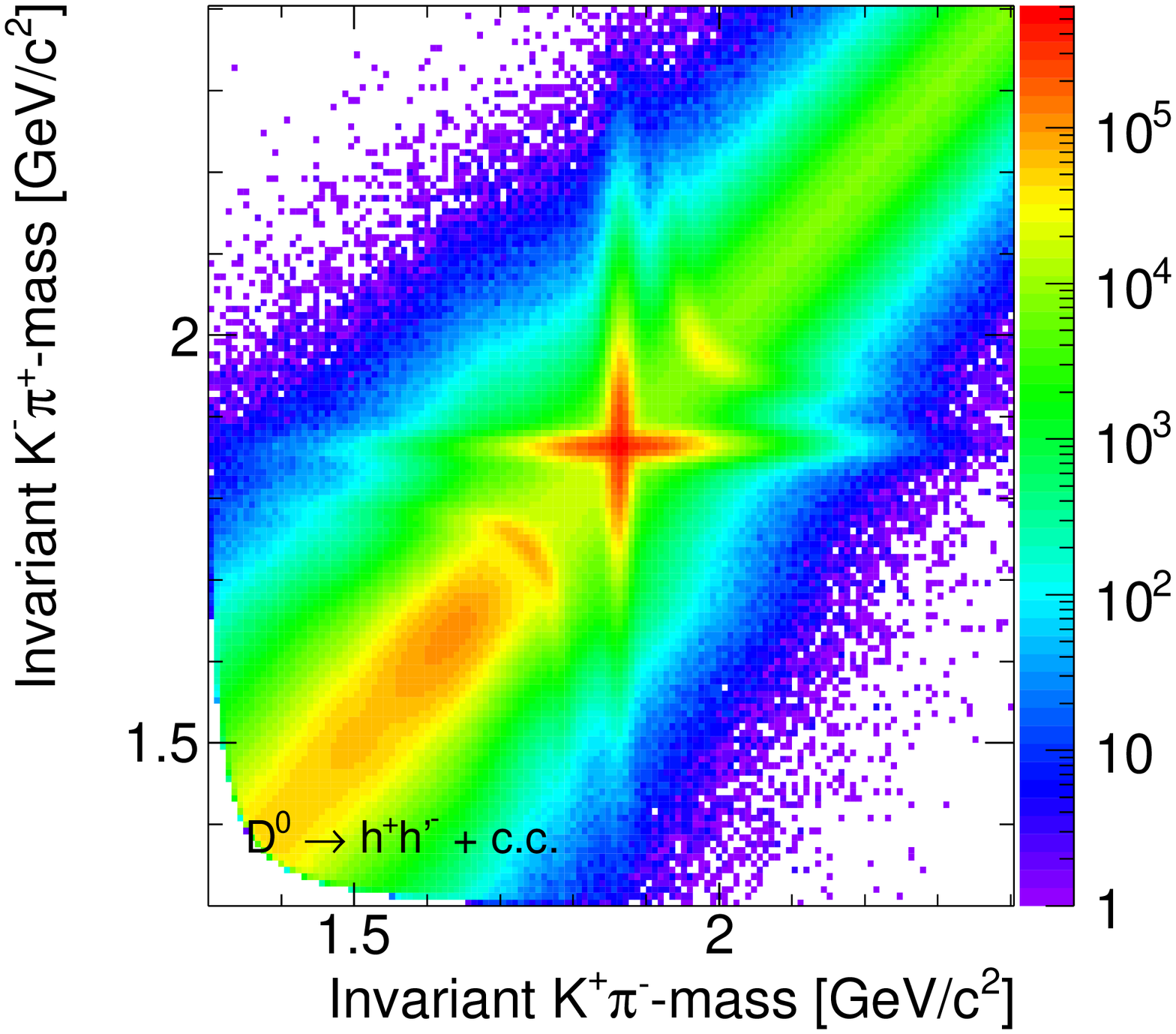}     
\put(30,49){\scriptsize\boldmath $K^-\pi^+$}
\put(53.5,61){\scriptsize \begin{rotate}{90}{\boldmath$K^+\pi^-$}\end{rotate}}
\put(54,58){\scriptsize \begin{rotate}{-43}{\boldmath$\pi^+\pi^-$}\end{rotate}}
\put(41,46){\scriptsize \begin{rotate}{-43}{\boldmath$K^+K^-$}\end{rotate}}
\put(26,22){\scriptsize \begin{rotate}{48}{\bfseries Multi-body}\end{rotate}}
\put(31,22){\scriptsize \begin{rotate}{48}{\bfseries decays}\end{rotate}}
\put(66,64){\scriptsize \begin{rotate}{48}{\bfseries Combinatorics}\end{rotate}}
\end{overpic}
\caption{Distribution of  $K^-\pi^+$--mass as a function of  $K^+\pi^-$--mass for the untagged sample. Note the logarithmic scale on $z$ axis.}\label{fig:mass_2d}
\end{figure}

Background is not subtracted from the distributions of the untagged sample. We simply select decays with $K^{+}\pi^{-}$ or $K^{-}\pi^{+}$--mass within  24~\massmev\ from the known $D^0$ mass, corresponding approximately to a cross-shaped $\pm 3\sigma$ range in the two-dimensional distribution (Fig.~\ref{fig:mass_2d}). The background contamination in this region is about 6\%. 
This contamination has a small effect on the final result. The observed asymmetries show a small dependence on the $D^0$ momentum, because detector-induced charge  asymmetries are tiny at transverse momenta greater than 2.2 GeV/$c$, as required for the $D^0$ decay products. Therefore any small imperfection in the reweighting of momentum spectra between tagged and untagged sample has a limited impact, if any. However, a systematic uncertainty is assessed for the possible effects of non-subtracted backgrounds (see Sec.~\ref{sec:syst}).
All entries in distributions shown in the remainder of this paper are reweighted according to the transverse momentum and pseudorapidity of the corresponding candidates unless otherwise stated.

\section{Determination of observed asymmetries\label{sec:fit}}
The asymmetries between observed numbers of $D^0$ and $\Dbar^0$ signal candidates are determined with fits of the $D^*$ (tagged samples) and $D^0$ (untagged sample)                 
 mass distributions. The mass resolution of the CDF tracker is sufficient to separate the different decay modes of interest. Backgrounds are modeled and included in the fits. In all cases we use a joint binned fit that minimizes a combined $\chi^2$ quantity, defined as  $\chi^2_{\rm tot} = \chi^{2}_{+} + \chi^{2}_{-},$
where $ \chi^{2}_{+}$  and  $\chi^{2}_{-}$ are the individual $\chi^2$ for the $D^0$ and $\Dbar^0$ distributions. 
Because we use copious samples, an unbinned likelihood fit would imply a substantially larger computational load without a significant improvement in statistical resolution.
The functional form that describes the mass shape is assumed to be the same for charm and anti-charm, although a few parameters are determined by the fit independently in the two samples. The functional form of the mass shape for all signals is extracted from simulation and the values of its parameters adjusted for the data.  The effect of this adjustment  is discussed in Sec.~\ref{sec:syst} where a systematic uncertainty is also assessed.

\subsection{Fit of tagged samples}
We extract the asymmetry of tagged samples by fitting the numbers of reconstructed $D^{*\pm}$ events in the $D^0\pi_s^+$ and  $\overline{D}^0\pi_s^-$ mass distribution.
Because all  \Dhh\ modes have the same $D^0\pi_s^+$ mass distribution,  we use a single shape to fit all tagged signals. We also assume that the shape of the background from random pions associated with a real neutral charm particle are the same. Systematic uncertainties due to variations in the shapes are
discussed later in Sec.\ \ref{sec:syst}.\par The general features of the signal distribution are extracted from simulated samples. The model is adjusted and finalized in a fit of the $D^0\pi_s$ mass of copious and pure tagged $K^-\pi^+$ decays. We fit the average histogram of the charm and anti-charm samples, $m = (m_{+}+m_{-})/2$, where   $m_{+}$ is the $D^{*+}$ mass distribution and $m_{-}$ the $D^{*-}$ one. The resulting signal shape is then used in the joint fit to measure the asymmetry between charm and anti-charm signal yields. The signal is described by a Johnson function \cite{johnson} (all functions properly normalized in the appropriate fit range),
\begin{equation}
J(x|\mu,\sigma,\delta,\gamma) =  \frac{e^{-\frac{1}{2}\left[\gamma~+~\delta~\text{sinh}^{-1}\left(\frac{x-\mu}{\sigma}\right)\right]^2}}{\sqrt{1+\left(\frac{x-\mu}{\sigma}\right)^2}}, \nonumber
\end{equation}
that accounts for the asymmetric tail of the distribution, plus two Gaussians, $\gauss(x|\mu,\sigma)$,
for the central bulk:
\begin{align}
\pdf_{\text{sig}}(m|\vec{\theta}_{sig}) =& f_J J(m|m_{D^*}+\mu_J,\sigma_J,\delta_J,\gamma_J) +(1-f_J)  \nonumber \\*
& \times [ f_{G1}\gauss(m|m_{D^*}+\mu_{G1},\sigma_{G1})  \nonumber\\*
& +(1-f_{G1})\gauss(m|m_{D^*}+\mu_{G2},\sigma_{G2}) ]. \nonumber
\end{align}
The signal parameters $\vec{\theta}_{sig}$ include the relative fractions between the Johnson and the Gaussian components;  the shift from the nominal $D^{*\pm}$ mass of the Johnson distribution's core, $\mu_J$,  and the two Gaussians,  $\mu_{G1(2)}$; the widths of the Johnson distribution's core, $\sigma_J$, and the two Gaussians,  $\sigma_{G1(2)}$; and the parameters $\delta_J$  and $\gamma_J$, which determine the asymmetry in the Johnson distribution's tails. For the random pion background we use an empirical shape form,
\begin{equation}
\pdf_{\text{bkg}}(m|\vec{\theta}_\text{bkg}) = \mathscr{B}(m|m_{D^0}+m_\pi,b_\text{bkg},c_\text{bkg}), \nonumber
\end{equation}
with $\mathscr{B}(x|a,b,c) = (x-a)^b e^{-c(x-a)}$ extracted from data by forming an artificial random combination made of a well-reconstructed $D^0$ meson from each event combined with pions from all other events.
The total function used in this initial fit is
\begin{equation}
N_{\text{sig}}\pdf_{\text{sig}}(m|\vec{\theta}_\text{sig}) + N_{\text{bkg}}\pdf_{\text{bkg}}(m|\vec{\theta}_\text{bkg}).\nonumber
\end{equation}
Each fit function is defined only above the threshold value of $m_{D^0}+m_{\pi}$.
\begin{figure}[t]
\centering
\includegraphics[width=8.6cm]{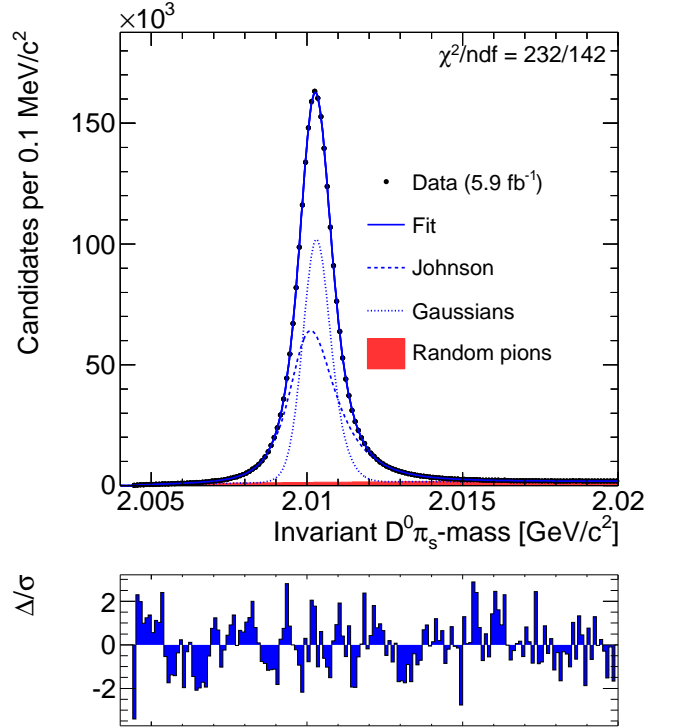}
\caption{Distribution of $D^0\pi_s$ mass of tagged $D^0\to K^-\pi^+$ decays with fit results overlaid. The total fit projection (blue)  is shown along with the double Gaussian bulk (dotted line), the Johnson tail (dashed line) and the background (full hatching).}\label{fig:preliminary-kpi*}
\end{figure}
Figure \ref{fig:preliminary-kpi*} shows the resulting fit which is used to determine the shape parameters for subsequent asymmetry fits. All parameters are free to float in the fit.
\begin{figure*}[t]
\centering
\begin{overpic}[width=5.9cm,grid=false]{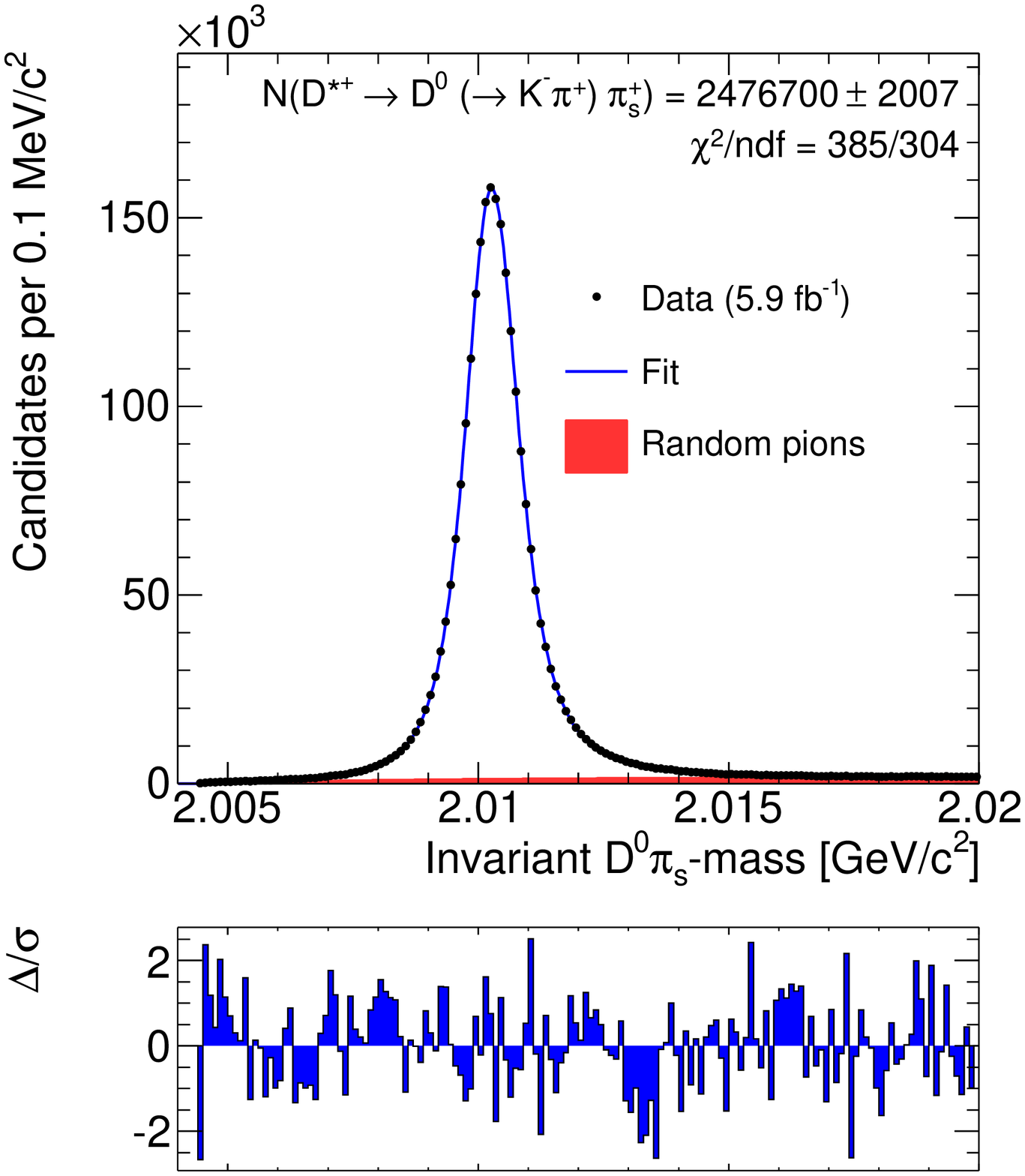}
\put(25,75){(a)} 
\end{overpic}\hfil
\begin{overpic}[width=5.9cm,grid=false]{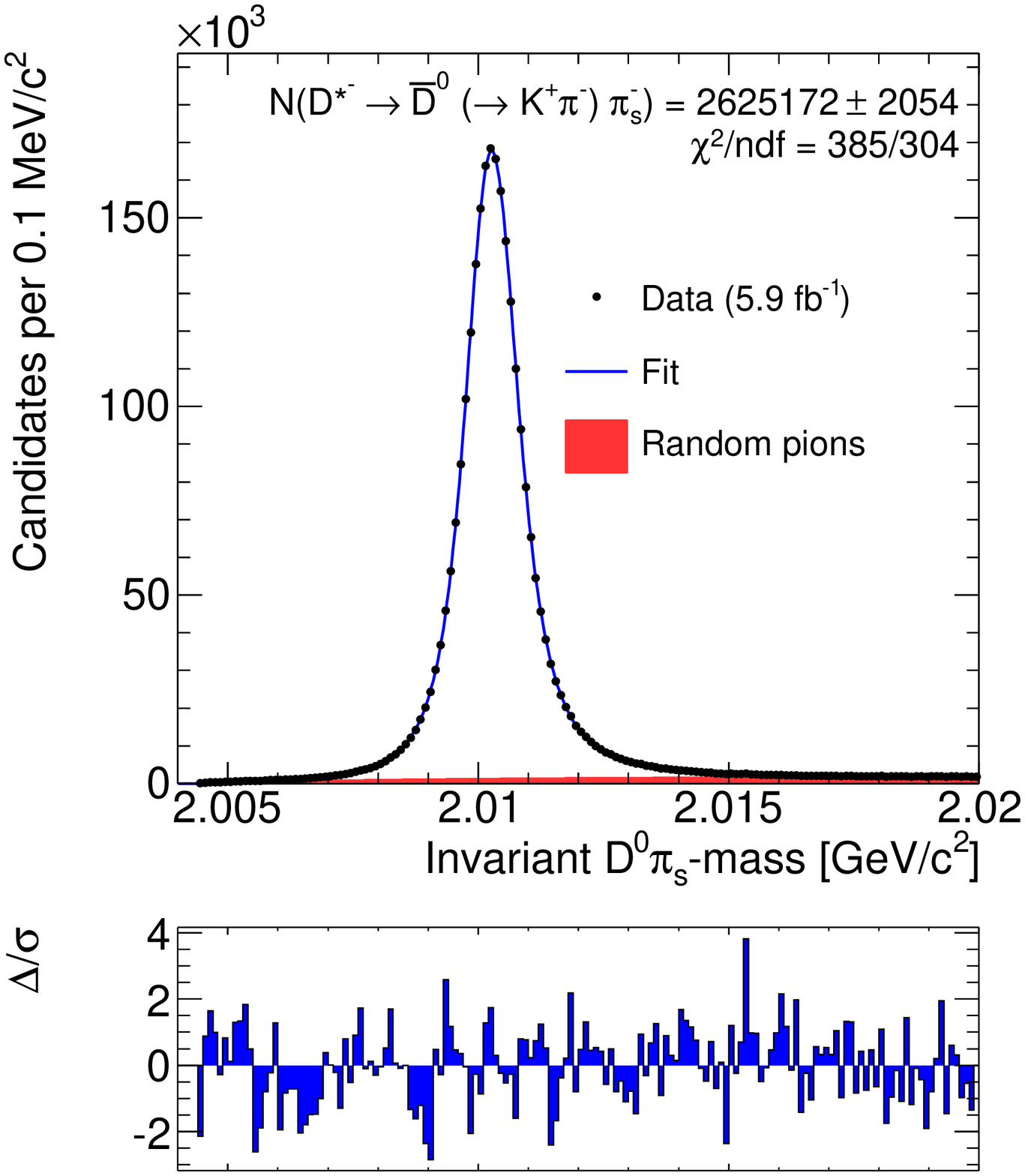}
\put(25,75){(b)} 
\end{overpic}\hfil
\begin{overpic}[width=5.9cm,grid=false]{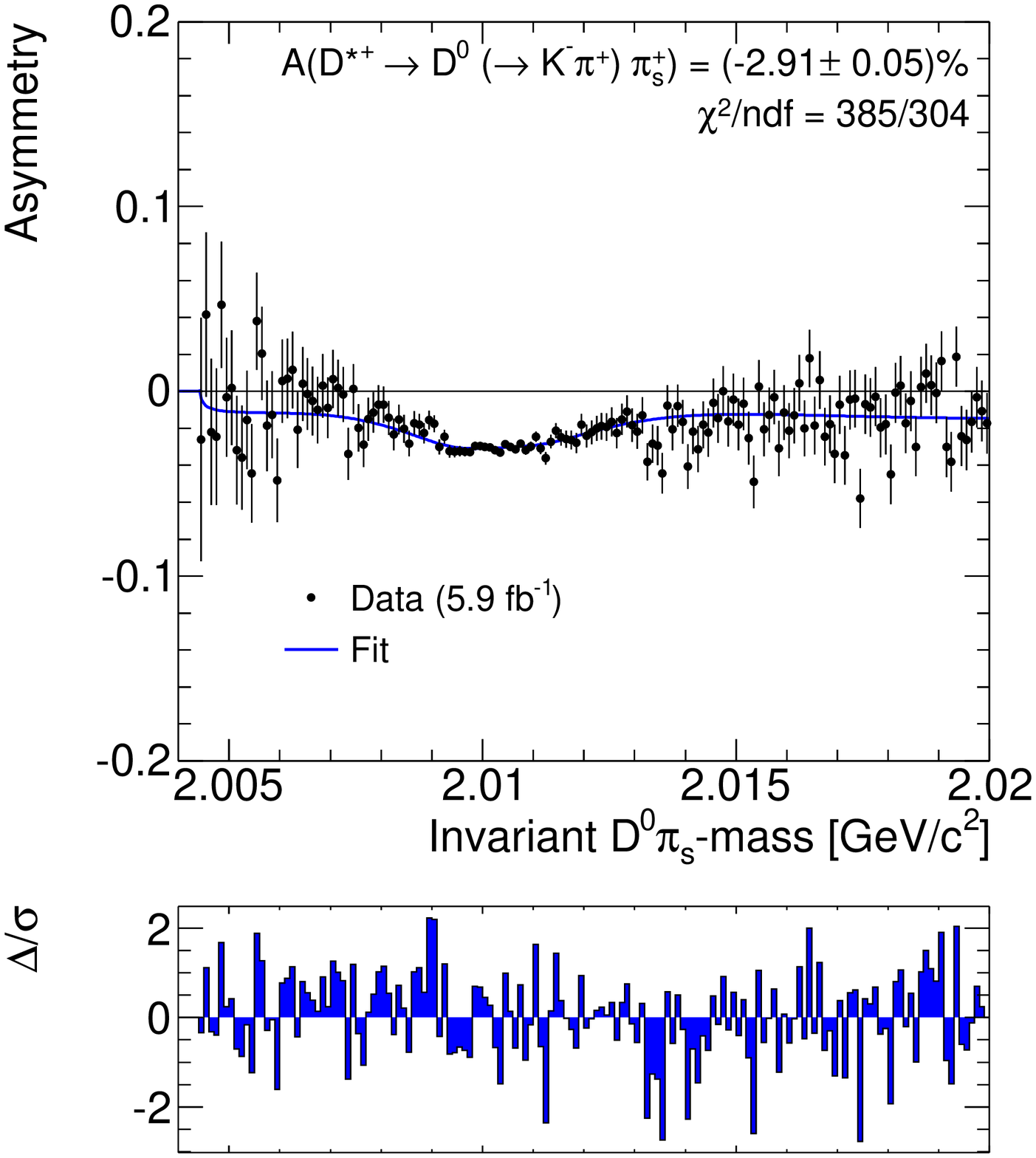}
\put(25,75){(c)}
\end{overpic}
\caption{Results of the combined fit of the tagged $D^0\to K^-\pi^+$ samples. Distribution of $D^{0}\pi_s$ mass for (a) charm, and (b) anti-charm decays,  and (c) asymmetry as a function of the mass. Fit results are overlaid.}\label{fig:acp-kpi*}
\end{figure*}

We then fix the signal parametrization and simultaneously fit the $D^0\pi_s$ mass distributions of $D^{*+}$ and $D^{*-}$ candidates with independent normalizations to extract the asymmetry. The parameter $\delta_J$ varies independently for charm and anti-charm decays. The background shape parameters are common in the two samples and are determined by the fit. Figures~ \ref{fig:acp-kpi*}~(a) and (b) show the projections of this simultaneous fit on the $D^{0}\pi_s$ mass distribution, for the tagged $D^0\to K^-\pi^+$ sample. Figures \ref{fig:acp-kpi*}~(c) shows the projection on the asymmetry distribution as a function of the $D^0\pi_s$ mass. The asymmetry distribution is constructed by evaluating bin-by-bin the difference and sum of the distributions in mass for charm ($m_+$ ) and anti-charm ($m_-$) decays to obtain $A = (m_{+}-m_{-})/(m_{+}+m_{-})$. The variation of the asymmetry as a function of mass indicates whether backgrounds with asymmetries different from the signal are present. As shown by the difference plots at the bottom of Fig.~\ref{fig:acp-kpi*}, the fits  correctly describe  the asymmetry across the whole mass range.\par
We allowed independent $\delta_J$ parameters in the charm and anti-charm samples because the $D^0\pi_s$ mass distribution for $D^{*+}$ candidates has slightly higher tails and a different width than the corresponding distribution for $D^{*-}$ candidates.  The relative difference between the resulting $\delta_J$ values does not exceed $0.5\%$. However, by allowing the parameter $\delta_J$ to vary independently the $\chi^2/$ndf value improves from $414/306$ to $385/304$. We do not expect the source of this difference to be asymmetric background because the difference is maximally visible in the signal region, where the kinematic correlation between $D^0\pi_s$ mass and $\pi_s$ transverse momentum is stronger. Indeed, small differences between  $D^{*+}$ and $D^{*-}$ shapes may be expected because the drift chamber has different resolutions for positive and negative low momentum particles. Independent $\delta_J$ parameters provide a significantly improved description of the asymmetry as a function of $D^0\pi_s$ mass in the signal region (Fig.~\ref{fig:acp-kpi*}~(c)). In Sec.~\ref{sec:syst:rawasy} we report a systematic uncertainty  associated with this assumption. No significant improvement in fit quality is observed when leaving other signal shape parameters free to vary independently for $D^{*+}$ and $D^{*-}$ candidates. 

\begin{figure*}[t]
\centering
\begin{overpic}[width=5.9cm,grid=false]{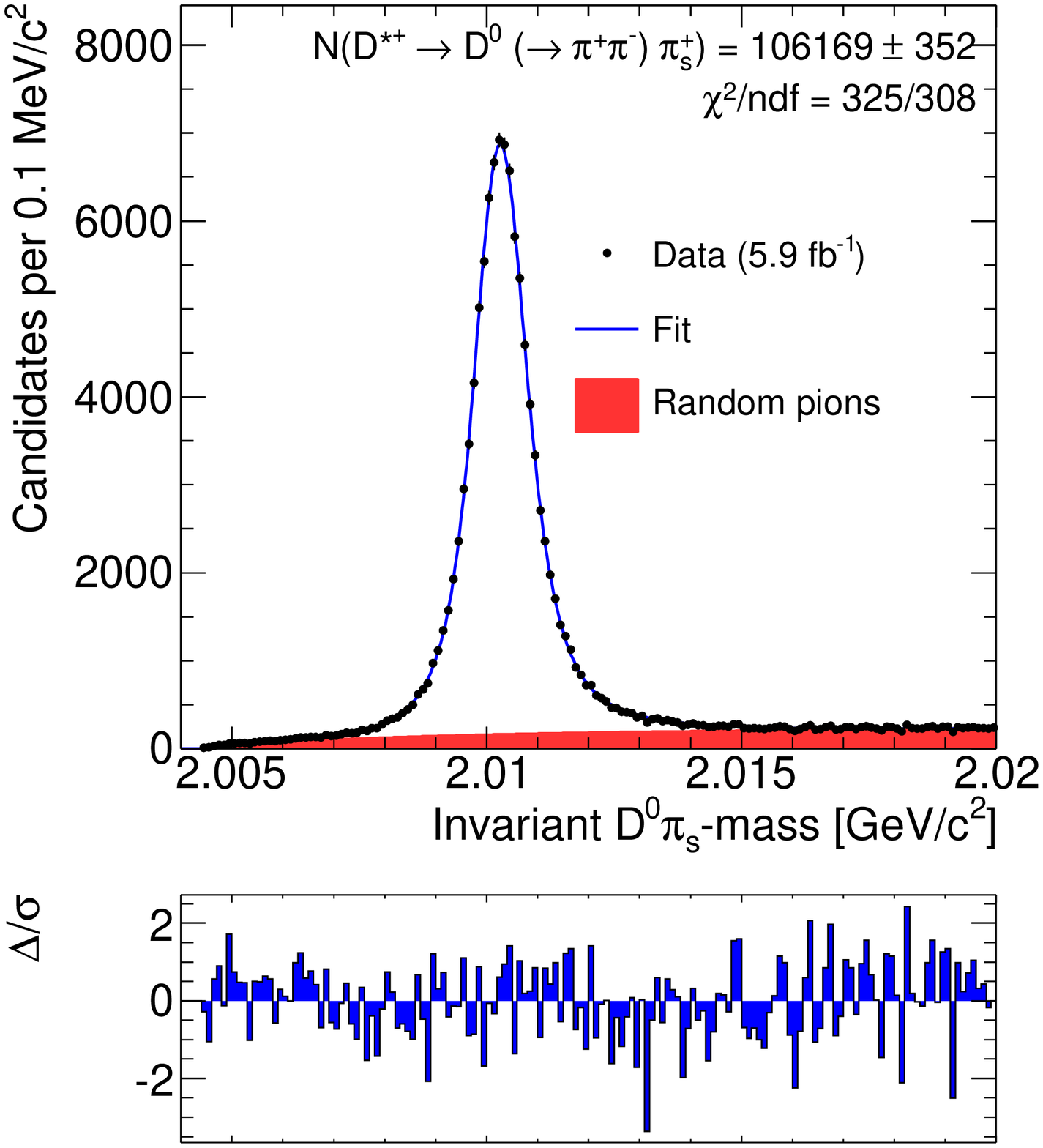}
\put(25,75){(a)}
\end{overpic}\hfil
\begin{overpic}[width=5.9cm,grid=false]{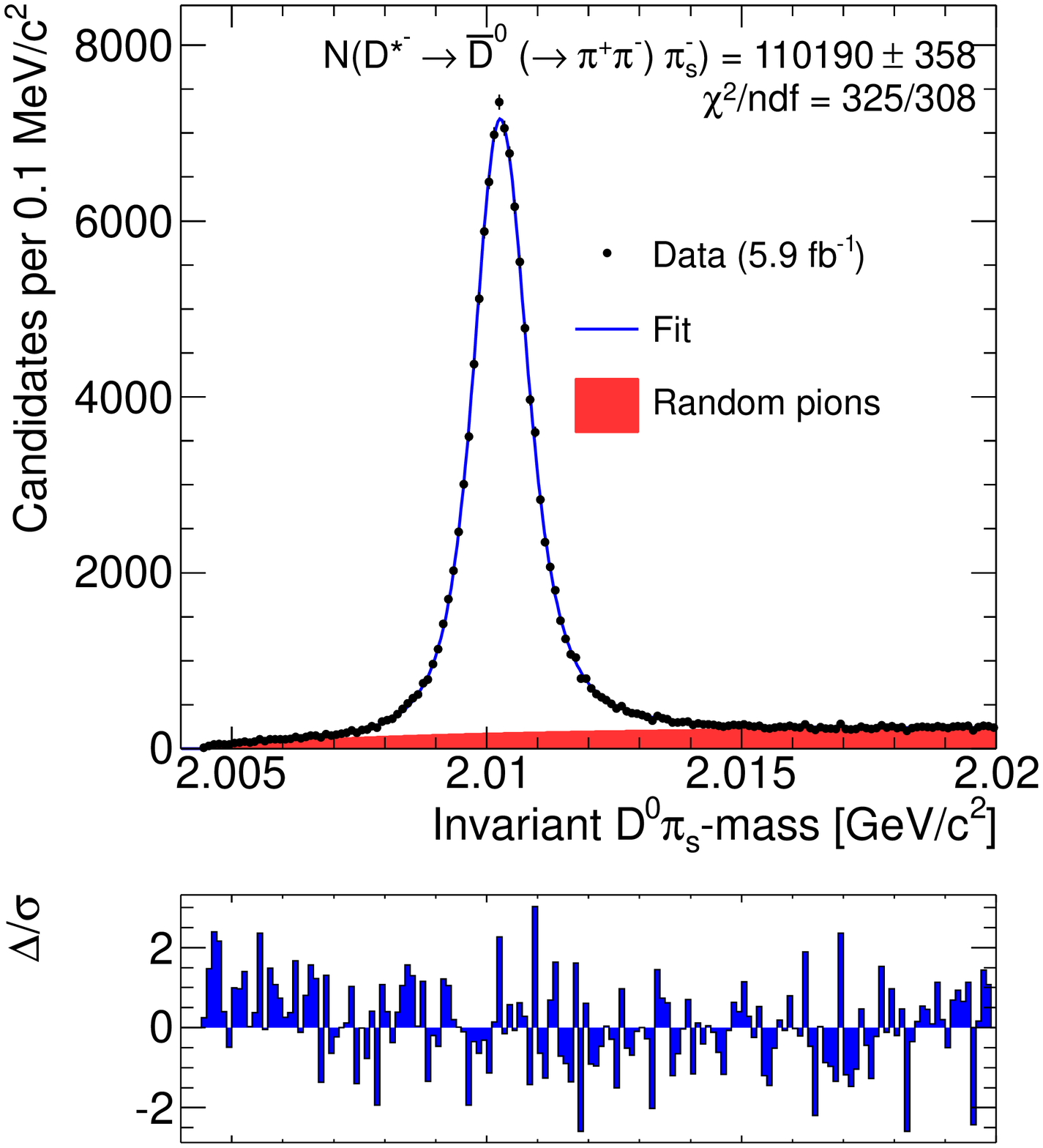}
\put(25,75){(b)}
\end{overpic}\hfil
\begin{overpic}[width=5.9cm,grid=false]{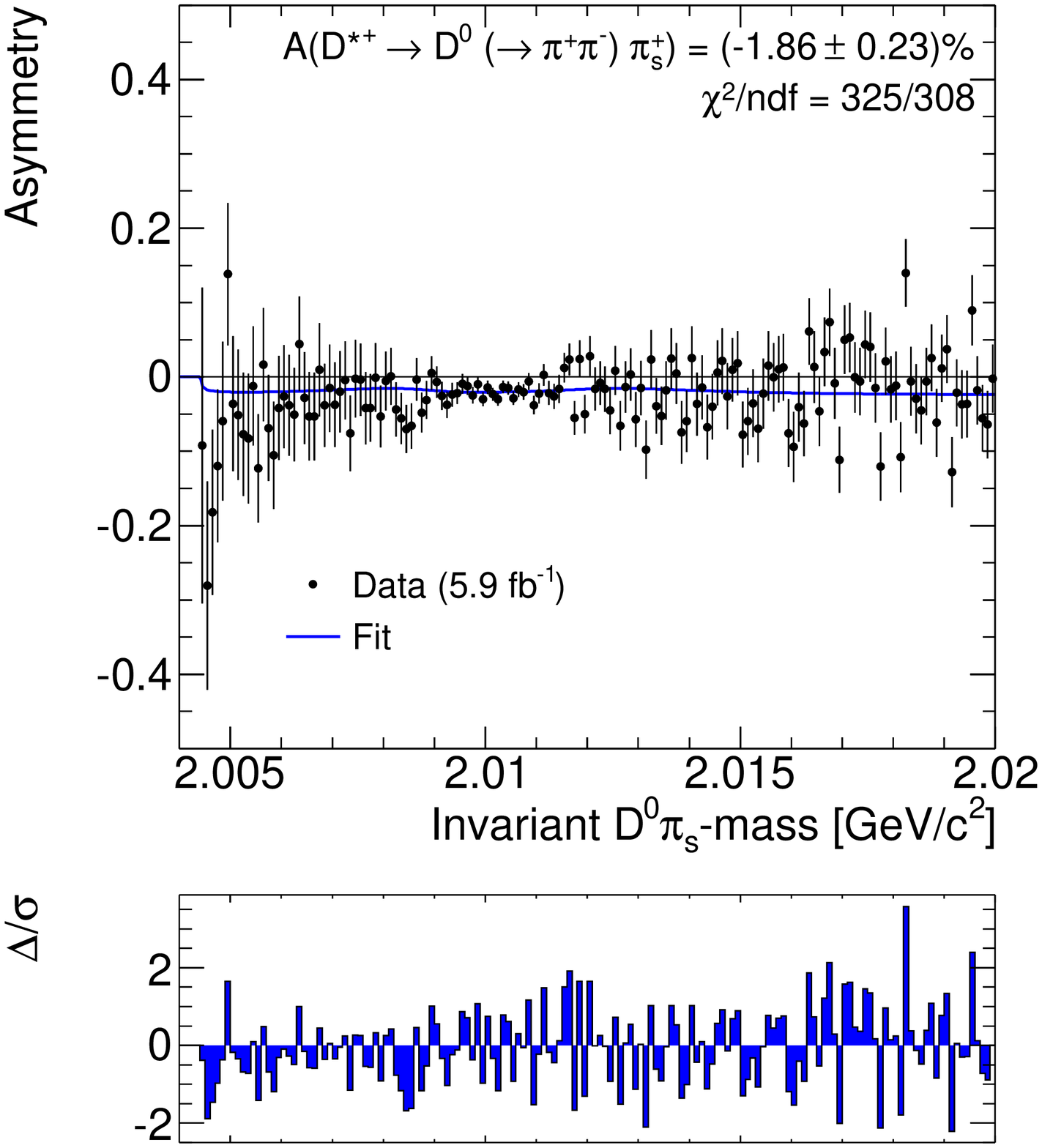}
\put(25,75){(c)}
\end{overpic}\\
\begin{overpic}[width=5.9cm,grid=false]{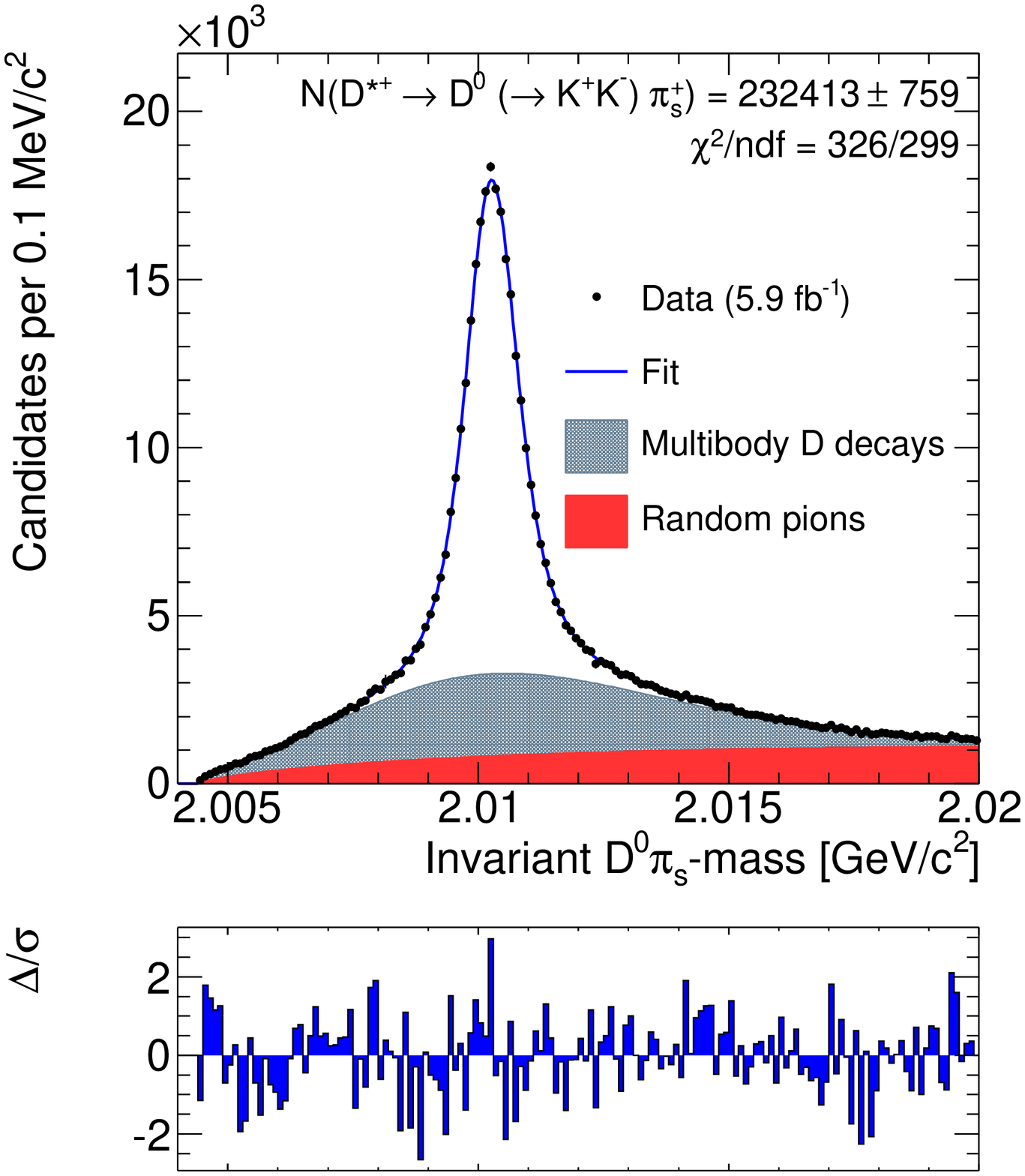}
\put(25,75){(d)}
\end{overpic}\hfil
\begin{overpic}[width=5.9cm,grid=false]{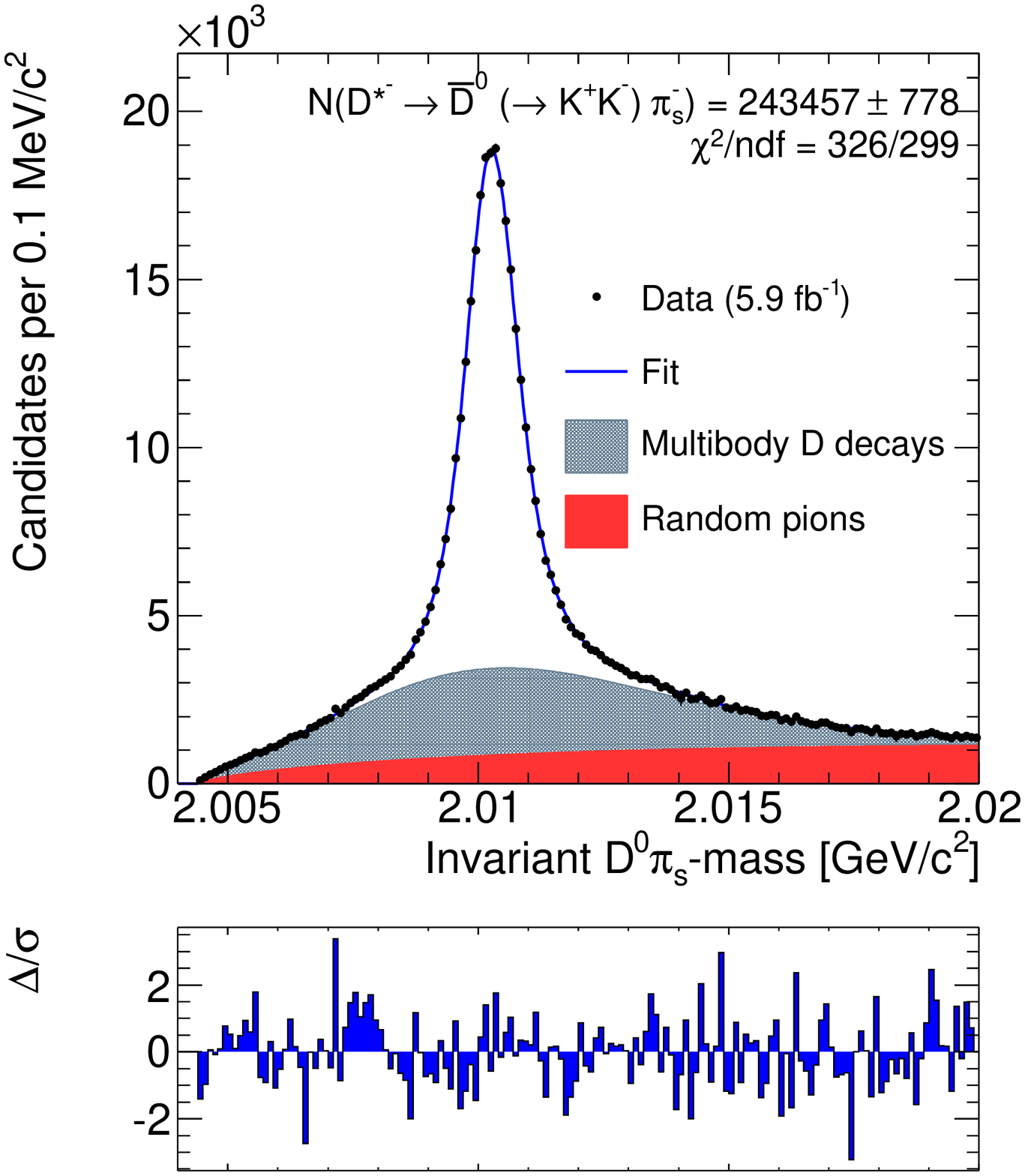}
\put(25,75){(e)}
\end{overpic}\hfil
\begin{overpic}[width=5.9cm,grid=false]{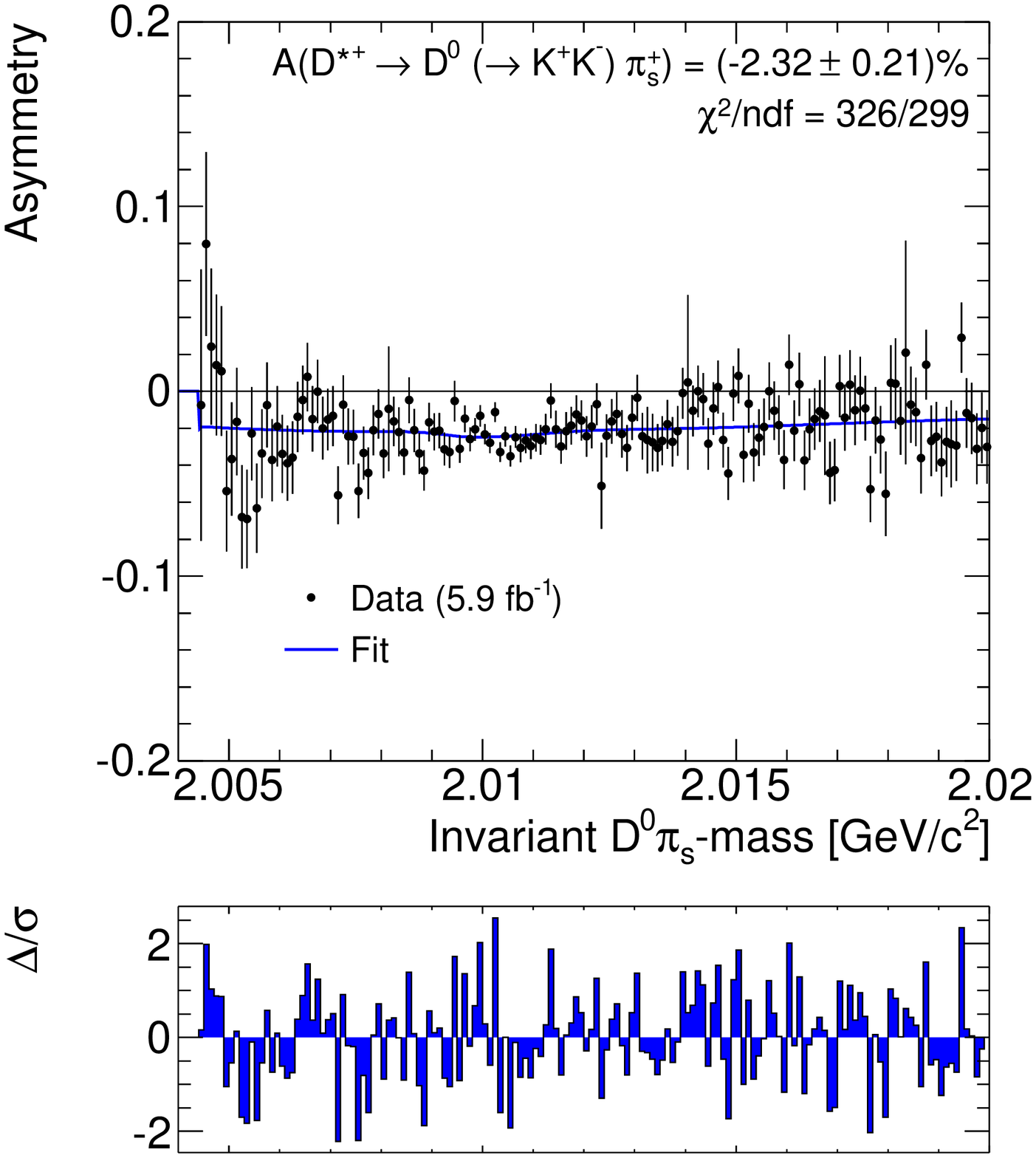}
\put(25,75){(f)}
\end{overpic}
\caption{Results of the combined fit of the tagged $D^0\to \pi^+\pi^-$ and  $D^0\to K^+K^-$ samples. Distribution of $D^{0}\pi_s$ mass for (a), (d) charm and (b), (e) anti-charm decays,  and (c), (f) asymmetry as a function of the mass (c,f). Fit results are overlaid.}\label{fig:fits-hh}
\end{figure*}

The plots in Fig.~\ref{fig:fits-hh} show the fit results for tagged $D^0\to\pi^+\pi^-$ and $D^0\to K^-K^+$ samples. In the $D^0\to K^+K^-$ fit we include an additional component from mis-reconstructed multibody decays. Because signal plus random pion shapes are fixed to those obtained by fitting the tagged $K\pi$ sample (Fig.~\ref{fig:acp-kpi*}), the shape of this additional multibody component is conveniently extracted from the combined fit to data and is described by 
\begin{align}
\pdf_{\text{mbd}}(m|\vec{\theta}_\text{mbd}) =& f_\text{mbd} J(m|m_{D^*}+\mu_\text{mbd},\sigma_\text{mbd},\delta_\text{mbd},\gamma_\text{mbd})\nonumber \\*
+&(1-f_\text{mbd}) \mathscr{B}(m|m_{D^0}+m_{\pi},b_\text{mbd},c_\text{mbd}).\nonumber
\end{align}
The total function used to fit the $KK^*$ sample is then
\begin{equation}
N_{\text{sig}}\pdf_{\text{sig}}(m|\vec{\theta}_\text{sig}) + N_{\text{bkg}}\pdf_{\text{bkg}}(m|\vec{\theta}_\text{bkg})
+N_\text{mbd}\pdf_\text{mbd}(m|\vec{\theta}_\text{mbd}). \nonumber
\end{equation}

We observe the following asymmetries in the three tagged samples:
\begin{align}
\label{eq:tagged-results}
A(\pi\pi^*) &= (-1.86\pm0.23)\%, \nonumber \\*
A(KK^*) &= (-2.32\pm0.21)\%,  \\*
A(K\pi^*) &= (-2.910\pm0.049)\% \nonumber.
\end{align}

\subsection{Fit of the untagged sample\label{sec:fit_untag_dkpi}}
In untagged $K\pi$ decays no soft pion is associated with the neutral charm meson to form a $D^*$ candidate so there is no identification of its charm or anti-charm content. We infer the flavor of the neutral charm meson on a statistical basis using the mass resolution of the tracker and the quasi--flavor-specific nature of neutral charm decays into $K\pi$ final states. The role of mass resolution is evident in Fig.~\ref{fig:mass_2d}, which shows the distribution of $K^-\pi^+$ mass as a function of $K^+\pi^-$ mass for the sample of untagged \Dhh\ decays. The cross-shaped structure at the center of the plot is dominated by $K\pi$ decays. In each mass projection the narrow component of the structure is due to decays where the chosen $K\pi$ assignment is correct. The broader component is due to decays where the  $K\pi$ assignment is swapped. In the momentum range of interest, the observed widths of these two components differ by roughly an order of magnitude. Because of the CKM hierarchy of couplings, approximately 99.6\% of neutral charm decays into a $K^-\pi^+$ final state are from Cabibbo-favored decays of $D^0$ mesons, with only 0.4\% from the doubly-suppressed decays of $\overline{D}^0$ mesons, and vice versa for $K^+\pi^-$ decays.
Therefore, the narrow (broad) component in the $K^-\pi^+$ projection is dominated by $D^0$ ($\overline{D}^0$) decays. Similarly,  the narrow (broad) component in the $K^+\pi^-$ projection is dominated by $\overline{D}^0$ ($D^0$) decays. \par

We extract the asymmetry between charm and anti-charm decays in the untagged sample from a simultaneous binned fit of the $K^+\pi^-$ and $K^-\pi^+$ mass distributions in two independent subsamples. We randomly divide the untagged sample into two independent subsamples, equal in size, whose events were collected in the same data-taking period (``odd'' and ``even'' sample). We arbitrarily choose to reconstruct the \piK\ mass for candidates of the odd sample and the \Kpi\ mass for candidates of the even sample. 
In the odd sample the decay \DKpi\ is considered ``right sign'' (RS) because it is reconstructed with proper mass assignment.  In the even sample it is considered a ``wrong sign'' (WS) decay, since it is reconstructed  with swapped mass assignment. The opposite holds for the \aDKpi\ decay.
The shapes used in the fit are the same for odd and even samples. The fit determines  the number of \DKpi\ (RS decays) from the odd sample and the number of \aDKpi\ (RS decays) from the even sample thus determining the asymmetry.
We split the total untagged sample in half to avoid the need to account for correlations.  The reduction in statistical power has little practical effect since half of the untagged $K\pi$ decays  are still 30 (67) times more abundant than the tagged $K^+K^-$ ($\pi^+\pi^-$) decays, and the corresponding  statistical uncertainty gives a negligible contribution to the uncertainty of the final result.\par 
The mass shapes used in the combined fit of the untagged sample are extracted from simulated events and adjusted by fitting the $K\pi$ mass distribution in data. All functions described in the following are properly normalized when used in fits.
The mass line shape of right-sign decays is parametrized using the following analytical expression:
\begin{align}
\pdf_{\rm RS}(m|\vec{\theta}_{\rm RS})  =&   f_{{\rm bulk}} [f_1 \gauss(m | m_{D^{0}}+\delta_{1},\sigma_1) \nonumber \\*
 &\quad + (1- f_1) \gauss(m|m_{D^{0}}+\delta_{2},\sigma_2) ] \nonumber  \\*
 &+ (1-f_{{\rm bulk}}) \tail(m | b,c,m_{D^{0}}+\delta_{1}), \nonumber
\end{align}
where
\begin{equation}
\tail(m|b,c,\mu) =   e^{b(m-\mu)} {\rm Erfc}(c(m-\mu)), \nonumber
\end{equation}
with ${\rm Erfc}(x) = (2/\sqrt{\pi})\int^{+\infty}_{x} e^{-t^{2}}dt$.
We use the sum of two Gaussians to parametrize the bulk of the distribution.  The function $\tail(m;b,c,\mu)$ 
describes the lower-mass tail due to the soft photon emission. The parameter 
$f_{{\rm bulk}}$ is the relative contribution of the double Gaussian. 
 The parameter $f_{1}$ is the fraction of dominant Gaussian, relative to the sum of the two Gaussians. 
The parameters $\delta_{1(2)}$ are possible shifts in mass  from the known $D^0$ mass \cite{pdg}. 
 Because the soft photon emission makes the mass distribution asymmetric, the means of the Gaussians cannot be assumed to be the same.
Therefore $m_{D^0}$ is fixed in the parametrization while $\delta_{1(2)}$ are determined by the fit.
The mass distribution of wrong-sign decays,  $\pdf_{\rm WS}(m;\vec{\theta}_{\rm WS})$,  is 
parametrized using the same functional form used to model RS decays.
The mass distribution of $D^0 \to \pi^+\pi^-$ decays is modeled using the following functional form:
\begin{align}
\pdf_{\pi\pi}(m|\vec{\theta}_{\pi\pi})  =&   f_{{\rm bulk}} [ f_1 \gauss(m|m_{0}+\delta_{1},\sigma_1) +  \nonumber \\*
 &\qquad (1- f_1) \gauss(m|m_{0}+\delta_{2},\sigma_2) ] \nonumber \\*
 &+  f_{t1} \tail(m|b_1,c_1,m_{1}) \nonumber \\*
 &+  (1-f_{{\rm bulk}}-f_{t1}) \tail(m|b_2,c_2,m_{2}).\nonumber
\end{align}
The bulk of the distribution is described by two Gaussians. Two tail functions $\tail(m;b,c,\mu)$ are added for
the low- and high-mass tails due to soft photon emission and incorrect mass assignment, respectively.
The shifts in mass, $\delta_{1(2)}$,  from the empirical value of the mass of $\pi\pi$ decays assigned the $K\pi$  mass, $m_{0}=1.96736~\massgev$, 
 are free to vary.
The mass distributions of the partially reconstructed multibody charm decays and combinatorial background are modeled using decreasing exponential functions with coefficients  $b_{\rm mbd}$ and $b_{\rm comb}$, respectively.

The function used in the fit is then
\begin{align}
&N_{\rm RS} \pdf_{\rm RS}(m|\vec{\theta}_{\rm RS})  + N_{\rm WS}\pdf_{\rm WS}(m|\vec{\theta}_{\rm WS}) \nonumber \\*
&+ N_{\pi\pi} \pdf_{\pi\pi}(m|\vec{\theta}_{\pi\pi})  + N_{\rm mbd}\pdf_{{\rm mbd}}(m|b_{\rm mbd})  \nonumber \\*
&+ N_{\rm comb}\pdf_{{\rm comb}}(m|b_{\rm comb}).\nonumber
\end{align}
where $N_{\rm RS}$, $N_{\rm WS}$, $N_{\pi\pi}$, $N_{\rm mbd}$,  $N_{\rm comb}$ are the event yields for right-sign decays, wrong-sign decays, $D^0 \to \pi^+\pi^-$ decays, partially reconstructed decays,  and combinatorial background, respectively.

\begin{figure}[!ht]
\centering
\includegraphics[width=8.6cm]{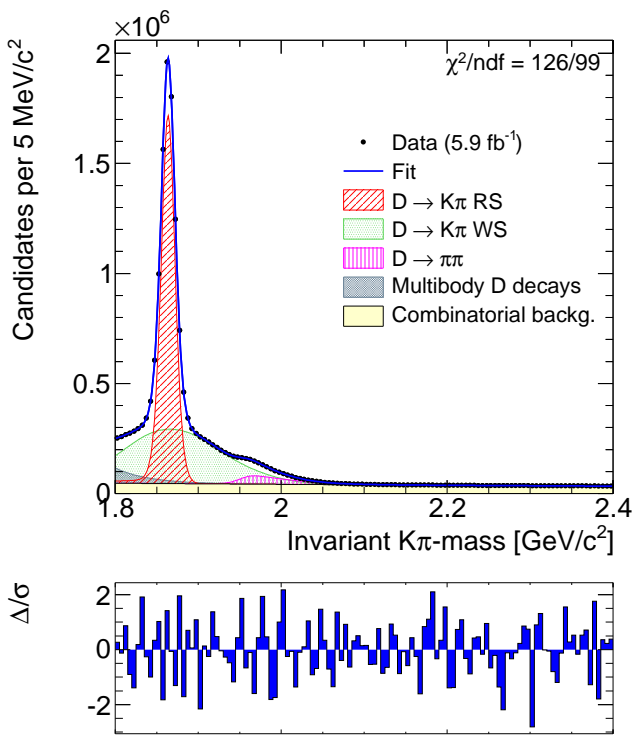}
\caption{Average ($m$) of the distribution of $K^+\pi^-$ mass in the even sample and $K^-\pi^+$ mass in the odd sample with fit projections overlaid.}
\label{fig:fit_mean}
\end{figure} 

The mass is fit in the range $1.8 < m <  2.4~\massgev$ to avoid the need for modeling most of the partially reconstructed charm meson decays. The ratio $N_{\rm RS}/N_{\rm mbd}$ and the parameter $b_{\rm mbd}$ are fixed from simulated inclusive $D^{0}$ and $D^{+}$ decays.  The contamination from partially reconstructed $D^{+}_{s}$ decays is negligible for masses greater that 1.8~\massgev.
The result of the fit to the distribution averaged between odd and even samples is shown in Fig.~\ref{fig:fit_mean}. In this preliminary fit we let vary the number of events in each of the various components, the parameters  of the two Gaussians describing the bulk of the $D^0\to h^+h'^-$ distributions, and the slope of the combinatorial background $b_{\rm comb}$. We assume that the small tails are described accurately enough by the simulation. This preliminary fit is used to extract all shape parameters that will be fixed in the subsequent combined fit for the asymmetry.\par
Odd and even samples are fitted simultaneously using the same shapes for each component to determine the asymmetry of RS decays. Because no asymmetry in $D^0 \to \pi^+\pi^-$ decays and combinatorial background is expected by construction, we include the following constraints: $N^{+}_{\pi\pi}=N^{-}_{\pi\pi}$ and $N^{+}_{\rm comb}=N^{-}_{\rm comb}$. The parameters  $N^{+}_{\rm RS}$, $N^{-}_{\rm RS}$, $N^{+}_{\rm WS}$, $N^{-}_{\rm WS}$, $N^{+}_{\rm mbd}$ and $N^{-}_{\rm mbd}$ are determined by the fit independently in the even and odd  samples. 
\begin{figure*}[!ht]
\centering
\begin{overpic}[width=5.9cm]{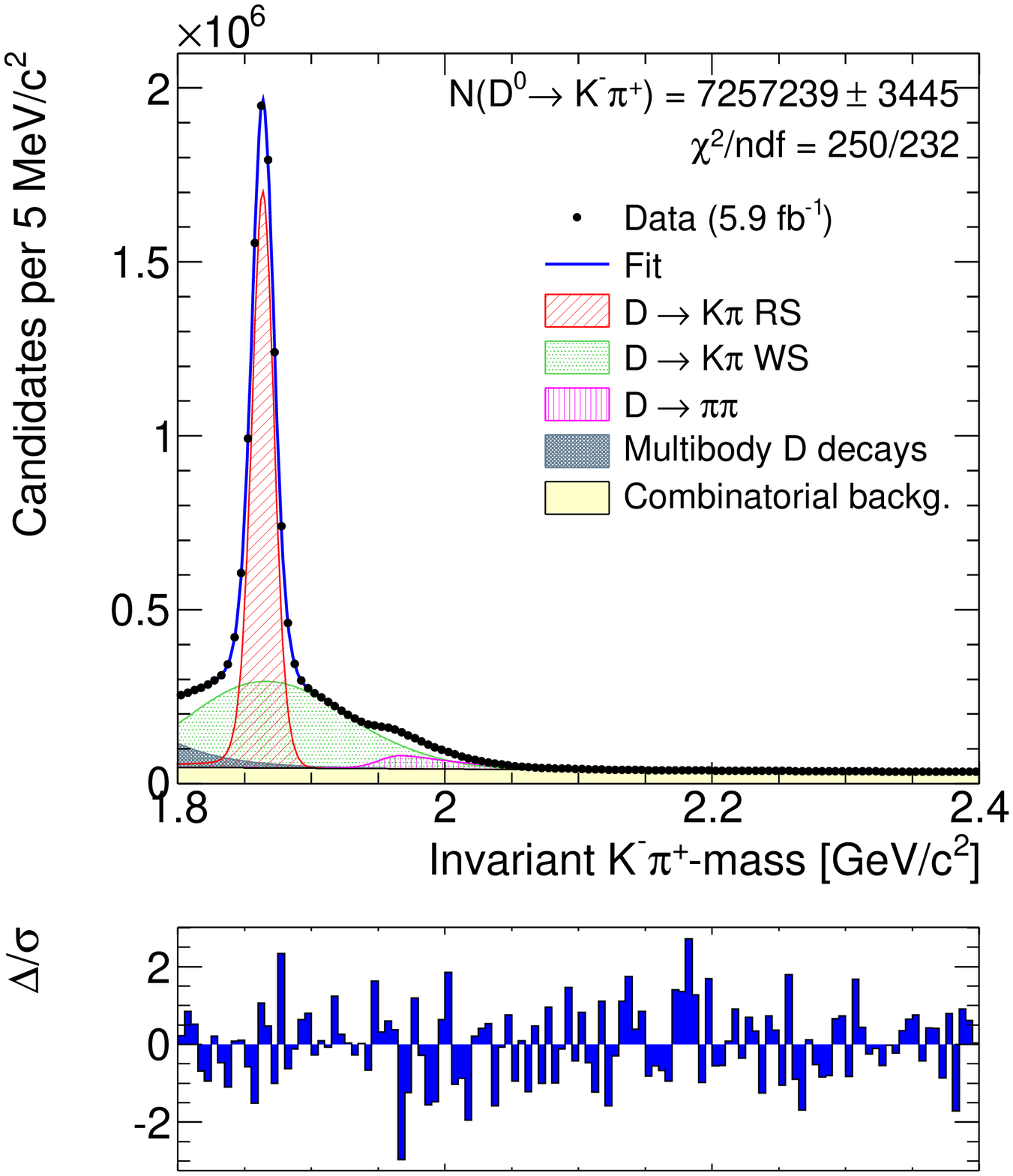}
\put(27,80){(a)}					        	  
\end{overpic}\hfil
\begin{overpic}[width=5.9cm]{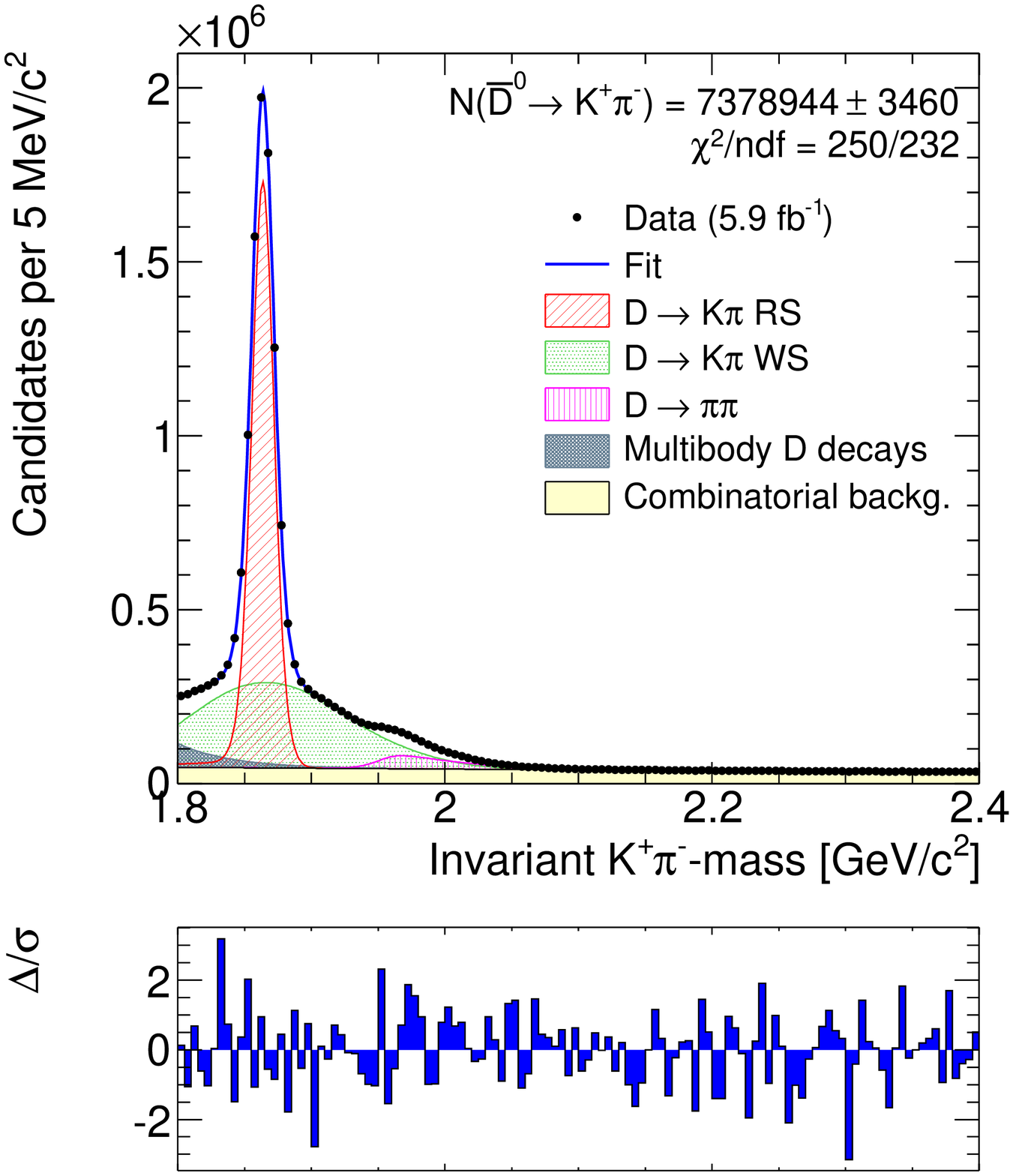}
\put(27,80){(b)}
\end{overpic}\hfill
\begin{overpic}[width=5.9cm]{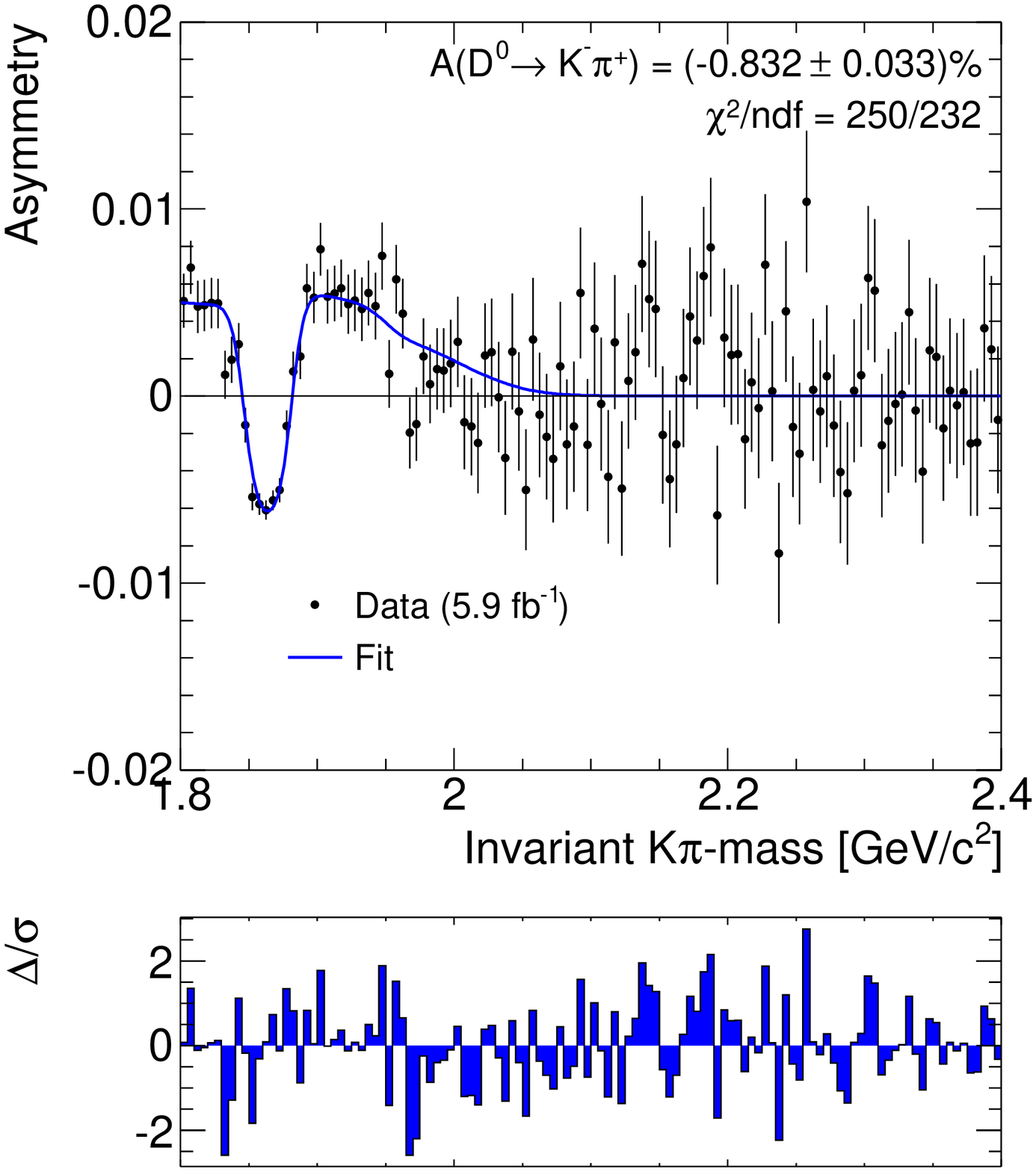}
\put(27,80){(c)}					        	  
\end{overpic}
\caption{Results of the combined fit of the untagged $D^0\to K^-\pi^+$ sample. Distribution of $D^{0}\pi_s$ mass for (a) charm, and (b) anti-charm decays,  and (c) asymmetry as a function of the mass. Fit results are overlaid.}
\label{fig:central_fit_proj}
\end{figure*}
Figures \ref{fig:central_fit_proj}~(a) and (b) show the fit projections for odd and even samples. Figure \ref{fig:central_fit_proj}~(c) shows the projection of the simultaneous fit on the asymmetry as a function of the $K\pi$ mass. The observed asymmetry for the $D^0 \to K^-\pi^+$~RS decays is 
\begin{equation}
\label{eq:untagged-results}
A(K\pi)=(-0.832 \pm 0.033)\%.
\end{equation} 

\section{Systematic uncertainties\label{sec:syst}}
The measurement strategy is designed to suppress systematic uncertainties.  However, we consider a few residual sources that can impact the results: approximations in the suppression of detector-induced asymmetries; production asymmetries; contamination from secondary $D$ mesons; assumptions and approximations in fits, which include 
specific choice of analytic shapes,  differences between distributions associated with charm and anti-charm decays, and contamination from unaccounted backgrounds; and, finally, assumptions and limitations of kinematic reweighting.

Most of the systematic uncertainties are evaluated by modifying the fit functions to include systematic variations and repeating the fits to data. The differences between results of modified fits and the central one are used as systematic uncertainties. This usually overestimates the observed size of systematic uncertainties, which include an additional statistical component. However, the additional uncertainty is negligible, given the size of the event samples involved. Sources of systematic uncertainty are detailed below. A summary of the most significant uncertainties is given in Table~\ref{tab:syst}.

\subsection{Approximations in the suppression of detector-induced effects}
\label{sec:sys-approx}
We check the reliability of the cancellation of all detector-induced asymmetries on simulated samples as described in Appendix \ref{sec:mcvalidation}. The analysis is repeated on several statistical ensembles in which we introduce known \CP--violating asymmetries in the $D^0\to h^+h^{(')-}$ decays and instrumental effects (asymmetric reconstruction efficiency for positive and negative soft pions and kaons) dependent on a number of kinematic variables (e.g.,  transverse momentum). These studies constrain  the size of residual instrumental effects that might not be fully cancelled by our method of linear subtraction of asymmetries. They also assess the impact of possible correlations between reconstruction efficiencies of $D^0$ decay-products and the soft pion, which are assumed negligible in the analysis. We further check this assumption on data  by searching for any variation of the observed asymmetry as a function of the proximity between the soft pion and the charm meson trajectories. No variation is found. \par Using the results obtained with realistic values for the simulated effects, we assess a $\Delta\Acp(hh)=0.009\%$ uncertainty. This corresponds to the maximum shift, increased by one standard deviation, observed in the results,  for true \CP--violating asymmetries in input ranging from $-5\%$ to $+5\%$.

\subsection{Production asymmetries}
Charm production in high-energy $p\bar{p}$ collisions is dominated by \CP--conserving $c\bar{c}$ production through the strong interaction. No production asymmetries are expected by integrating over the whole phase space. However, the CDF acceptance covers a limited region of the phase space, where \CP\ conservation may not be exactly realized.
Correlations with the $p\overline{p}$ initial state may induce pseudorapidity--dependent  asymmetries between the number of produced charm and anti-charm (or positive-- and negative--charged) mesons. These asymmetries are constrained by \CP\ conservation to change sign for opposite values of $\eta$. The net effect is expected to vanish if the pseudorapidity distribution of the sample is symmetric. 

To set an upper limit to the possible effect of small residual $\eta$ asymmetries of the samples used in this analysis, we repeat the fits enforcing a perfect $\eta$ symmetry by reweighting. We observe variations of $\Delta\Acp(KK)=0.03\%$ and $\Delta\Acp(\pi\pi)=0.04\%$ between the fit results obtained with and without re-weighting. We take these small differences as an estimate of the size of possible residual effects. 
 The cancellation of production asymmetries achieved in $p\bar{p}$ collisions (an initial \CP--symmetric state) recorded with a polar-symmetric detector provide a significant advantage in high-precision \CP-violation measurements over experiments conducted in  $pp$ collisions.

\subsection{Contamination of $D$ mesons from $B$ decays\label{sec:dzero_da_B}}

A contamination of charm mesons produced in $b$--hadron decays could bias the results. Violation of \CP\ symmetry in $b$--hadron decays may result in asymmetric production of charm and anti-charm mesons. This may be large for a single exclusive mode, but the effect is expected to vanish for inclusive $B \to D^0 X$ decays~\cite{gronau}. However, we use the impact parameter distribution of $D^0$ mesons to statistically separate primary and secondary mesons and assign a systematic uncertainty. Here,  by ``secondary" we mean any $D^0$  originating from the decay of any $b$ hadron regardless of the particular decay chain involved. In particular we do not distinguish whether the $D^0$ meson is coming from a $D^{*\pm}$ or not. \par
If $f_{B}$ is the fraction of secondary $D^0$ mesons in a given sample, the corresponding observed asymmetry $A$ can be written as a linear combination of the  asymmetries for primary and secondary $D^0$ mesons:
\begin{equation}\label{eq:acpB1}
A = f_B A(D^0\ \text{secondary}) + (1-f_B) A(D^0\ \text{primary}).
\end{equation}
The asymmetry observed for secondary $D^0$ mesons can be expressed, to first order, as the sum of the asymmetry you would observe for a primary $D^0$ sample, plus a possible \CP--violating asymmetry in inclusive $B\to D^0X$ decays,
\begin{equation}\label{eq:acpB2}
A(D^0\ \text{sec.}) = \Acp(B\to D^0 X) + A(D^0\ \text{prim.}).
\end{equation}
Hence, combining Eq.\ (\ref{eq:acpB1}) and  Eq.\ (\ref{eq:acpB2}), the asymmetry observed in each sample is given by
\begin{equation}\label{eq:acpB3}
A =  f_B\Acp(B\to D^0 X) + A(D^0\ \text{primary}).
\end{equation}
Because the fraction of secondary $D^0$ mesons is independent of their decay mode, we assume 
$f_B(\pi\pi^*)=f_B(KK^*)=f_B(K\pi^*)$. The contribution of \CP\ violation in $b$--hadron decays to the final asymmetries is written as
\begin{equation}
A(hh) = f_B(K\pi) \Acp(B\to D^0X) + \Acp(D^0\to hh),
\label{eq:central_asymmetry}
\end{equation}
where $f_B$ is estimated in the untagged $K^-\pi^+$ sample because the two terms arising 
from the tagged components cancel in the subtraction provided by Eq.\ (\ref{eq:formula}).
\begin{figure}[th]
\centering
\includegraphics[width=8.6cm]{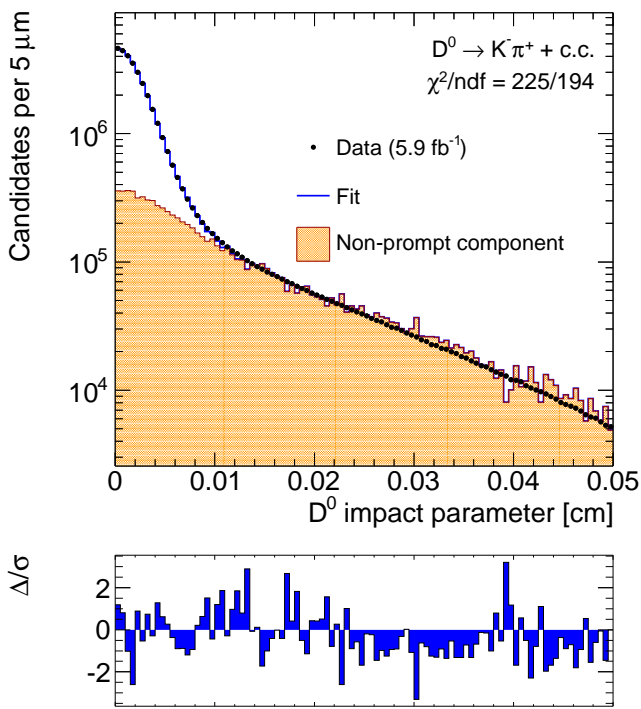}
\caption{Impact parameter distribution of $D^{0}$ candidates in the \DKpi\ signal region. Top plot with data and fit projections overlaid uses a logarithmic scale vertically. Bottom plot shows fractional difference between data and the fit on a linear scale.}
\label{fig:contamination_daB}
\end{figure}
In this analysis, the contamination from secondary $D^0$ decays is reduced by requiring the impact 
parameter of the $D^0$ candidate, $d_0(D^0)$,  not to exceed $100~\mum$. 
The fraction $f_B$ of residual $D^0$ mesons originating from $B$ decays 
has been determined by fitting the distribution of the impact parameter  
of untagged $D^0\to K^-\pi^+$ decays selected within $\pm 24$ MeV/$c^2$ of the known $D^0$ mass~\cite{pdg}.
We use two Gaussian distributions to model the narrow peak from primary $D^0$ mesons and a binned histogram, extracted from a simulated sample of inclusive
$B\to D^0X$ decays, to model the secondary component.
Figure~\ref{fig:contamination_daB} shows the data with the fit projection overlaid. 
A residual contamination of 16.6\% of $B \to D^0X$ decays with impact parameter lower than $100~\mum$ is estimated.
To constrain the size of the effect from \mbox{$\Acp(B\to D^0X)$}
we repeat the analysis inverting the impact parameter selection, namely requiring 
$d_0(D^0) >100~\mum$. This selects an almost pure sample of  \DKpi\ decays from $B$ decays ($f_B = 1$).
We reconstruct about 900~000 decays with an asymmetry, $A(K\pi)= (-0.647 \pm  0.172)\%$, consistent with $(-0.832 \pm  0.033)\%$, the value used in  our measurement.
Using Eq.\  (\ref{eq:acpB2}) we write the difference between the above asymmetry and the asymmetry observed in the central analysis (Eq.\ (\ref{eq:central_asymmetry})), $A(d_{0}>100~\mum) - A(d_{0}<100~\mum)$,   as
\begin{equation}\label{eq:acp_diff}
(1-f_B) A_{\rm CP}(B\to D^0X)  = (-0.18 \pm 0.17)\%.
\end{equation}
Using $f_B=16.6\%$ we obtain $\Acp(B\to D^0X)  = (-0.21 \pm 0.20)\%$ showing that no evidence for  a bias induced by secondary
$D^0$ mesons is present.  Based on Eq.\ (\ref{eq:central_asymmetry}), we assign a conservative systematic uncertainty evaluated as
$f_B A_{\rm CP}(B\to D X)=  f_B/(1-f_B)  \Delta= 0.034\%$,  where $f_B$ equals 16.6\% and $\Delta$ corresponds to the $0.17\%$ standard deviation of the difference in Eq.\ (\ref{eq:acp_diff}).

\subsection{Assumptions in the fits of tagged samples\label{sec:syst:rawasy}}

\subsubsection{Shapes of fit functions\label{sec:sys-tagged-shapes}}

The mass shape extracted from simulation has been adjusted using data for a more accurate description of the observed signal shape. A systematic uncertainty is associated with the finite accuracy of this tuning and covers the effect of possible mis-modeling of the shapes of the fit components.

\begin{figure}[!ht]
\centering
\includegraphics[width=8.6cm]{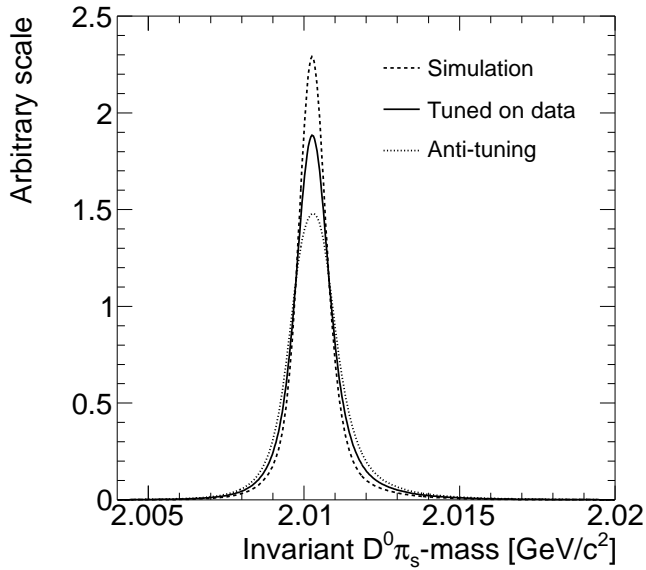}   
\caption{Shape of $D^0\pi_s$ mass as extracted from simulation without tuning, with data tuning and with anti-data tuning.}
\label{fig:syst_shape*}
\end{figure}

\Fig{syst_shape*} shows a comparison between the shape extracted from the simulation and the templates used in the fit after the tuning. It also shows an additional template, named ``anti-tuned'',  where the corrections that adjust the simulation to data have been inverted. If $f(m)$ is the template tuned on data, and $g(m)$ is the template extracted from the simulation, the anti-tuned template is constructed as  $h(m) = 2f(m)-g(m)$. We repeat the measurement using the templates extracted from the simulation without any tuning, and those corresponding to the anti-tuning. The maximum variations from the central fit results,  $\Delta\Acp(\pipi)=0.009\%$ and $\Delta\Acp(\KK)=0.058\%$, are assigned as systematic uncertainties. The larger effect observed in the $D^0\to K^+K^-$ case comes from the additional degrees of freedom introduced in the fit by the multibody-decays component.

In addition, we perform a cross-check of the shape used for the background of real $D^0$ mesons associated with random tracks. In the analysis, the shape parameters of $D^0\to h^+h^-$ fits are constrained to the values obtained in the higher-statistics tagged $D^0\to K^-\pi^+$ sample. If the parameters are left floating in the fit,  only a negligible variation on the final result ($<0.003\%$) is observed.

\subsubsection{Charge-dependent mass distributions}
\label{sec:sys-tagged-charges}

We observe small differences between distributions of $D^0\pi_s$ mass  for positive and negative $D^{*}$ candidates. These are ascribed to possible differences in tracking resolutions between low-momentum positive and negative particles. Such differences may impact our results at first order and would not be corrected by our subtraction method. To determine a systematic uncertainty, we repeat the fit in several configurations where various combinations of signal and background parameters are independently determined for positive and negative  $D^{*}$ candidates. The largest effects are observed by leaving the background shapes to vary independently and constraining the parameter $\delta_J$ of the Johnson function to be the same~\cite{tesi-angelo}. The values of the shape parameters in $D^0\to h^+h^-$ fits are always fixed to the ones obtained from the $D^0\to K^-\pi^+$ sample. The maximum variations with respect to the central fits,  $\Delta\Acp(\pipi)=0.088\%$ and $\Delta\Acp(\KK)=0.027\%$,  are used as systematic uncertainties. 

\subsubsection{Asymmetries from residual backgrounds}
\label{sec:sys-tagged-bckg}

A further source of systematic uncertainty is the approximations used in the subtraction of physics backgrounds. In the $K^+K^-$ sample we fit any residual background contribution, hence this uncertainty is absorbed in the statistical one. However, in the $\pi^+\pi^-$ and $K^-\pi^+$ cases  we assume the residual backgrounds to be negligible.  Using simulation we estimate that a $0.22\%$ and $0.77\%$  contamination from physics backgrounds enters the $\pm 24$ MeV/$c^2$ $\pi^+\pi^-$ and $K^-\pi^+$ signal range, respectively. The contamination in the $\pi^+\pi^-$ sample is dominated by the high mass tail of the $D^0\to K^-\pi^+$ signal. The asymmetry of this contamination is determined from a fit of the tagged $K^-\pi^+$ sample. The contamination of the $K^-\pi^+$ sample is dominated by the tail from partially reconstructed $D^0$ decays.  The fit of the tagged $K^+K^-$ sample provides an estimate of the asymmetry of this contamination. In both cases we assign a systematic uncertainty that is the product of the contaminating fraction times the additional asymmetry of the contaminant. This yields a maximum effect of $0.005\%$ on the measured asymmetries for both $D^0\to\pi^+\pi^-$ and $D^0\to K^+K^-$ cases.

\subsection{Assumptions in the fits of untagged samples}

\subsubsection{Shapes of fit functions\label{sec:mass_shape}}
We follow the same strategy used for the tagged case to assign the systematic uncertainty associated with possible mis-modeling of the shapes in fits of the untagged sample.

\begin{figure*}[!ht]
\begin{center}
\begin{overpic}[width=5.9cm,grid=false,tics=1]{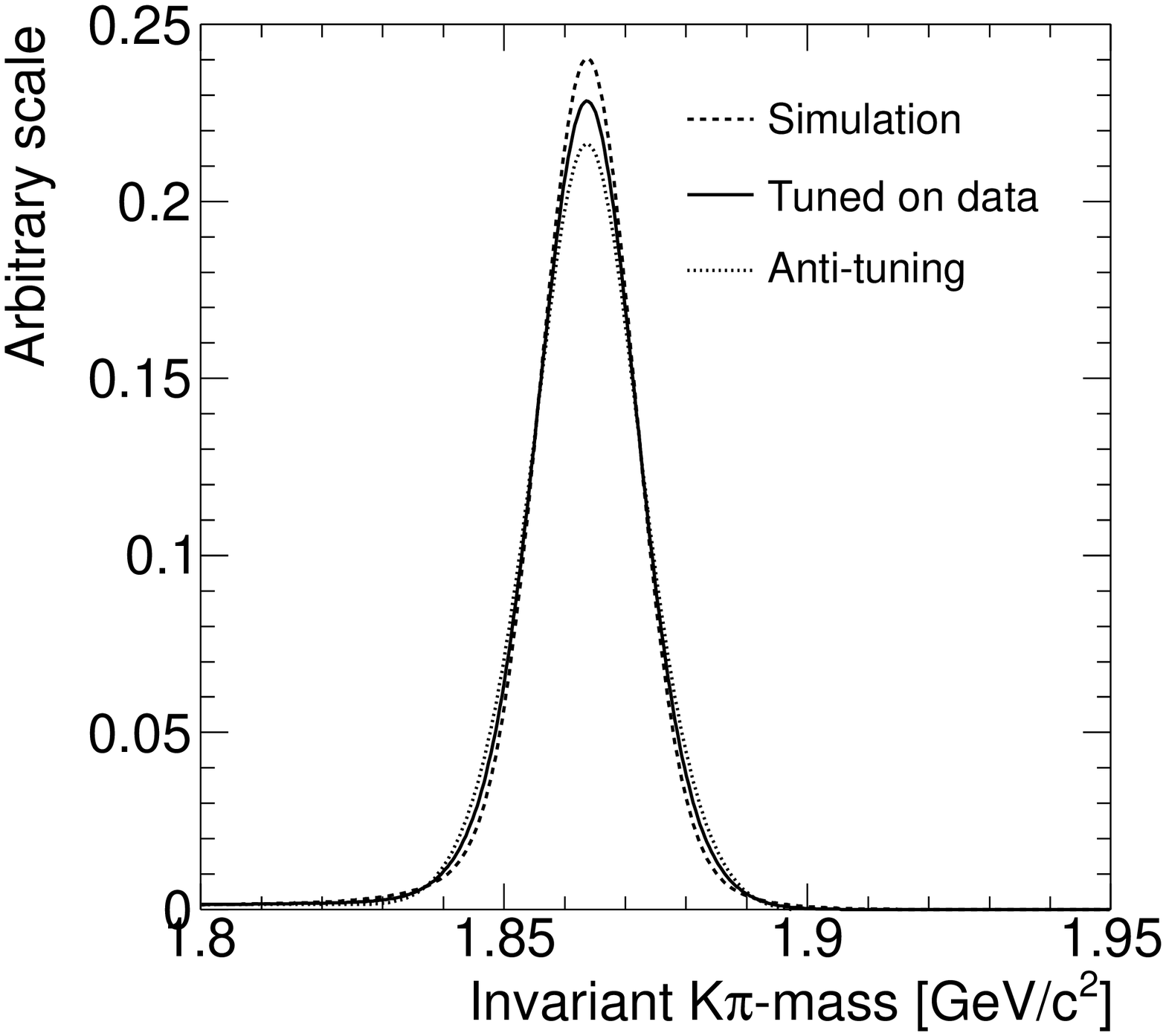}   
\put(25,78){(a)}					        	  
\end{overpic}
\begin{overpic}[width=5.9cm,grid=false,tics=1]{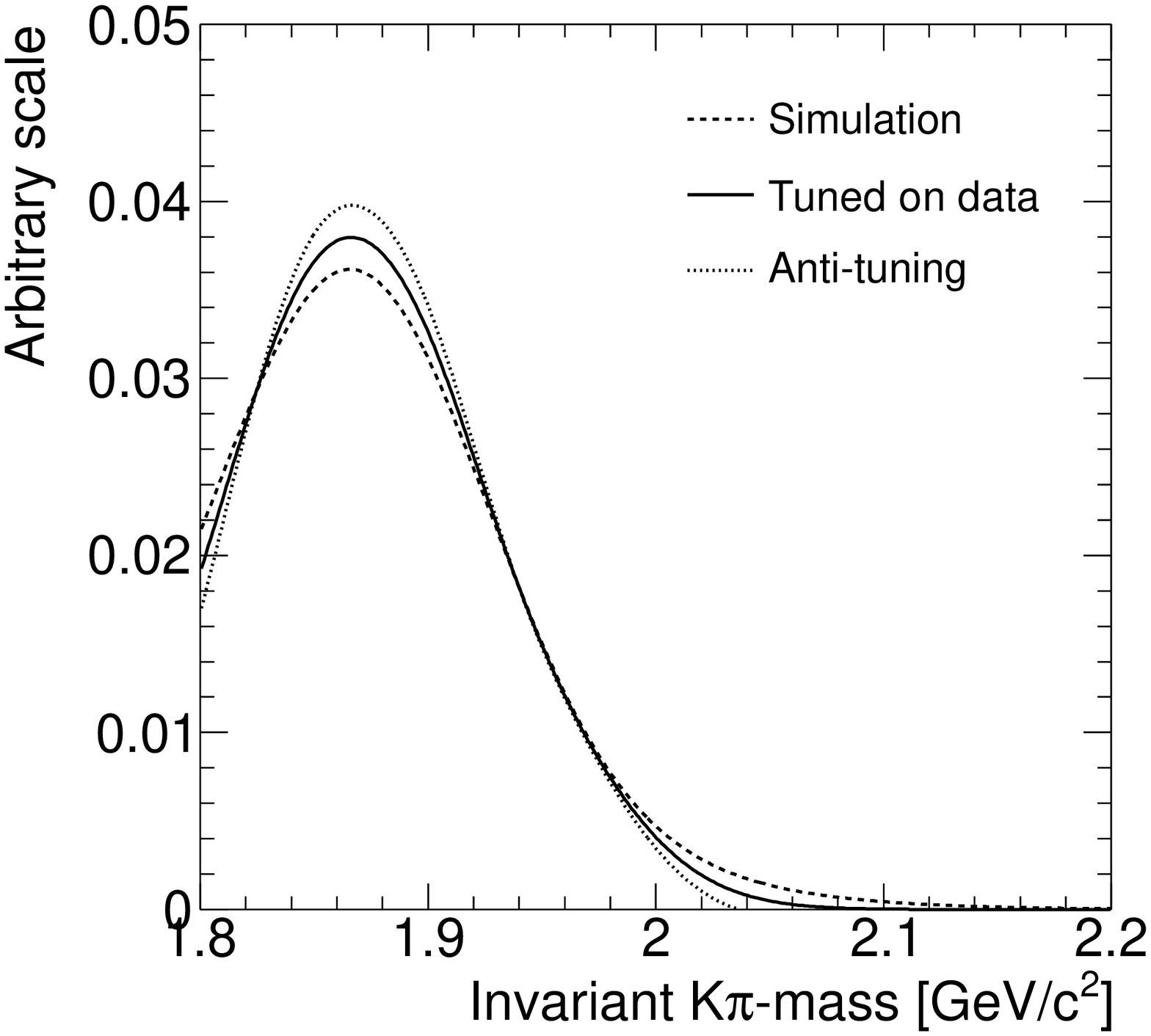}     
\put(25,78){(b)}
\end{overpic}						                  
\begin{overpic}[width=5.9cm,grid=false,tics=1]{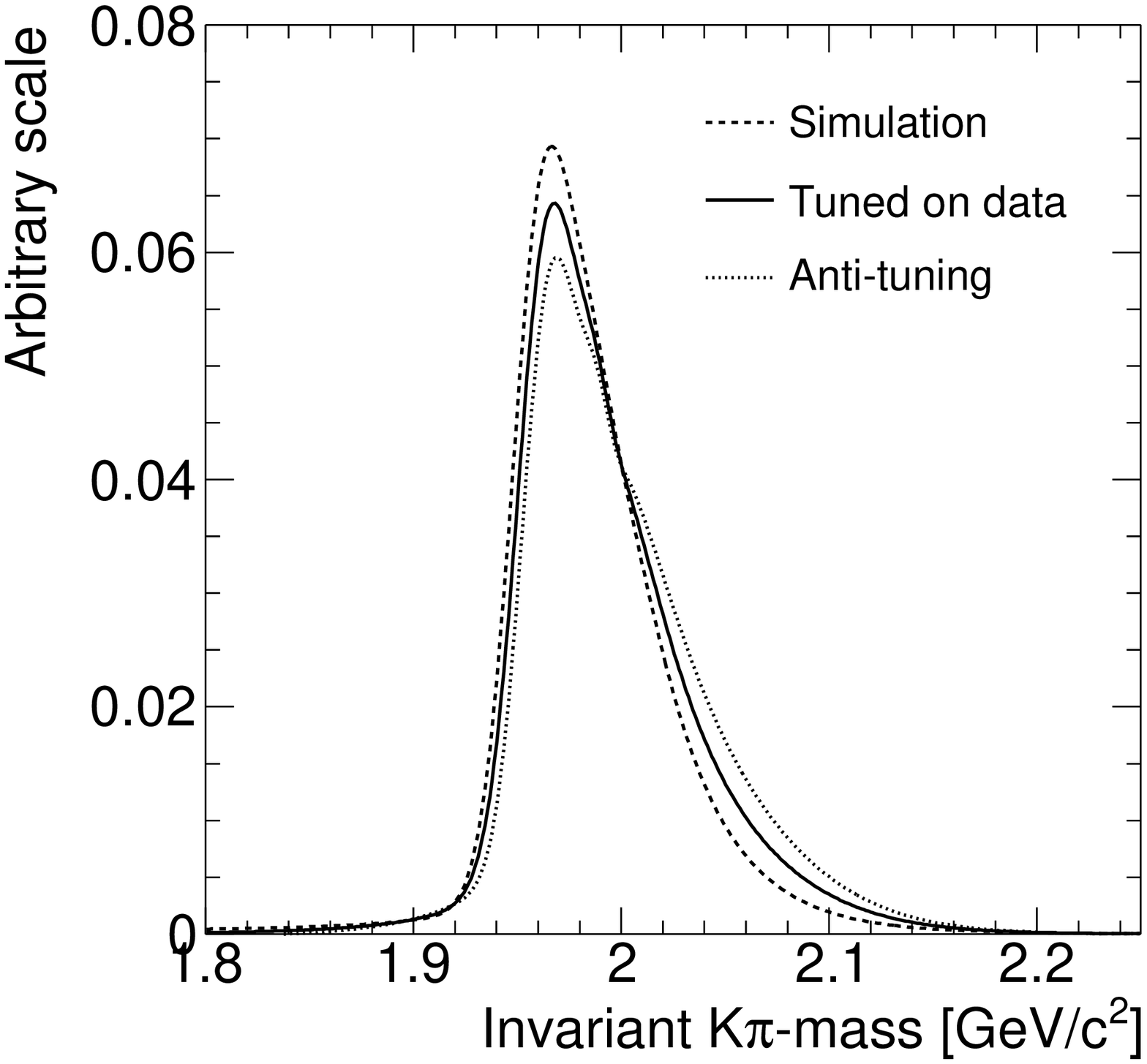}
\put(25,78){(c)}
\end{overpic}  
\end{center}
\caption{Shapes of $K^\pm\pi^\mp$ mass  from simulation without tuning, with data tuning, and with anti-data tuning for (a) right-sign and (b) wrong-sign $K^\pm\pi^\mp$ decays,  and for (c) $\pi^+\pi^-$ decays.}
\label{fig:syst_shape}
\end{figure*}

\Fig{syst_shape} shows the comparison between templates extracted from 
the simulation without any tuning, those tuned to data (and used in the central fit), and the anti-tuned ones.
We repeat the fit using the templates from simulation and the anti-tuned ones. 
The maximum variation from the central fit, $\Delta A(K\pi)=0.005\%$, is used as the systematic uncertainty.

\subsubsection{Charge-dependent mass distributions\label{sec:syst_pos_neg_template}}
In the untagged case we expect  the mass shapes of all components to be the same for charm and anti-charm samples. However, we repeat the simultaneous fit under different assumptions to assign the systematic uncertainty associated with possible residual  differences.
The parameters of the Gaussian distributions used to model the bulk of the mass distributions are left free to vary independently for the charm and anti-charm  samples, and separately for the right-sign, wrong-sign, and $D \to \pi^+\pi^-$ components. 
We assume no difference between mass distributions of combinatorial background and partially reconstructed decays.
The differences between estimated shape parameters in charm and anti-charm samples do not exceed $3\sigma$, showing compatibility between the shapes. 
A systematic uncertainty of $0.044\%$ is obtained by summing in quadrature the shifts from the central values of the estimated asymmetries in the three different cases. 

\subsubsection{Asymmetries from residual physics backgrounds}
In the measurement of the asymmetry of Cabibbo-favored \DKpi\ decays, we neglect the contribution from the small, but  irreducible,  component of doubly-Cabibbo-suppressed (DCS)  $D^0\to K^+\pi^-$ decays. Large \CP\ violation in DCS decays may bias the charge asymmetry we attribute to $D^0\to K^-\pi^+$ decays. We assign a systematic uncertainty corresponding to
$f_{DCS} A_{\it CP}(D^0\to K^+\pi^-) = f_{DCS} \Delta= 0.013\%$, where $f_{DCS}=0.39\%$ is the known~\cite{pdg} fraction of DCS decays with respect to Cabibbo-favored decays and $\Delta=2.2\%$ corresponds to one standard deviation of the current measured limit on the  \CP--violating asymmetry $\Acp(D^0\to K^+\pi^-)$ as reported in \refcita{pdg}.

In the central fit for the untagged $D^0\to K^-\pi^+$ sample, no asymmetry in \Dpipi\ decays or combinatorial background is included, as expected by the way the untagged sample is defined. We confirm the validity of this choice by fitting the asymmetry with independent parameters for these two shapes in the charm and anti-charm samples. The result corresponds to a $\Delta A(K\pi)=0.011\%$ variation from the central fit. 

\subsection{Limitations of kinematic reweighting}
The tagged event samples are reweighted after subtracting the background, sampled in signal mass sidebands. 
We constrain the size of possible residual systematic uncertainties by repeating the fit of tagged $D^0\to h^+h^-$ after a reweighting without any sideband subtraction. The variation in observed asymmetries is found to be negligible with respect to other systematic uncertainties.\par
In reweighting the untagged sample we do not subtract the background. The signal distributions are extracted by selecting a mass region corresponding approximately to a cross-shaped window of $\pm 3\sigma$ in the two-dimensional space ($M(K^+\pi^-), M(K^-\pi^+)$).
To assign a systematic uncertainty we extract the signal distributions and reweight the data using a smaller cross-shaped region 
of  $\pm 2\sigma$ (i.e. within 16~\massmev\ from the nominal $D^0$ mass). The background contamination decreases from $6\%$ to $4\%$. 
We repeat the analysis and find $A(K\pi)= (-0.831 \pm 0.033)\%$ corresponding to a variation from the central fit of $<0.001\%$, thus negligible with respect to other systematic uncertainties.

\subsection{Total systematic uncertainty}
Table~\ref{tab:syst} summarizes the most significant systematic uncertainties considered in the measurement.  Assuming them independent and summing in quadrature, we obtain a total systematic uncertainty of $0.11\%$ on the observed \CP--violating asymmetry of $D^0\to\pi^+\pi^-$ decays and $0.09\%$ in $D^0\to K^+K^-$ decays.
Their sizes  are approximately half of the statistical uncertainties.
 
\begin{table*}[t]
\centering
\caption{Summary of most significant systematic uncertainties. The uncertainties reported for the last three sources result from the sum in quadrature of the contributions in the tagged and untagged fits.}\label{tab:syst}
\begin{tabular}{lccc}
\hline\hline
Source  								& $\Acp(\pi^+\pi^-)$ [\%]&	& $\Acp(K^+K^-$) [\%]\\
\hline
Approximations in the suppression of detector-induced effects				& $0.009$ & &$0.009$\\
Production asymmetries					& $0.040$ & &$0.030$ \\
Contamination of secondary $D$ mesons	& $0.034$ & &$0.034$ \\
Shapes assumed in fits 					& $0.010$ & &$0.058$ \\
Charge-dependent mass distributions 		& $0.098$ & &$0.052$ \\
Asymmetries from residual backgrounds 		& $0.014$ & &$0.014$\\
Limitations of sample reweighting 			& $<0.001$ & &$<0.001$\\
\hline
Total 								& $0.113$ & &$0.092$\\
\hline\hline
\end{tabular}
\end{table*}

 \section{Final result\label{sec:final}}
Using the observed asymmetries from Eqs.\ (\ref{eq:tagged-results}) and (\ref{eq:untagged-results}) in the relationships of Eq.\ (\ref{eq:acpraw}), we determine the time-integrated \CP--violating asymmetries  in $D^0\to\pi^+\pi^-$ and  $D^0\to K^+K^-$ decays to be 
\begin{align}
\Acp(\pi^+\pi^-) &= \bigl(+0.22\pm0.24\stat\pm0.11\syst\bigr)\% \nonumber \\*
\Acp(K^+K^-) &= \bigl(-0.24\pm0.22\stat\pm0.09\syst\bigr)\%,\nonumber
\end{align}
corresponding to \CP\ conservation in the time-evolution of these decays. These are the most precise determinations of these quantities to date, and significantly improve the 
world's average values. The results are also in agreement with theory predictions~\cite{Bigi:1986dp,Golden:1989qx,Buccella:1994nf, Xing:1996pn,Du:2006jc,Grossman:2006jg}.\par 
A useful comparison with results from other experiments is achieved by expressing the observed asymmetry as a linear combination  (Eq.\  (\ref{eq:acp3})) of a direct component,  $\Acp^{\rm{dir}}$,  and an indirect component, $\Acp^{\rm{ind}}$,  through a coefficient that is the mean proper decay time of charm mesons in the data sample.  The direct component corresponds to a difference in width between charm and anti-charm decays into the same final state. The indirect component is due to the probability for a charm meson to oscillate into an anti-charm meson being different from the probability for an anti-charm meson to oscillate into a charm meson.
\begin{figure*}[t]
\centering
\begin{overpic}[width=8.6cm,grid=false]{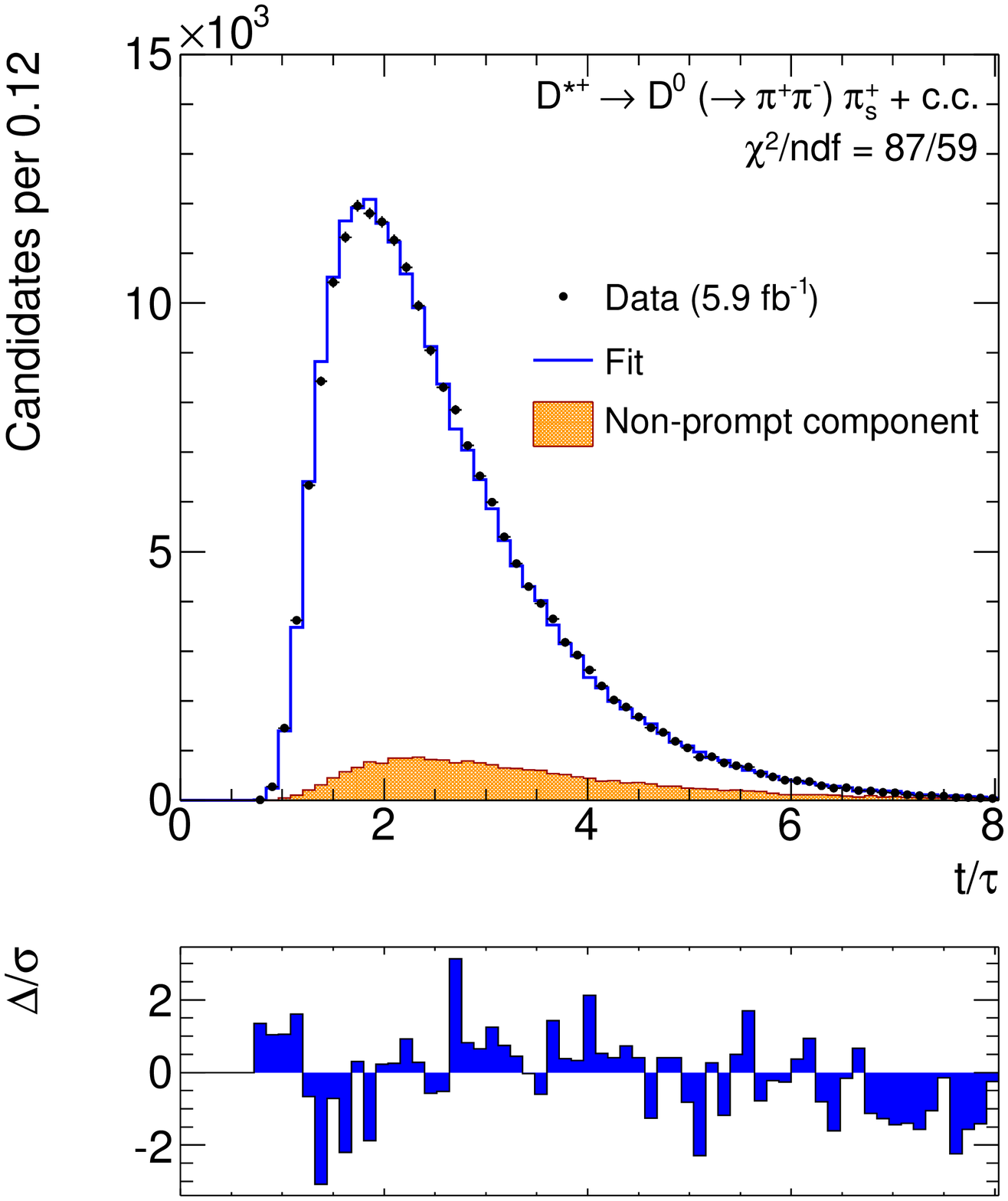}
\put(20,85){(a)} 
\end{overpic}\hfil
\begin{overpic}[width=8.6cm,grid=false]{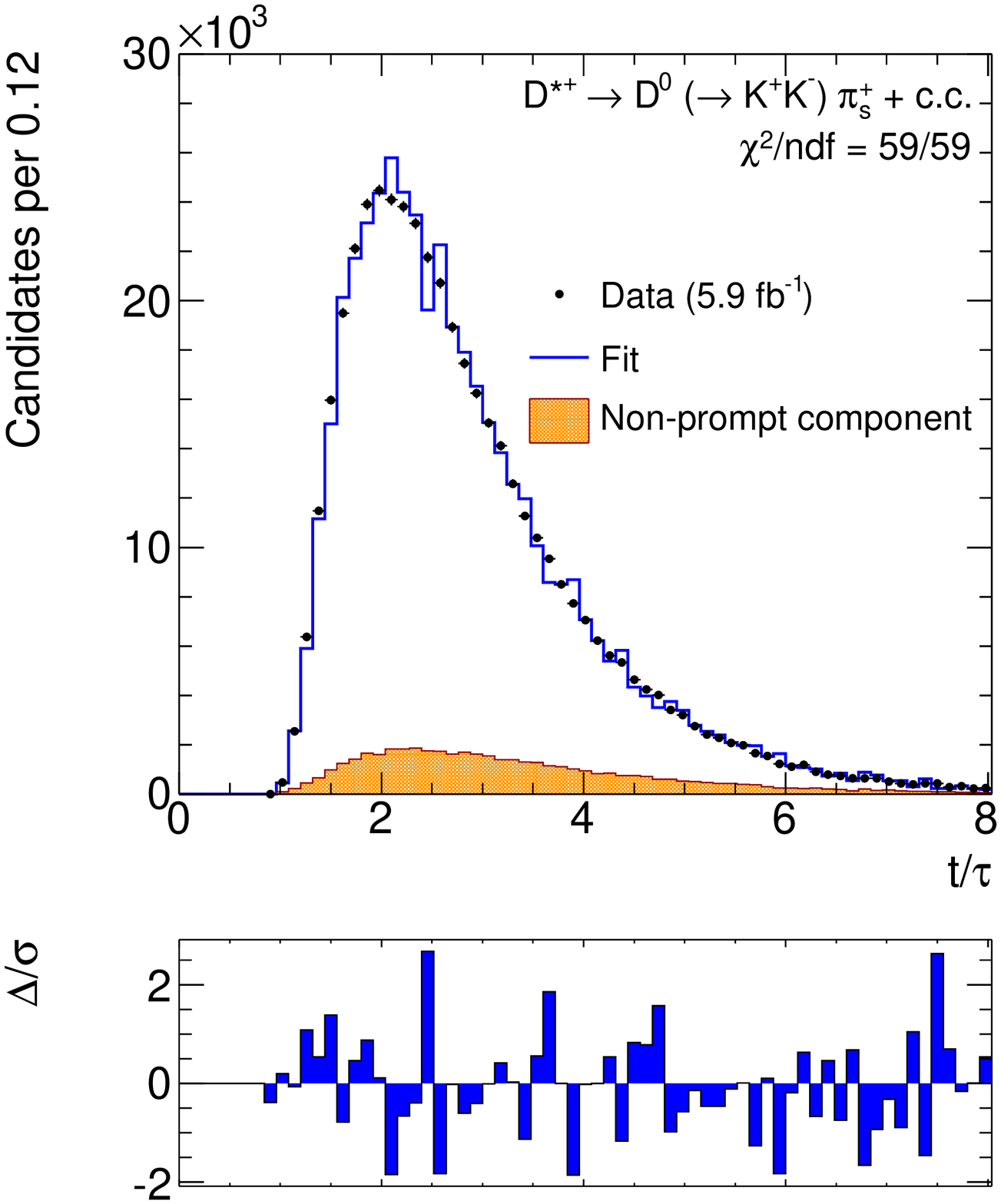}
\put(20,85){(b)} 
\end{overpic}
\caption{Distribution of proper decay time (in units of $D^0$ lifetime) for sideband-subtracted tagged (a) $D^0\to\pi^+\pi^-$ and (b) $D^0\to K^+K^-$ data. Fit results are overlaid including the component from secondary charmed mesons  (red).}\label{fig:propertime}
\end{figure*}
\par The decay time of each $D^0$ meson, $t$, is determined as 
\begin{equation}\label{eq:t_from_Lxy}
t = \frac{L_{xy}}{c \left( \beta \gamma\right)_T} 
  = L_{xy} \ \frac{m_{D^0}}{c\ p_T},\nonumber
\end{equation}
where $(\beta \gamma )_T = p_T/m_{D^0}$ is the transverse Lorentz factor. This is an unbiased estimate of the actual decay time only for primary charmed mesons. For secondary charm, the decay time of the parent $B$ meson should be subtracted. The mean decay times of our signals are determined from a fit to the proper decay time distribution of sideband-subtracted tagged decays  (Fig.~\ref{fig:propertime}). The fit includes components for primary and secondary $D$ mesons, whose shapes are modeled from simulation. The simulation is used to extract the information on the mean decay time of secondary charmed decays, using the known true decay time. The proportions between primary and secondary are also determined from this fit and are consistent with results of the fit to the $D^0$ impact parameter in data (Sec.\ \ref{sec:dzero_da_B}). We determine a mean decay time of $2.40\pm0.03$ and $2.65\pm0.03$, in units of $D^0$ lifetime, for \mbox{$D^0\to\pi^+\pi^-$} and \mbox{$D^0\to K^+K^-$} decays, respectively. The uncertainty is the sum in quadrature of statistical and systematic contributions. The small difference in the two samples is caused by the slightly different kinematic distributions of the two decays, which impacts their trigger acceptance. 

\begin{figure*}[t]
\centering
\begin{overpic}[width=8.6cm,grid=false]{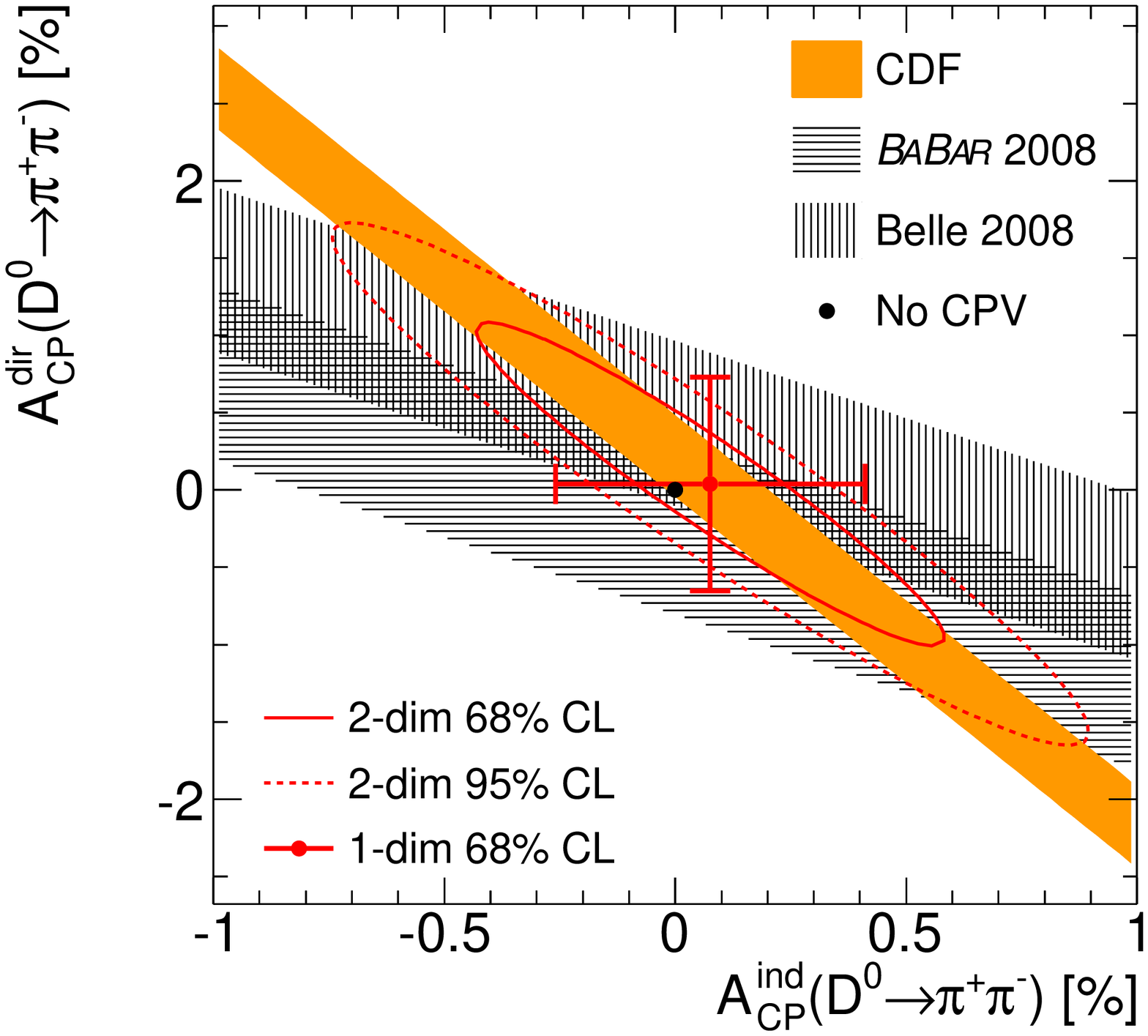}
\put(23,34){(a)} 
\end{overpic}\hfil
\begin{overpic}[width=8.6cm,grid=false]{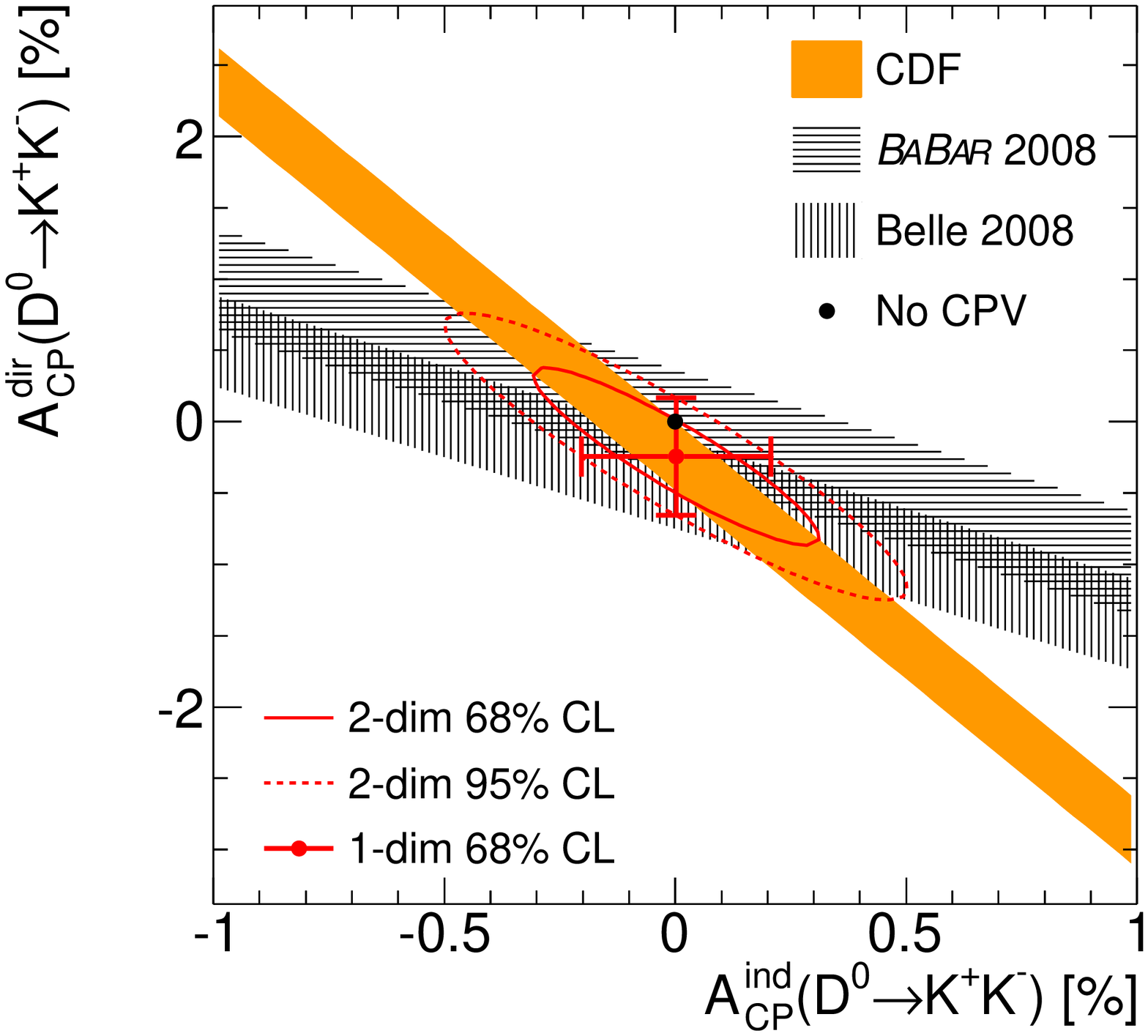}
\put(23,34){(b)}
\end{overpic}
\caption{Comparison of the present results with Belle and \babar\ measurements of time-integrated \CP--violating asymmetry in (a) $D^0\to \pi^+\pi^-$  and (b) $D^0\to K^+K^-$  decays displayed in the $(\Acp^{\rm{ind}},\Acp^{\rm{dir}})$ plane.  The point with error bars denotes the central value of the combination of the three measurements with one-dimensional 68\% confidence level uncertainties.}\label{fig:combination}
\end{figure*}

Each of our measurements defines a band in the  $(\Acp^{\rm{ind}},\Acp^{\rm{dir}})$  
plane with slope  $-\left<t\right>/\tau$ (Eq.\ (\ref{eq:acp3})). The same holds for \babar\ and Belle measurements, with slope $-1$ \cite{Aubert:2007wf,Staric:2007dt}, due to unbiased acceptance in decay time. The results of this measurement and the most recent $B$-factories' results are shown in Fig.~\ref{fig:combination},  which displays their relationship. The bands represent $\pm 1\sigma$ uncertainties and show that all measurements are compatible with \CP\ conservation (origin in the two-dimensional plane). The results of the three experiments can be combined assuming Gaussian uncertainties. We construct combined confidence regions in the $(\Acp^{\rm{ind}},\Acp^{\rm{dir}})$ plane, denoted with  $68\%$ and $95\%$ confidence level ellipses. The corresponding values for the asymmetries are $\Acp^{\rm{dir}}(D^0\to\pi^+\pi^-) = (0.04 \pm 0.69)\%$,   $\Acp^{\rm{ind}}(D^0\to\pi^+\pi^-) = (0.08 \pm 0.34)\%$,  $\Acp^{\rm{dir}}(D^0\to K^+K^-) = (-0.24 \pm 0.41)\%$, and  $\Acp^{\rm{ind}}(D^0\to K^+K^-) = (0.00\pm 0.20)\%$, 
 in which the uncertainties represent one-dimensional 68\% confidence level intervals.

\subsection{CP violation from mixing only}

Assuming negligible direct \CP\ violation in both decay modes, the observed asymmetry is only due to mixing, $\Acp(h^+h^-) \approx  \Acp^{\rm{ind}}\ \langle t \rangle /  \tau$, 
yielding
\begin{align}
\Acp^{\rm{ind}}(\pi^+\pi^-) &= \bigl(+0.09\pm0.10\stat\pm0.05\syst\bigr)\% \nonumber \\*
\Acp^{\rm{ind}}(K^+K^-) &= \bigl(-0.09\pm0.08\stat\pm0.03\syst\bigr)\%. \nonumber
\end{align}
Assuming that no large weak phases from non-SM contributions appear in the decay amplitudes, $\Acp^{\rm{ind}}$ is independent of the final state. Therefore the two measurements can be averaged,  assuming correlated systematic uncertainties,  to obtain a precise determination of \CP\ violation in charm mixing:
\begin{equation}
\Acp^{\rm{ind}}(D^0) = \bigl(-0.01\pm0.06\stat\pm0.04\syst\bigr)\%. \nonumber
\end{equation}
This corresponds to the following upper limits on \CP\ violation in charm mixing: 
\begin{equation}
|\Acp^{\rm{ind}}(D^0)| < 0.13~(0.16)\% \mbox{ at the 90 (95)\% C.L}.\nonumber
\end{equation}

\begin{figure*}[t]
\centering
\begin{overpic}[width=8.6cm,grid=false]{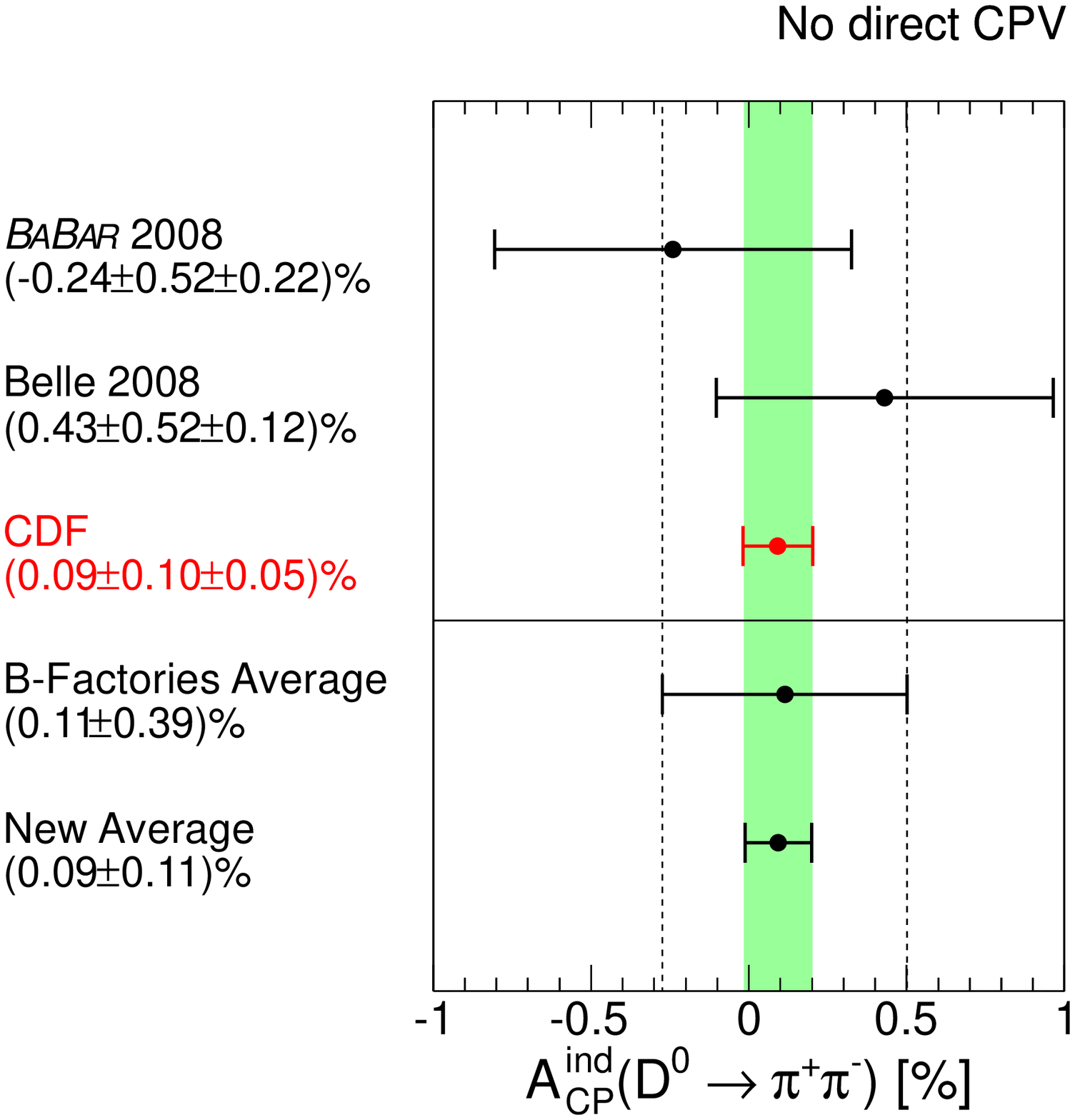}
\put(85,75){(a)} 
\end{overpic}\hfil
\begin{overpic}[width=8.6cm,grid=false]{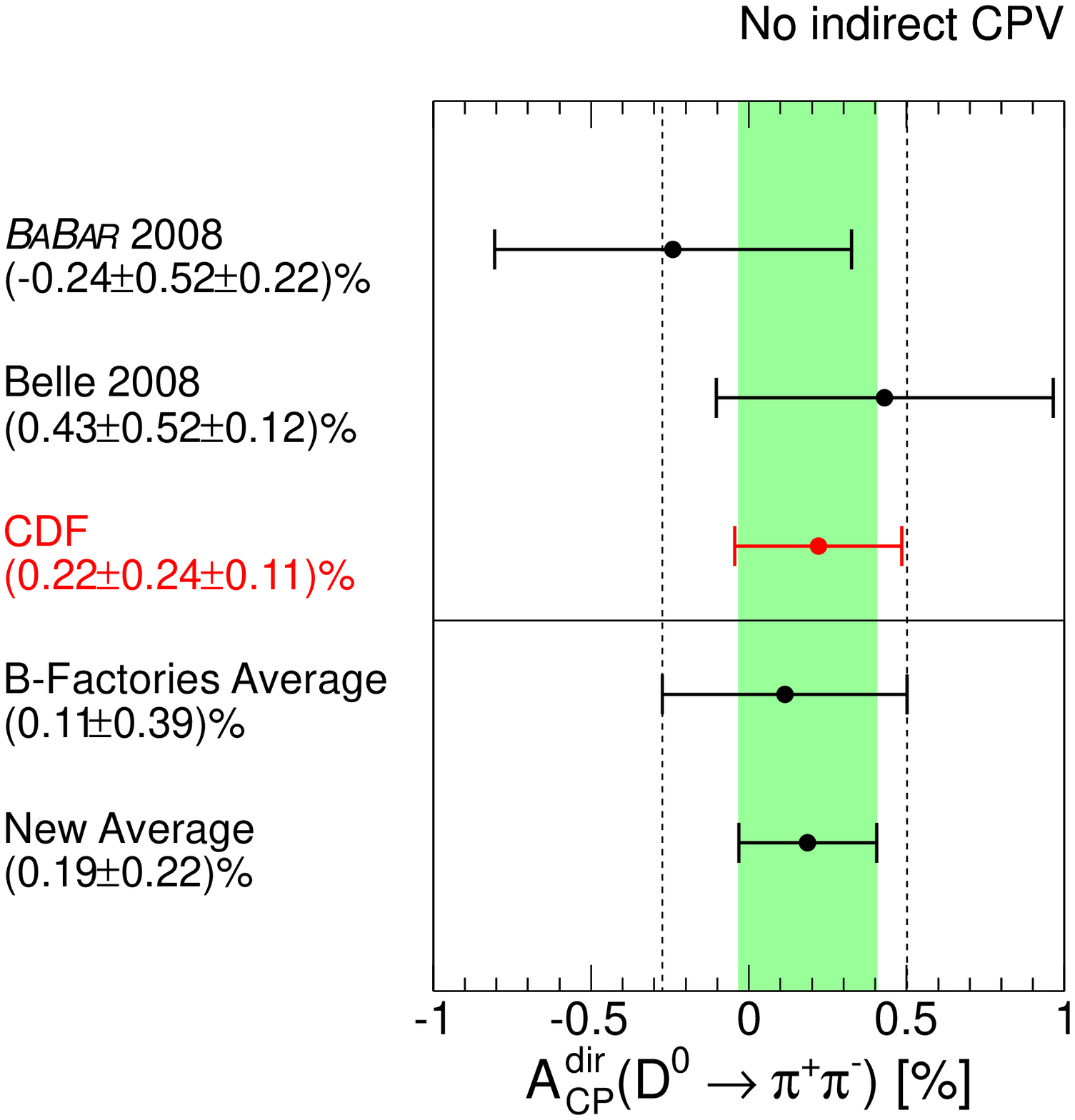}
\put(85,75){(b)}
\end{overpic}\\
\begin{overpic}[width=8.6cm,grid=false]{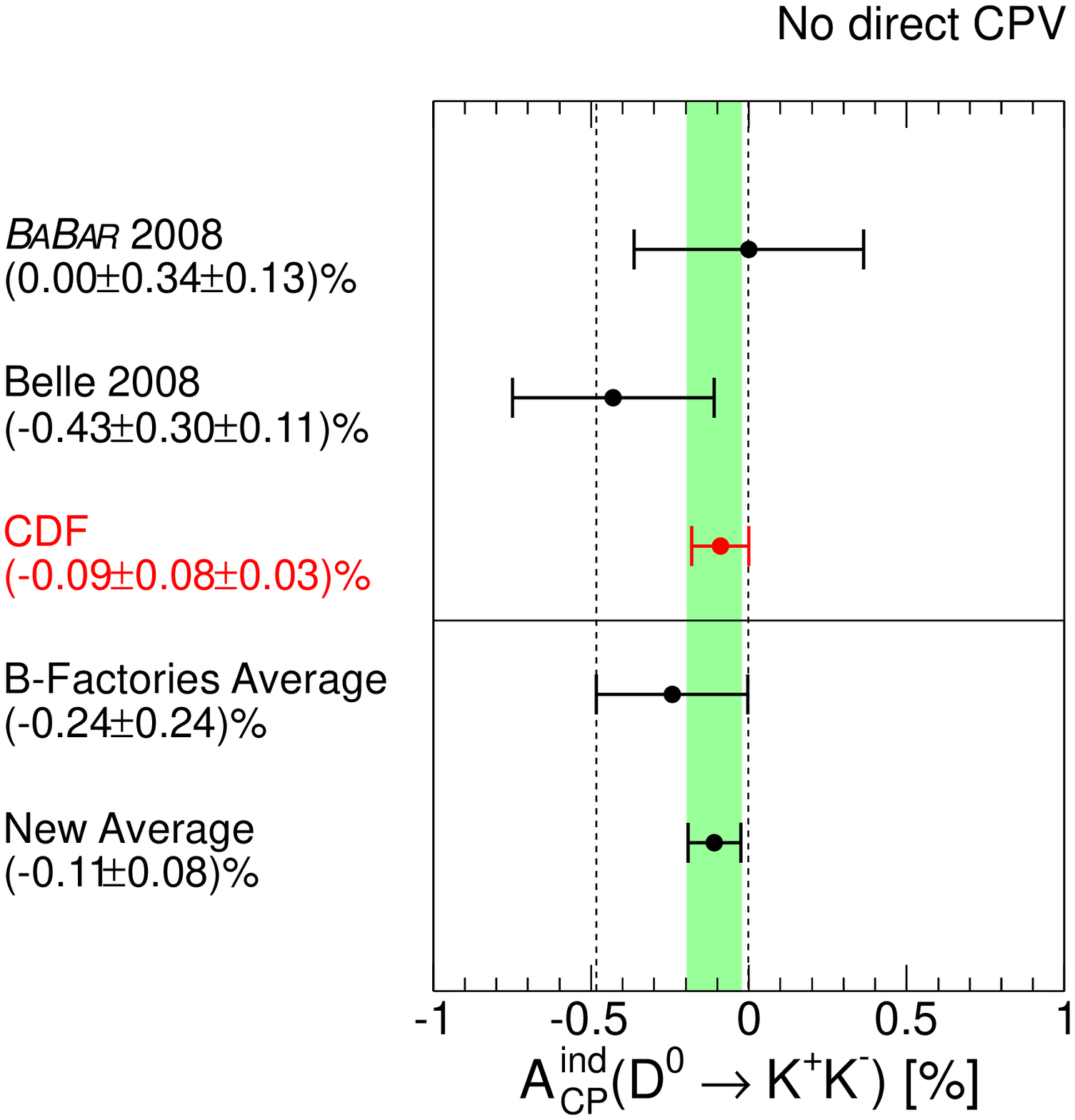}
\put(85,75){(c)} 
\end{overpic}\hfil
\begin{overpic}[width=8.6cm,grid=false]{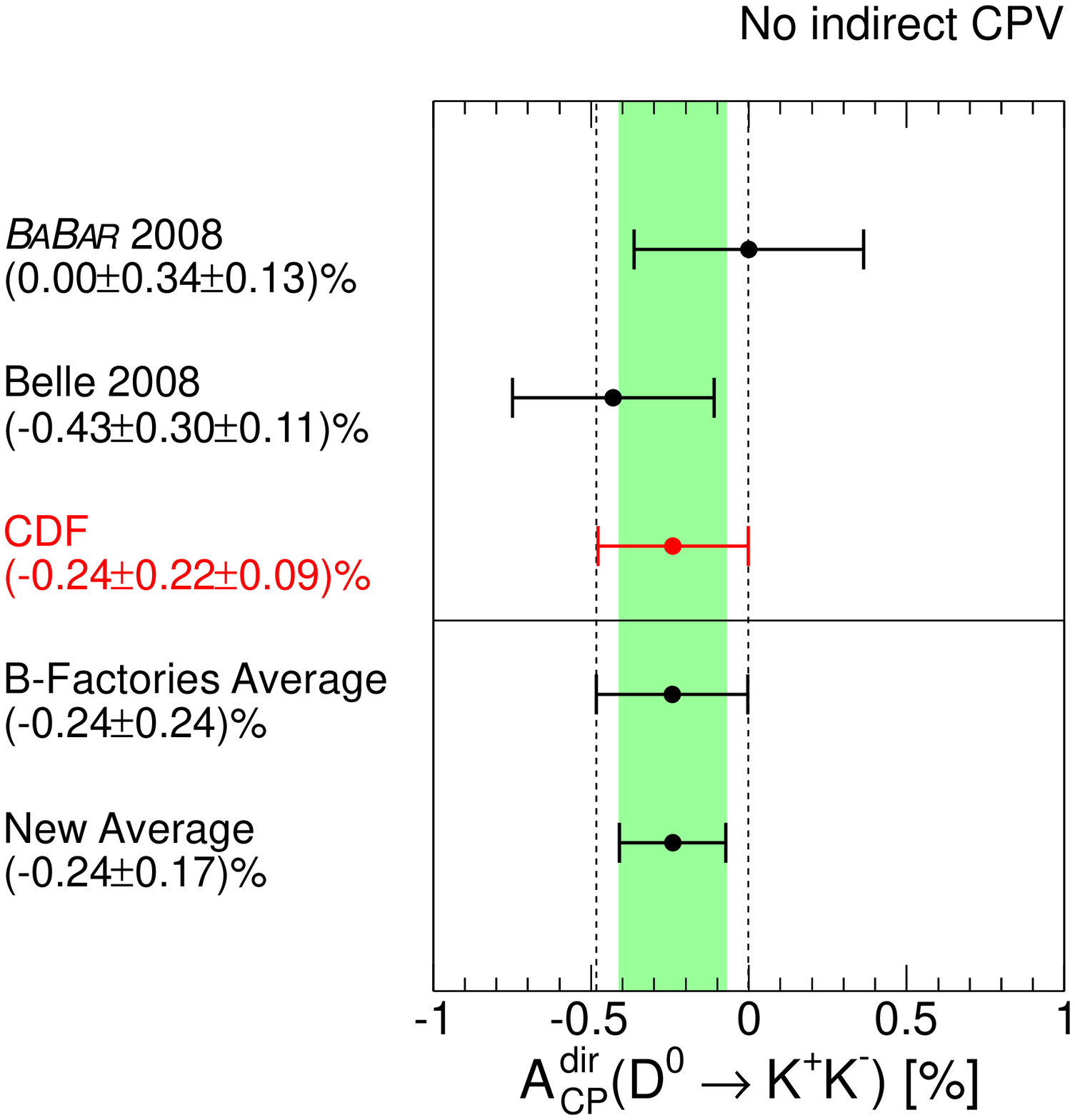}
\put(85,75){(d)}
\end{overpic}
\caption{Comparison of the present results  
with results from Belle and \babar\ assuming (a), (c) no direct,  or (b), (d) no indirect \CP\ violation. In each plot the $1\sigma$ band of the $B$-factories' average is displayed in blue, while the new average that includes the CDF result is shown in green.}\label{fig:direct_and_indirect}
\end{figure*}
The bias toward longer-lived decays of the CDF sample offers a significant advantage over $B$-factories in sensitivity to the time-dependent component, as shown in Figs.~\ref{fig:direct_and_indirect}~(a), (c).

\subsection{Direct CP violation only}

Assuming that \CP\ symmetry is conserved in charm mixing, our results are readily comparable to measurements obtained at $B$-factories; $\Acp(\pi^+\pi^-)= (0.43\pm0.52\stat\pm0.12\syst)\%$ and $\Acp(K^+K^-)= (-0.43\pm0.30\stat\pm0.11\syst)\%$ from Belle,  and  $\Acp(\pi^+\pi^-)= (-0.24\pm0.52\stat\pm0.22\syst)\%$ and $\Acp(K^+K^-)= (0.00\pm0.34\stat\pm0.13\syst)\%$ from \babar\  (Figs.~\ref{fig:direct_and_indirect}~(b)-(d)). The CDF result is the world's most precise.

\subsection{Difference of asymmetries}

A useful comparison with theory predictions is achieved by calculating the difference between the asymmetries observed in the $D^0 \to K^+ K^-$ and  $D^0 \to \pi^+ \pi^-$ decays ($\Delta\Acp$). Since the difference in decay-time acceptance is small, $\Delta\langle t \rangle/\tau = 0.26 \pm 0.01$, most of the indirect \CP-violating asymmetry cancels in the subtraction, assuming that no large \CP-violating phases from non-SM contributions enter the decay amplitudes. Hence $\Delta\Acp$ approximates the difference in direct \CP-violating asymmetries of the two decays.  Using the observed asymmetries from Eq.\ (\ref{eq:tagged-results}), we determine 
\begin{align}
\Delta\Acp =& \Acp(K^+K^-) - \Acp(\pi^+\pi^-) \nonumber \\*
		 =& \Delta\Acp^{\rm{dir}} +  \Acp^{\rm{ind}}\Delta\langle t \rangle/\tau \nonumber \\*
		 =& A(KK^*) - A(\pi\pi^*)  \nonumber \\*
		 =&  \bigl(-0.46 \pm 0.31\stat  \pm 0.12 \syst \bigr)\%. \nonumber
\end{align}
The systematic uncertainty is dominated by the 0.12\% uncertainty from the shapes assumed in the mass fits, and their possible dependence on the charge of the $D^*$ meson.
This is determined by combining the difference of shifts observed in Secs.\ \ref{sec:sys-tagged-shapes} and~\ref{sec:sys-tagged-charges} including correlations: $(0.058 - 0.009)\% = 0.049\%$ and $(-0.027 - 0.088)\% = 0.115\%$.
Smaller contributions include a 0.009\% from the finite precision associated to the suppression of detector-induced effects  (Sec.\ \ref{sec:sys-approx}), and a 0.005\% due to the  0.22\% background we ignore under the $D^0\to \pi^+\pi^-$ signal (Sec.\ref{sec:sys-tagged-bckg}). The effects of production asymmetries and contamination from secondary charm decays cancel in the difference. \par
\begin{figure}[t]
\centering
\includegraphics[width=8.6cm,grid=false]{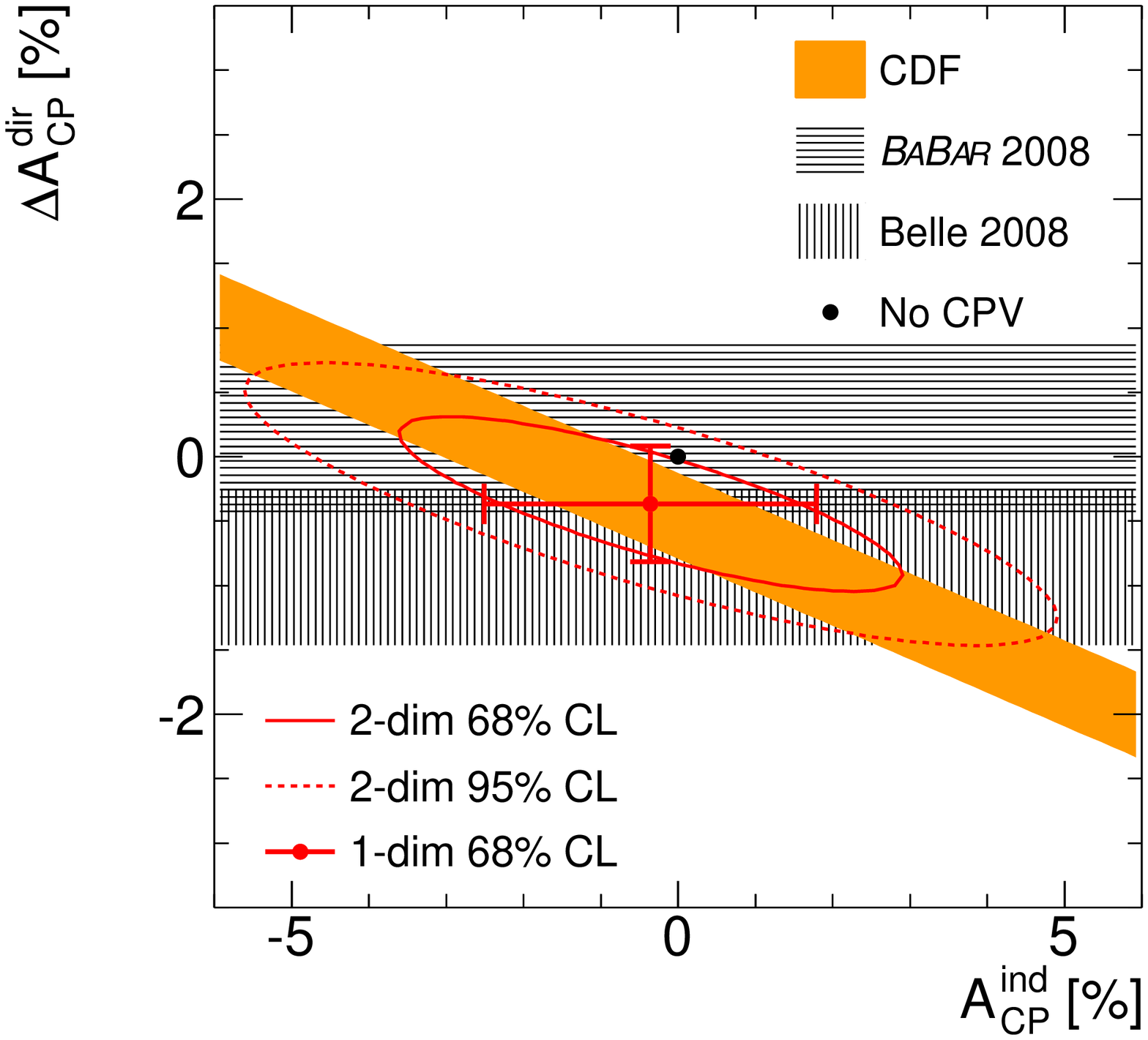}
\caption{Difference between direct \CP--violating asymmetries in the $K^+K^-$ and $\pi^+\pi^-$ final states as a function of the indirect asymmetry. Belle and \babar\ measurements are also reported for comparison. The point with error bars denotes the central value of the combination of the three measurements with one-dimensional 68\% confidence level uncertainties.}\label{fig:difference}
\end{figure}

 We see no evidence of a difference in \CP\ violation between $D^0\to K^+K^-$ and $D^0 \to \pi^+\pi^-$ decays. Figure \ref{fig:difference} shows the difference in direct asymmetry ($\Delta\Acp^{\rm{dir}}$) as a function of the indirect asymmetry compared with experimental results from \babar\ and Belle~\cite{Aubert:2007wf,Staric:2007dt}. The bands represent $\pm 1\sigma$ uncertainties. The measurements, combined assuming Gaussian uncertainties, provide  $68\%$ and $95\%$ confidence level regions in the $(\Delta\Acp^{\rm{dir}}, \Acp^{\rm{ind}})$ plane, denoted with ellipses. The corresponding values for the asymmetries are $\Delta\Acp^{\rm{dir}}  = (-0.37 \pm 0.45)\%$,   $\Acp^{\rm{ind}} = (-0.35 \pm 2.15)\%$.

\section{Summary\label{sec:theend}}

In summary, we report the results of the most sensitive search for \CP\ violation in singly-Cabibbo--suppressed 
$D^0\to\pi^+\pi^-$ and $D^0\to K^+K^-$ decays. We reconstruct  signals of $\mathcal{O}(10^5)$ $D^*$--tagged decays in an event sample of $p\bar{p}$ collision data corresponding to approximately 5.9~fb$^{-1}$ of integrated luminosity collected by a trigger on displaced tracks. A fully data-driven method to cancel instrumental effects provides
effective suppression of systematic uncertainties to the 0.1\% level, approximately half the magnitude of the statistical uncertainties.\par
We find no evidence of \CP\ violation and measure $\Acp(D^0\to\pi^+\pi^-) = \bigl(+0.22\pm0.24\stat\pm0.11\syst\bigr)\%$  and \mbox{$\Acp(D^0\to K^+K^-) = \bigl(-0.24\pm0.22\stat\pm0.09\syst\bigr)\%$}. These are the most precise determinations from a single experiment to date,  and supersede the corresponding results of Ref.~\cite{Acosta:2004ts}.
The average decay times of the charmed mesons used in these measurements are $2.40 \pm 0.03$ units of $D^0$ lifetime in the $D^0\to \pi^+\pi^-$ sample and  $2.65 \pm 0.03$ units of $D^0$ lifetime in the $D^0\to K^+K^-$ sample. Assuming negligible \CP\ violation in \mbox{$D^0\to\pi^+\pi^-$} and \mbox{$D^0\to K^+K^-$} decay widths (direct \CP\ violation),  the above results, combined with the high-valued average proper decay time of the charmed mesons in our sample,  provide a stringent general constraint on \CP\ violation in $D^0$ mixing,  $|\Acp^{\rm{ind}}(D^0)|< 0.13\%$ at the 90\% confidence level. 
The results probe significant regions of the parameter space of charm phenomenology where discrimination between SM and non-SM dynamics becomes possible \cite{Bigi:2011re,Bigi:2011em}.

\begin{acknowledgments}
We thank Y.~Grossmann, A.~Kagan, A.~Petrov, and especially I.~I.~Bigi and A.~Paul for useful discussions. We thank the Fermilab staff and the technical staffs of the participating institutions for their vital contributions. This work was supported by the U.S. Department of Energy and National Science Foundation; the Italian Istituto Nazionale di Fisica Nucleare; the Ministry of Education, Culture, Sports, Science and Technology of Japan; the Natural Sciences and Engineering Research Council of Canada; the National Science Council of the Republic of China; the Swiss National Science Foundation; the A.P. Sloan Foundation; the Bundesministerium f\"ur Bildung und Forschung, Germany; the Korean World Class University Program, the National Research Foundation of Korea; the Science and Technology Facilities Council and the Royal Society, UK; the Russian Foundation for Basic Research; the Ministerio de Ciencia e Innovaci\'{o}n, and Programa Consolider-Ingenio 2010, Spain; the Slovak R\&D Agency; and the Academy of Finland. 
\end{acknowledgments}

\appendix
\section{Method to suppress detector asymmetries\label{sec:method_math}}
A mathematical derivation of the concepts described in Sec.\ \ref{sec:method} follows. We measure the \CP--violating asymmetry by determining the asymmetry between number of detected particles of opposite charm content $A = (N_+-N_-)/(N_++N_-)$, where $N_+$ and $N_-$ are the number of  $D^0$ and $ \Dbar^0$ decays
found in three different data samples: $D^*$-tagged $D^0\to h^+h^-$ decays (or simply $hh^*$), $D^*$-tagged $D^0\to K^-\pi^+$ decays ($K\pi^*$) and untagged $D^0\to K^-\pi^+$ decays ($K\pi$). We show that the combination of asymmetries
measured in these three samples yields an unbiased estimate of the physical value of \Acp\ with a high degree of 
suppression of systematic uncertainties coming from detector asymmetries. 
In the discussion we always refer to the \emph{true} values of kinematic variables of particles.
The \emph{measured} quantities, affected by experimental uncertainties, play no role here since we are only
interested in counting particles and all detection efficiencies are assumed to be dependent on true quantities only.

\subsection {$D^*$--tagged $D^0\to h^+h^-$}

Assuming factorization of efficiencies for reconstructing the neutral charmed meson and the soft pion, 
we write 
\begin{align}
N_\pm = & \frac{N_*}{2}    B_{D\pi}^*   \int\!\!dp_* dp_s dp_{h^+} dp_{h^-}  \rho_{*\pm}(p_* )  B_{hh}^\pm   \nonumber\\
             &\times \rho_{hh^*}( p_{h^+}, p_{h^-},p_s \: \vert \: p_*) \varepsilon_{hh} (p_{h^+},p_{h^-} )  \varepsilon_{s\pm} (p_{s} )  ,  \nonumber 
\end{align}
where 
$N^*$ is the total number of $D^{*+}$ and $D^{*-}$ mesons;  $p_*, p_s, p_{h^+}, p_{h^-}$ are the three-momenta of the $D^*$, soft $\pi$, $h^+$, and $h^-$, respectively; $\rho_{*+}$ and $\rho_{*-}$ are the densities in phase space of $D^{*+}$ and  $D^{*-}$ mesons (function of the production cross sections and experimental acceptances and efficiencies); $\rho_{hh^*}$ is the density in phase space of the soft pion and $h^+h^-$ pair from $D^0$ decay; $B_{hh}^+$ and $B_{hh}^-$ are the branching fractions of $D^0\to h^+h^-$ and $\Dbar^0\to h^+h^-$; $B_{D\pi}^*$ is the branching fraction of $D^{*+} \rightarrow  D^0\pi^+$ and $D^{*-} \rightarrow  \Dbar^0\pi^-$, assumed to be charge--symmetric; $\varepsilon_{hh}$ is the detection efficiency of the $h^+h^-$ pair from the $D^0$ decay; and $\varepsilon_{s+}$ and  $\varepsilon_{s-}$ are the detection efficiencies of the positive and negative soft pion, respectively. Conservation of four-momenta is implicitly assumed in all densities.
Densities are normalized as $\int dp_* \rho_{*\pm} (p_*) = 1 = \int  dp_s dp_{h^+} dp_{h^-}  \rho_{hh^*}( p_{h^+}, p_{h^-},p_s \: \vert \: p_*)$ for each $p_*$. The difference between event yields is therefore
\begin{align}
N_+ - N_-  = & \frac{N_*}{2}   B_{D\pi}^*   \int\!\! dp_* dp_s dp_{h^+} dp_{h^-}   \nonumber \\
                  & \times \rho_{hh^*}( p_{h^+}, p_{h^-},p_s \: \vert \: p_*)  \varepsilon_{hh} (p_{h^+},p_{h^-} )    \nonumber\\
                  & \times \{  \rho_{*+}(p_* ) B_{hh}^+  \varepsilon_{s+} (p_{s} )   -  \rho_{*-}(p_* ) B_{hh}^-  \varepsilon_{s-} (p_{s} )  \}\nonumber\\
                  \nonumber \\*
	     = & \frac{N_*}{2}   B_{D\pi}^*   \int\!\! dp_* dp_s dp_{h^+} dp_{h^-}  \varepsilon_{hh} (p_{h^+},p_{h^-} ) \nonumber \\
                 & \times    \rho_{hh^*}( p_{h^+}, p_{h^-},p_s \: \vert \: p_*) \rho_{*}(p_* ) B_{hh} \varepsilon_{s} (p_{s} )   \nonumber\\
                &\times [ (1+\delta \rho_{*}(p_* )) \left( 1+A_{\CP} \right) (1+\delta\varepsilon_{s} (p_{s} ) ) \nonumber\\
               & \quad - \    (1-\delta \rho_{*}(p_* )) \left( 1-A_{\CP} \right) (1-\delta\varepsilon_{s} (p_{s} ) ) ], \nonumber
\end{align}
where we have defined the following additional quantities:
$\rho_*  =  (1/2)\left(\rho_{*+} + \rho_{*-}\right)$, $\delta \rho_*  =  (\rho_{*+} - \rho_{*-})/(\rho_{*+} + \rho_{*-})$,
$B_{hh}  =  (1/2) (B_{hh}^+ + B_{hh}^-)$,  $A_{\CP} \equiv A_{\CP}(hh)= (B_{hh}^+ - B_{hh}^-)/(B_{hh}^+ + B_{hh}^-)$, 
$\varepsilon_s  =  (1/2)( \varepsilon_{s+} + \varepsilon_{s-})$, and $\delta \varepsilon_s  =  (\varepsilon_{s+} - \varepsilon_{s-})(\varepsilon_{s+} + \varepsilon_{s-})$.
Expanding the products we obtain
\begin{align}
	N_+ - N_-  = &  N_*    B_{D\pi}^*   B_{hh}   \int\!\! dp_* dp_s dp_{h^+} dp_{h^-}  \rho_{*}(p_* )  \varepsilon_{s} (p_{s} )  \nonumber\\
   	  & \times  \rho_{hh^*}( p_{h^+}, p_{h^-},p_s \: \vert \: p_*) \varepsilon_{hh} (p_{h^+},p_{h^-} )   \nonumber\\
            & \times  [A_{\CP}  + \delta \rho_{*}(p_* ) + \delta\varepsilon_{s} (p_{s} ) \nonumber  \\
      	& \quad + A_{\CP}  \delta \rho_{*}(p_* )  \delta\varepsilon_{s} (p_{s} )   ]. \nonumber
\end{align}
Since the \CP\ symmetry of the $p\bar{p}$ initial state ensures that $\delta\rho_*(p_*) = -  \delta\rho_*(-p_*)$, the second and fourth term in brackets vanish when integrated over a $p_*$ domain symmetric in $\eta$. In a similar way we obtain
\begin{align}
	N_+ + N_-  = & N_*    B_{D\pi}^*   B_{hh}    \int\!\! dp_* dp_s dp_{h^+} dp_{h^-}   \rho_{*}(p_* )  \varepsilon_{s} (p_{s} )\nonumber\\
   	  & \times  \rho_{hh^*}( p_{h^+}, p_{h^-},p_s \: \vert \: p_*) \varepsilon_{hh} (p_{h^+},p_{h^-} ) \nonumber\\
          & \times [ 1 + A_{\CP} \delta\varepsilon_{s} (p_{s} ) A_{\CP} \delta \rho_{*}(p_* )\nonumber \\
         & \quad   + \delta\varepsilon_{s} (p_{s}) \delta \rho_{*}(p_* ) ].    \nonumber
\end{align}
The second term in brackets is small with respect to $A_{\CP} $ and can be neglected, while
the third and fourth terms vanish once integrated over a $p_*$ domain symmetric in $\eta$.
Hence the observed asymmetry is written as
\begin{widetext}
\begin{align}
A(hh^*) & = \left( \frac{N_+ - N_- }{N_+ + N_- }\right)^{hh^*} = A_{\CP}(h^+h^-) + \int dp_s h^{hh^*}_s(p_s) \delta\varepsilon_s(p_s), \text{where} \\ 
h^{hh^*}_s(p_s) & = \frac{ \int\!\!  dp_* dp_{h^+} dp_{h^-}     \rho_{*}(p_* )   \rho_{hh^*}( p_{h^+}, p_{h^-},p_s \: \vert \: p_*)    \varepsilon_{hh} (p_{h^+},p_{h^-} )  \varepsilon_{s} (p_{s} )}
 {  \int\!\! dp_* dp_{h^+} dp_{h^-}  dp_s    \rho_{*}(p_* )   \rho_{hh^*}( p_{h^+}, p_{h^-},p_s \: \vert \: p_*)   \varepsilon_{hh} (p_{h^+},p_{h^-} )  \varepsilon_{s} (p_{s} )} \label{eq:normalized-densities}
 \end{align}
 \end{widetext}
is the normalized density in phase space of the soft pion for the events included in our sample.

\subsection {$D^*$-tagged $D^0\to K^-\pi^+$}
Assuming factorization of efficiencies for reconstructing the neutral charmed meson and the soft pion, 
we write
\begin{align}
N_\pm  = & \frac{N_*}{2}    B_{D\pi}^*  \int\!\! dp_* dp_s dp_{\pi} dp_{K}  \rho_{*\pm}(p_* )   B_{K\pi}^\pm       \nonumber\\
               & \times \rho_{K\pi^*}( p_{K}, p_{\pi},p_s \: \vert \: p_*) \varepsilon_{K\mp\pi\pm} (p_{K} ,p_{\pi} ) \varepsilon_{s\pm} (p_{s} ),  \nonumber 
 \end{align}
where $p_\pi$ and $p_K$ are the three-momenta of the pion and kaon, $\rho_{K\pi}^*$ is the density in phase space of the soft pion and $K\pi$ pair from the $D^0$ decay, $B_{K\pi}^+$ and $B_{K\pi}^-$ are the branching fractions of $D^0\to K^-\pi^+$ and $\Dbar^0\to K^+ \pi^-$, and $\varepsilon_{K-\pi+}$ and  $\varepsilon_{K+\pi-}$ are the detection efficiencies of the $K^-\pi^+$ and $K^+\pi^-$ pairs from  $D^0$ and $\Dbar^0$ decay.
The difference between charm and anti-charm event yields is written as 
\begin{align}
N_+ - N_-  = & \frac{N_*}{2}    B_{D\pi}^*  \int\!\! dp_* dp_s dp_{\pi} dp_{K}  \rho_{K\pi^*}( p_{K}, p_{\pi},p_s \: \vert \: p_*) \nonumber \\
          &\times [  \rho_{*+}(p_* )   B_{K\pi}^+   \varepsilon_{K-\pi+} (p_{K} ,p_{\pi} ) \varepsilon_{s+} (p_{s} )  \nonumber\\
        	& \quad -   \rho_{*-}(p_* )   B_{K\pi}^-   \varepsilon_{K+\pi-} (p_{K} ,p_{\pi} ) \varepsilon_{s-} (p_{s} )  ] \nonumber\\
	\nonumber \\*
	= & \frac{N_*}{2}   B_{D\pi}^*   B_{K\pi}   \int\!\! dp_* dp_s dp_{\pi} dp_{K}  \rho_{*}(p_* )   \varepsilon_{s} (p_{s} )    \nonumber \\*
                 &\times  \rho_{K\pi^*}( p_{K}, p_{\pi},p_s \: \vert \: p_*)  \varepsilon_{K\pi}(p_K,p_{\pi} )\nonumber\\
    	&\times  \{ (1+\delta \rho_{*}(p_* ) )  (1+A_{\CP})  \nonumber \\
	& \times  (1+\delta \varepsilon_{K\pi} (p_K,p_{\pi} ) )  (1+\delta  \varepsilon_{s} (p_{s} ) )  \nonumber\\
     	& \qquad -   \: (1-\delta \rho_{*}(p_* ) )   (1-A_{\CP})  \nonumber \\*
	& \times  (1-\delta \varepsilon_{K\pi} (p_K,p_{\pi} ) ) (1-\delta  \varepsilon_{s} (p_{s} ) ) \},\nonumber
\end{align}
where we have defined the following additional quantities:
$B_{K\pi}  =  (1/2)(B_{K\pi}^+ + B_{K\pi}^-)$,  $A_{\CP}  \equiv  A_{\CP}(K\pi) =  (B_{K\pi}^+ - B_{K\pi}^-)/(B_{K\pi}^+ + B_{K\pi}^-)$,
$\varepsilon_{K\pi}  =  (1/2)( \varepsilon_{K-\pi+} + \varepsilon_{K+\pi-})$, and $\delta \varepsilon_{K\pi}  =  (\varepsilon_{K-\pi+} - \varepsilon_{K+\pi-})/(\varepsilon_{K-\pi+} + \varepsilon_{K+\pi-})$. Expanding the products and observing that all terms in $\delta \rho_{*}(p_* )$ vanish upon integration over a symmetric $p_*$ domain, we obtain
\begin{align}
N_+ - N_-  = & N_*    B_{D\pi}^*   B_{K\pi}   \int\!\! dp_* dp_s dp_{\pi} dp_{K}\rho_{*}(p_* )  \varepsilon_{s} (p_{s} )     \nonumber\\
                 &\times    \rho_{K\pi^*}( p_{K}, p_{\pi},p_s \: \vert \: p_*)    \varepsilon_{K\pi} (p_{K},p_{\pi} )   \nonumber\\
        	& \times \{ A_{\CP} +\delta \varepsilon_{K\pi} (p_K,p_{\pi}) +\delta  \varepsilon_{s} (p_{s} )  +  \ldots \}, \nonumber
\end{align}
where we have neglected one term of order $A_{\CP}   \delta^2$. Similarly,
\begin{align}
N_+ + N_-  =	& N_*    B_{D\pi}^*   B_{K\pi}   \int\!\! dp_* dp_s dp_{\pi} dp_{K}  \rho_{*}(p_* )  \varepsilon_{s} (p_{s} )    \nonumber\\
                 		& \times  \rho_{K\pi^*}( p_{K}, p_{\pi},p_s \: \vert \: p_*)    \varepsilon_{K\pi} (p_{K},p_{\pi} )   \nonumber\\
        			& \times [ 1 + A_{\CP} \delta \varepsilon_{K\pi} (p_K,p_{\pi}) +A_{\CP} \delta  \varepsilon_{s} (p_{s} ) \nonumber \\
        			& \quad + \delta \varepsilon_{K\pi} (p_K,p_{\pi})\delta \varepsilon_{s}(p_{s} ) ]. \nonumber
\end{align}
If we neglect all terms of order $A_{\CP}   \delta$ and $\delta^2$, we finally obtain
\begin{widetext}
\begin{align}
A(K\pi^*) = \left( \frac{N_+ - N_- }{N_+ + N_- }\right)^{K\pi^*} =&  A_{\CP}(K^-\pi^+) 
+ \int dp_\pi h^{K\pi^*}_{K\pi}(p_K,p_\pi) \delta\varepsilon_{K\pi}(p_K,p_\pi) 
+ \int dp_s h^{K\pi^*}_s(p_s) \delta\varepsilon_s(p_s), \\
 \text{where} \qquad h^{K\pi^*}_{K\pi}(p_K,p_\pi) =& \frac{ \int\!\! dp_*  dp_s    \rho_{*}(p_* )   \rho_{K\pi^*}( p_{K}, p_{\pi},p_s \: \vert \: p_*) 
   \varepsilon_{K\pi} (p_{K},p_{\pi} ) \varepsilon_{s} (p_{s} ) }
{  \int  dp_*  dp_{\pi} dp_{K}  dp_s    \rho_{*}(p_* )   \rho_{K\pi^*}( p_{K}, p_{\pi},p_s \: \vert \: p_*) 
  \varepsilon_{K\pi} (p_{K},p_{\pi} ) \varepsilon_{s} (p_{s} ) }, 
 \end{align}
 \end{widetext}
and $h^{K\pi^*}_s(p_s)$ (the $K\pi$ analogous to $h^{hh^*}_s(p_s)$ in Eq.\ \ref{eq:normalized-densities})  are the normalized densities in phase space of $\pi$,$K$ and soft $\pi$,  respectively, for the events included in our sample.

\subsection{Untagged $D^0\to K^-\pi^+$}

In this case
\begin{align}
N_\pm = & \frac{N_0}{2} \int  dp_0 dp_\pi dp_K  \rho_{0\pm}(p_0) B^\pm_{K\pi} \nonumber \\
&\times \rho^0_{K\pi}(p_K,p_\pi \:\vert\: p_0)   \varepsilon_{K\mp\pi\pm}(p_K,p_\pi) \nonumber
\end{align}
\begin{align}
N_+ - N_-  = & \frac{N_0}{2}  B_{K\pi}  \int  dp_0 dp_\pi dp_K \nonumber \\
& \times \rho_{0}(p_0) \rho^0_{K\pi}(p_K,p_\pi \:\vert\: p_0)   \varepsilon_{K\pi}(p_K,p_\pi)   \nonumber\\
& \times  \{ (1+\delta\rho_0(p_0))   (1+A_{\CP})   (1+\delta\varepsilon_{K\pi}(p_K,p_\pi)) \nonumber\\
& \quad -  (1-\delta\rho_0(p_0))   (1-A_{\CP})   (1-\delta\varepsilon_{K\pi}(p_K,p_\pi))\} \nonumber
\end{align}
where we have defined the following quantities $\rho_0  =  (1/2)\left(\rho_{0+} + \rho_{0-}\right)$ and $\delta \rho_0  =  (\rho_{0+} - \rho_{0-})/(\rho_{0+} + \rho_{0-})$.
Assuming $\eta$ symmetry of the $p_0$ integration region, 
\begin{align}
N_+ - N_- = & N_0   B_{K\pi}  \int  dp_0 dp_\pi dp_K \rho_{0}(p_0) \rho^0_{K\pi}(p_K,p_\pi \:\vert\: p_0) \nonumber \\
& \times   \varepsilon_{K\pi}(p_K,p_\pi)    [ A_{\CP} - \delta\varepsilon_{K\pi}(p_K,p_\pi)]. \nonumber
\end{align}
Similarly we obtain
\begin{align}
N_+ +N_-  = & N_0   B_{K\pi}  \int  dp_0 dp_\pi dp_K \rho_{0}(p_0) \rho^0_{K\pi}(p_K,p_\pi \:\vert\: p_0)  \nonumber \\
& \times  \varepsilon_{K\pi}(p_K,p_\pi)  [ 1+ A_{\CP} \delta\varepsilon_{K\pi}(p_K,p_\pi)],  \nonumber
\end{align}
and neglecting the second term in brackets, 
\begin{widetext}
\begin{align}
A(K\pi) & = \left( \frac{N_+ - N_- }{N_+ + N_- }\right)^{K\pi} = A_{\CP}(K^-\pi^+) 
+ \int dp_\pi dp_K h^{K\pi}_{K\pi}(p_K, p_\pi) \delta\varepsilon_{K\pi}(p_K,p_\pi), \text{where} \nonumber \\
h^{K\pi}_{K\pi}(p_K,p_\pi) & = \frac{ \int dp_0     \rho_{0}(p_0)   \rho_{K\pi}^0( p_{K}, p_{\pi}\: \vert \: p_0) 
   \varepsilon_{K\pi} (p_K, p_{\pi} ) }
{  \int  dp_0  dp_{\pi} dp_{K}     \rho_{0}(p_0)   \rho_{K\pi}^0( p_{K}, p_{\pi} \: \vert \: p_0)     \varepsilon_{K\pi} (p_K, p_{\pi} ) }
\end{align}
\end{widetext}
is the normalized density in phase space of the $K\pi$ system in the events included in our sample.

\subsection{Combining the asymmetries\label{sec:AllTogether}}

By combining the asymmetries measured in the three event samples we obtain
\begin{widetext}
\begin{align}
A(hh^*)-&A(K\pi^*)+A(K\pi) = A_{\CP}(h^+h^-) + \int dp_s h^{hh^*}_s(p_s) \delta\varepsilon_s(p_s) \nonumber \\
&- A_{\CP}(K^-\pi^+) -\int dp_K dp_\pi h^{K\pi^*}_{K\pi}(p_K,p_\pi) \delta\varepsilon_{K\pi}(p_K,p_\pi)-\int dp_s h^{K\pi^*}_s(p_s) \delta\varepsilon_s(p_s) \nonumber \\
&+ A_{\CP}(K^-\pi^+) 
  +\int dp_K dp_\pi h^{K\pi}_{K\pi}(p_K, p_\pi) \delta\varepsilon_{K\pi}(p_K,p_\pi) = A_{\CP}(h^+h^-),
\end{align}
\end{widetext}
where we assumed
$h^{K\pi^*}_s(p_s) =  h^{hh^*}_s(p_s)$, and $h^{K\pi^*}_{K\pi}(p_K,p_\pi) = h^{K\pi}_{K\pi}(p_K, p_\pi)$.
The last two equalities are enforced by appropriate kinematic reweighing of the event samples.
We need to equalize distributions with respect to the true momenta while we only access the 
distributions with respect to the measured momenta. Hence the assumption
that event samples that have the same distribution with respect to the measured quantities
also have the same distribution with respect to the true quantities is needed. 
\par The mathematical derivation shows that for small enough physics and detector-induced asymmetries,  the linear combination of the observed 
asymmetries used in this measurement achieves an accurate cancellation of the instrumental effects with minimal impact on systematic uncertainties.

\section{Monte Carlo test of the analysis technique\label{sec:mcvalidation}}
We tested the suppression of instrumental effects by repeating the analysis in simulated samples in which known instrumental and physics asymmetries were introduced.  Many different configurations for the input asymmetries were tested, covering a rather extended range,  to ensure the reliability of the method independently of their actual size in our data. For each configuration, $\mathcal{O}(10^6)$ decays were simulated to reach the desired 0.1\% sensitivity.  Only the $D^0\to \pi^+\pi^-$ sample was tested although the results are valid for the $D^0\to K^+K^-$ case as well.

\begin{figure}[t]
\centering
\includegraphics[width=0.5\textwidth]{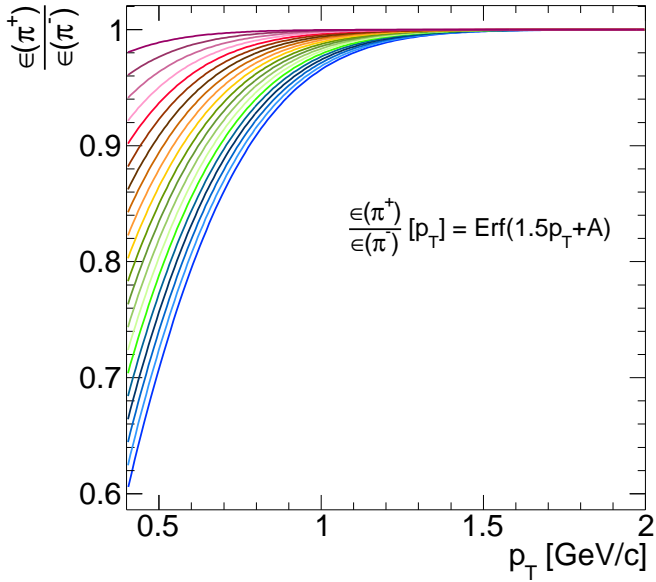}
\caption{Curves corresponding to simulated ratios of efficiencies for reconstructing positive versus negative pions as a function of transverse momentum.}\label{fig:curves}
\end{figure}

We test cancellation of instrumental effects arising from different reconstruction efficiencies between positive and negative particles, which in general depend on the particle species and momentum. Furthermore, the reliability of the suppression should not depend on the actual size of \CP\ violation in $D^0\to K^-\pi^+$  and $D^0\to \pi^+\pi^-$ decays.\par  We repeated the measurement on statistical ensembles where the above effects are known and arbitrarily varied using 
a combination of event-specific weights applied to the true values of simulated quantities.  Each ensemble consists of approximately one thousand trials. We compare the resulting observed asymmetry $\Acp^\text{obs}(\pi\pi)$ to the one given in input, $\Acp^\text{true}(\pipi)$, by inspecting the distribution of the residual,  $\Delta\Acp(\pi\pi) = \Acp^\text{obs}(\pipi)-\Acp^\text{true}(\pipi).$\par
We first investigate the individual impact of each effect. We scan the value of a single input parameter across a range that covers larger variations than expected in data and assume all other effects are zero. First a $p_T$-dependent function that represents the dependence observed in data (see Fig.~\ref{fig:soft}) is used to parametrize the soft pion reconstruction efficiency ratio as $\epsilon(\pi^+)/\epsilon(\pi^-)= \text{Erf}\left(1.5\cdot p_T + A\right)$, where $p_T$ is in GeV/$c$ and various values of the constant $A$ have been tested so that the efficiency ratio at $0.4$ GeV/$c$ spans the 0.6--1 GeV/$c$ range as shown in Fig.~\ref{fig:curves}. Then, the kaon reconstruction efficiency ratio $\epsilon(K^-)/\epsilon(K^+)$ is varied similarly in the  0.6--1 GeV/$c$ range. Finally, a range $-10\%< \Acp< 10\%$ is tested for the physical \CP--violating asymmetry in $D^0\to K^-\pi^+$  and  $D^0\to \pi^+\pi^-$ decays. 

\begin{figure*}[t]
\centering
\includegraphics[width=8.6cm]{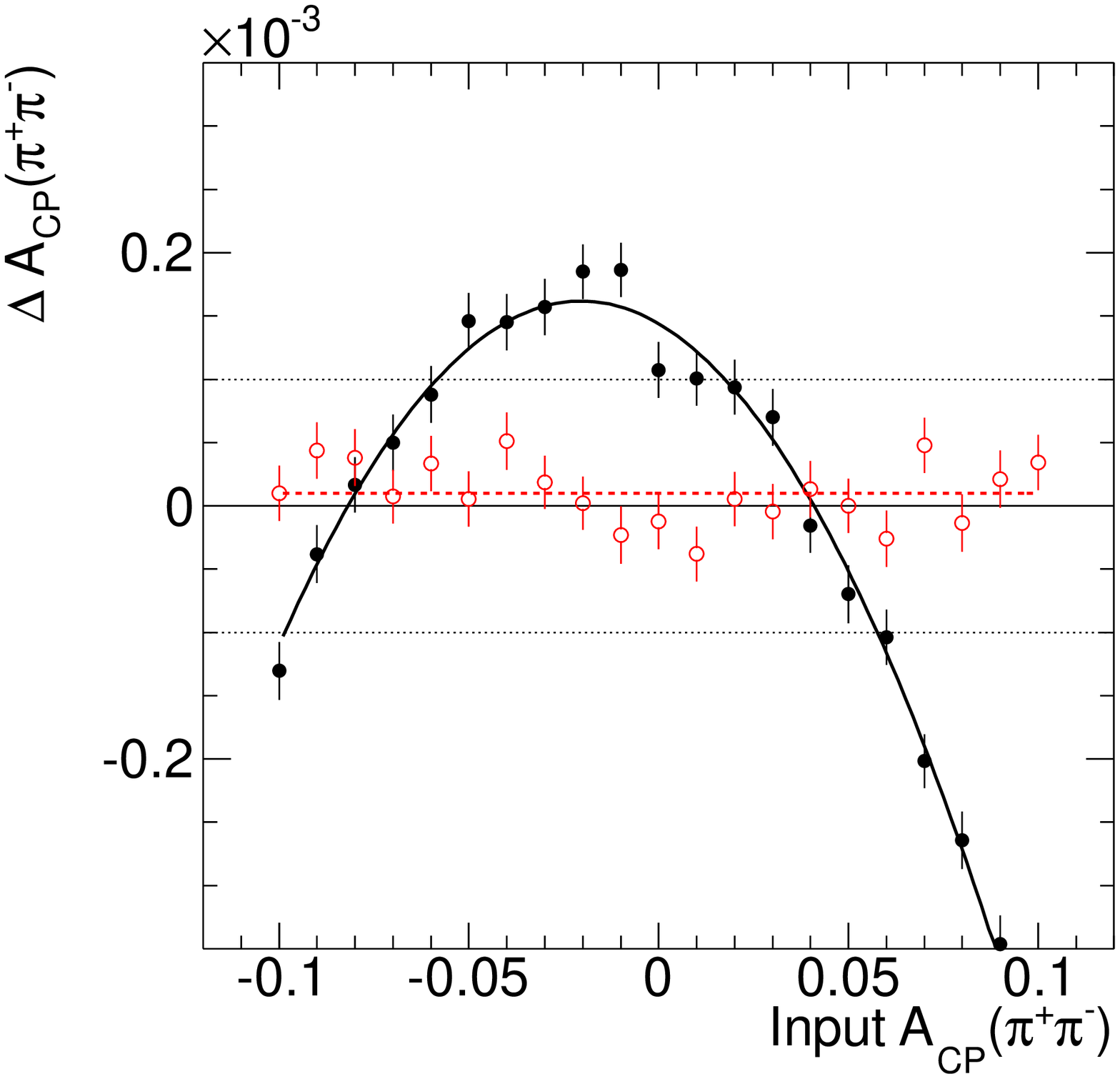}
\includegraphics[width=8.6cm]{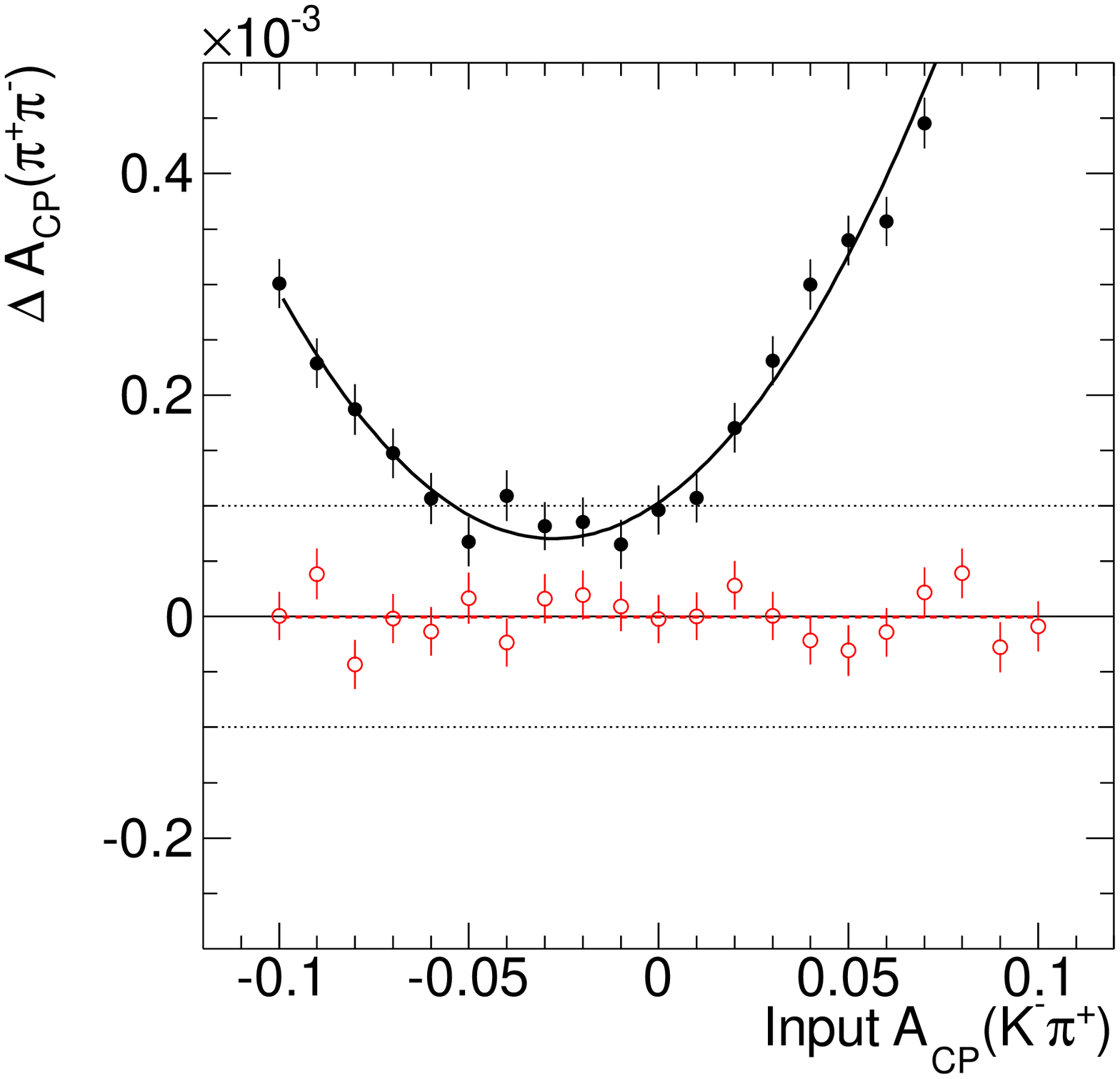}\\
\includegraphics[width=8.6cm]{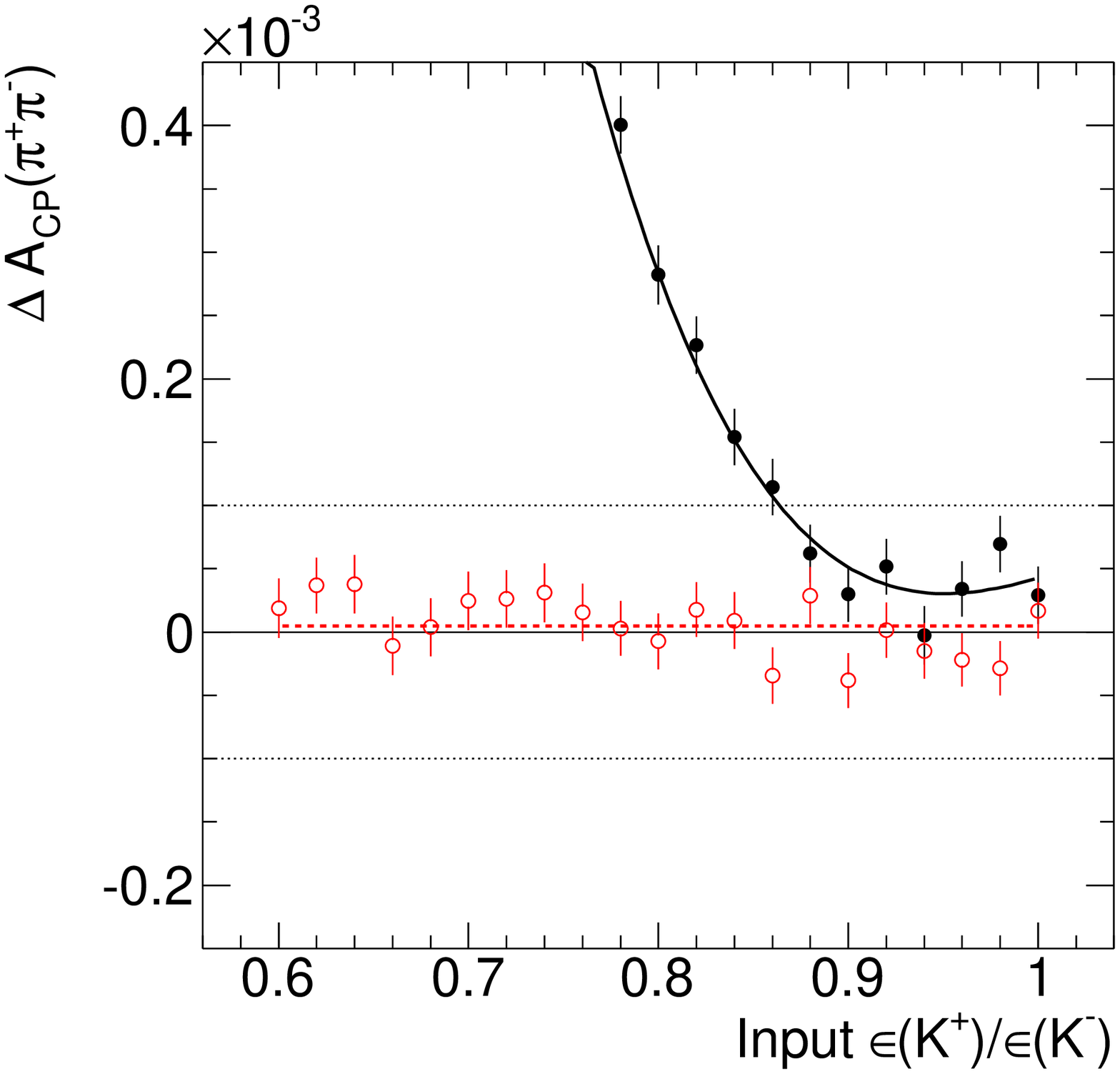}
\includegraphics[width=8.6cm]{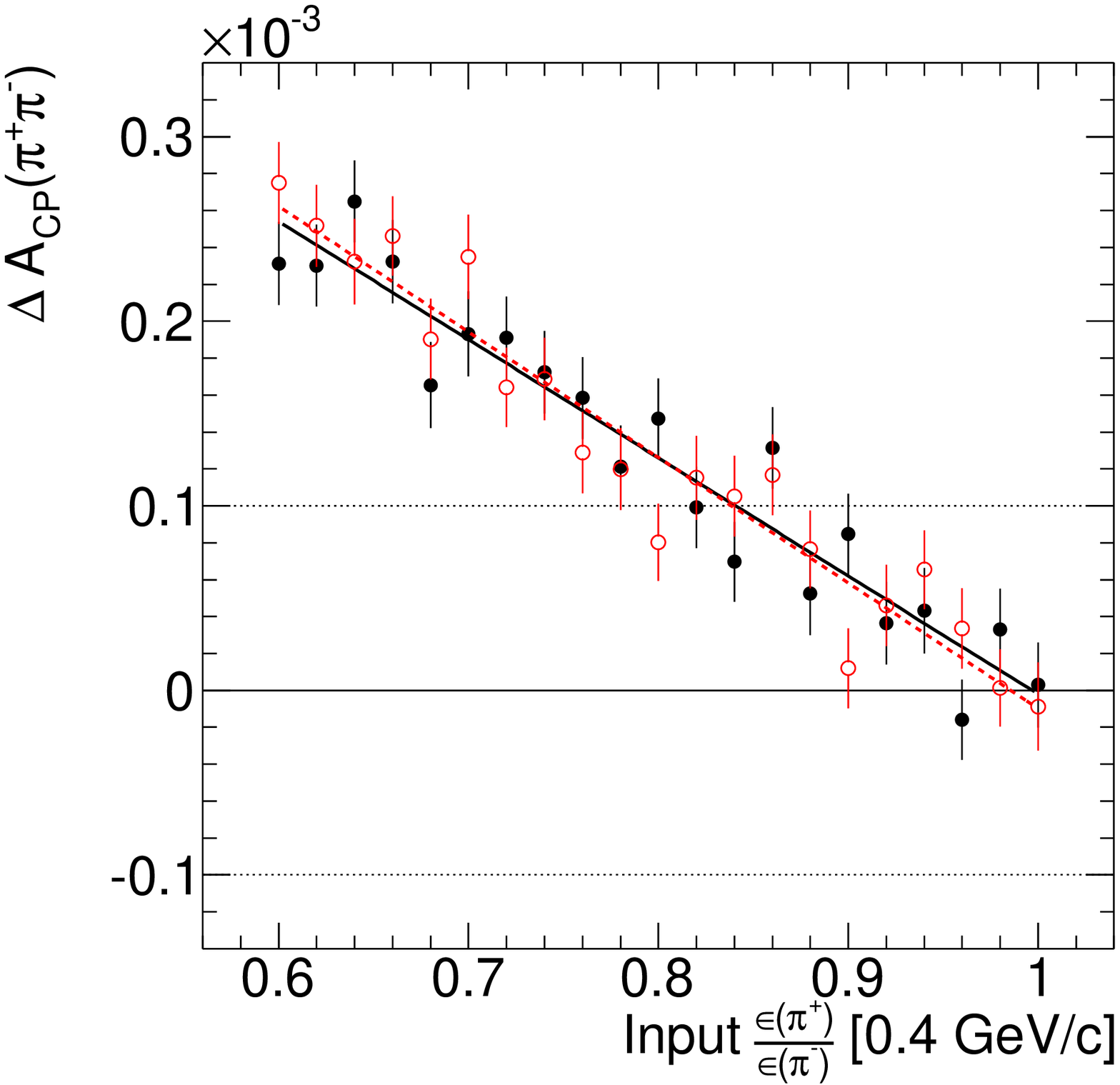}
\caption{Asymmetry residual as a function of the input quantity varied. Other effects are assumed zero (empty dots) or different from zero (filled dots).}\label{fig:c0b}
\end{figure*}

The results are shown in Fig.~\ref{fig:c0b} (empty dots). The cancellation of instrumental asymmetries is realized at the sub-per mil  level even with input effects of size much larger than expected in data.


Figure~\ref{fig:c0b} (filled dots) shows the results of a more complete test in which other effects are simulated,  in addition to the quantities varied in the single input parameter scan: a $p_T$-dependent relative efficiency $\epsilon(\pi^+)/\epsilon(\pi^-)$, corresponding to 0.8 at $0.4$ GeV/$c$, $\epsilon(K^-)/\epsilon(K^+)=98\%$, $\Acp(K\pi) = 0.8\%$ and $\Acp(\pi\pi) = 1.1\%$. Larger variations of the residual are observed with respect to the previous case. This is expected because mixed higher-order terms corresponding to the product of different effects are not canceled and become relevant.

\begin{figure}[t]
\centering
\includegraphics[width=8.6cm]{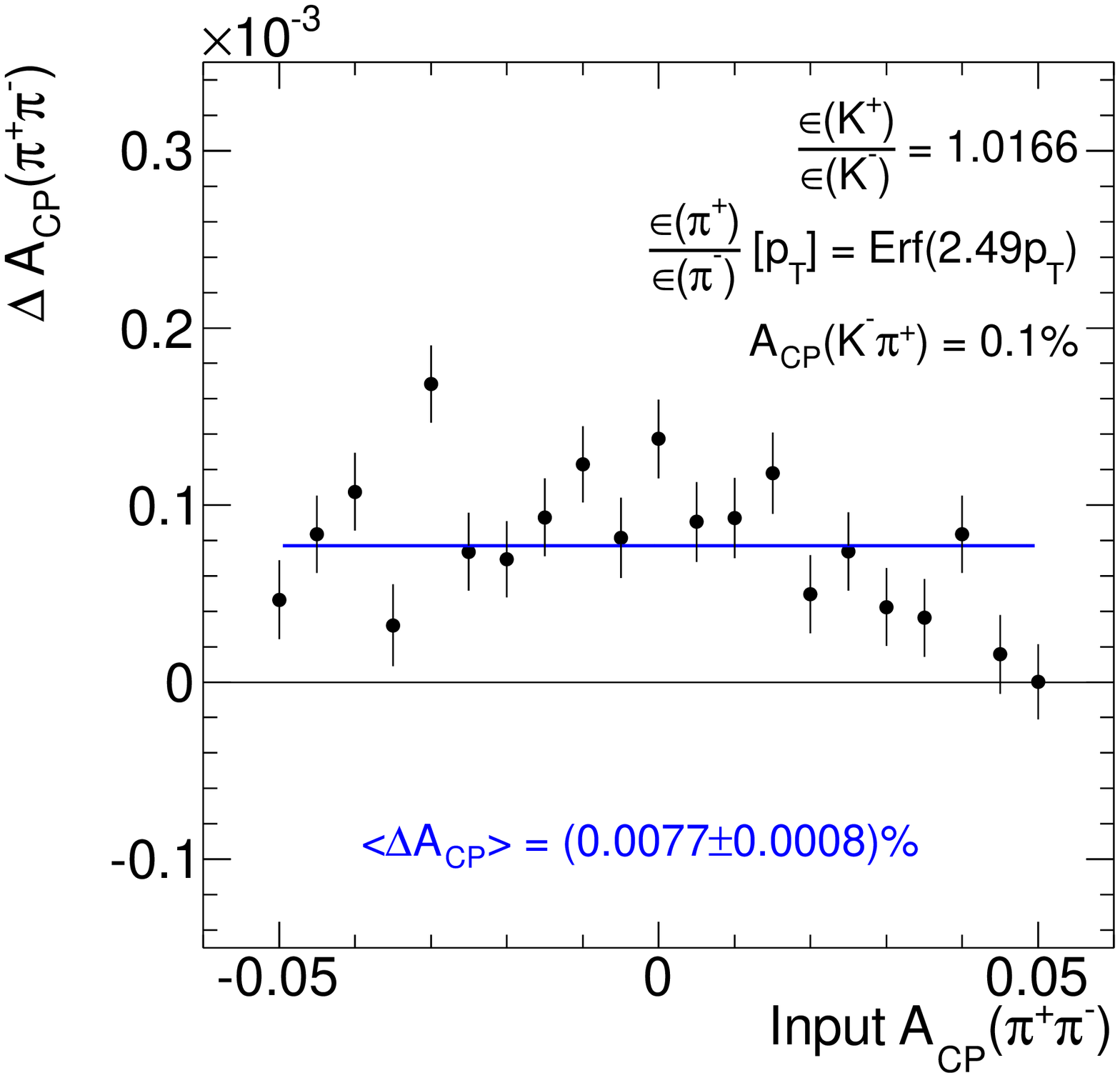}
\caption{Asymmetry residual as a function of the physical \CP--violating asymmetry in $D^0\to \pi^+\pi^-$ decays. Realistic effects other than shown in the scan are also simulated. The line represents the value averaged over the $-5\% < \Acp(\pi\pi)< 5\%$ range.}\label{fig:realt}
\end{figure}

Finally we tested one case with more realistic values for the input effects. The $p_T$ dependence of $\epsilon(\pi^+)/\epsilon(\pi^-)$ is extracted from fitting  data (Fig.~\ref{fig:soft}) to be distributed as $\text{Erf}\left(2.49\ p_T \right)$, with $p_T$ in GeV/$c$. We used \mbox{$\epsilon(K^+)/\epsilon(K^-) \approx \epsilon(K^+\pi^-)/\epsilon(K^-\pi^+) = 1.0166$}, in which the approximation holds assuming equal efficiency for reconstructing positive and negative pions at $p_T>2$ GeV/$c$ \cite{bhh-paper}. We assume  $\Acp(K\pi) = 0.1\%$, ten times larger than the current experimental sensitivity. A $-5\% < \Acp(\pi\pi) < 5\%$ range is tested in steps of $0.5\%$ for the physical asymmetry to be measured. 
The results are shown in Fig.~\ref{fig:realt}.  The maximum observed bias is of the order of $0.02\%$, one order of magnitude smaller than the statistical resolution on the present  measurement.  The observed bias is $(0.0077\pm0.0008)\%$ averaged over the $\Acp(\pi\pi)$ range probed. These results, which extend to the $K^+K^-$ case,  demonstrate the reliability of our method in extracting a precise and unbiased measurement of \CP\ violation in $D^0$ meson decays into $K^+K^-$ and $\pi^+\pi^-$ final states, even in the presence of sizable instrumental asymmetries. \par The results discussed in this appendix are used in Sec.~\ref{sec:syst} to estimate a systematic uncertainty on the final results due to neglecting higher order terms in Eq.\ (\ref{eq:formula}), including possible non-factorization of $h^+h^{'-}$ and $\pi_s$ reconstruction efficiencies.






\begin{thebibliography}{0}
\expandafter\ifx\csname natexlab\endcsname\relax\def\natexlab#1{#1}\fi
\expandafter\ifx\csname bibnamefont\endcsname\relax
  \def\bibnamefont#1{#1}\fi
\expandafter\ifx\csname bibfnamefont\endcsname\relax
  \def\bibfnamefont#1{#1}\fi
\expandafter\ifx\csname citenamefont\endcsname\relax
  \def\citenamefont#1{#1}\fi
\expandafter\ifx\csname url\endcsname\relax
  \def\url#1{\texttt{#1}}\fi
\expandafter\ifx\csname urlprefix\endcsname\relax\def\urlprefix{URL }\fi
\providecommand{\bibinfo}[2]{#2}
\providecommand{\eprint}[2][]{\url{#2}}

\end{thebibliography}


\begin{thebibliography}{99}

\bibitem{Antonelli:2009ws}
M.~Antonelli  {\it et al.}, Phys. Rept. {\bf 494}, 197 (2010).

\bibitem{Bianco:2003vb}
  S.~Bianco, F.~L.~Fabbri, D.~Benson, and I.~I.~Bigi,  Riv.\ Nuovo Cim.\  {\bf 26N7}, 1 (2003).

\bibitem{pdg} 
K.~Nakamura {\it et al.} (Particle Data Group), J.\ Phys.\ G {\bf 37}, 075021 (2010) and 2011 partial update for the 2012 edition.

\bibitem{hfag}
D. Asner  {\it et al.},  arXiv:1010.1589 and online update at http://www.slac.stanford.edu/xorg/hfag.

\bibitem{Artuso:2008vf}
 M.~Artuso, B.~Meadows, and A.~A.~Petrov, Ann.\ Rev.\ Nucl.\ Part.\ Sci.\  {\bf 58}, 249 (2008).

\bibitem{Shipsey:2006zz}
I.~Shipsey, Int.\ J.\ Mod.\ Phys.\  A {\bf 21}, 5381 (2006).
 
\bibitem{Burdman:2003rs}
G.~Burdman and I.~Shipsey, Ann.\ Rev.\ Nucl.\ Part.\ Sci.\  {\bf 53}, 431 (2003).

\bibitem{Nir:1993mx}
  Y.~Nir and N.~Seiberg, Phys.\ Lett.\  B {\bf 309}, 337 (1993).

\bibitem{Ciuchini:2007cw}
 M.~Ciuchini {\it et al.}, Phys.\ Lett.\  B {\bf 655}, 162 (2007).
 
 \bibitem{Aubert:2007wf}
  B.~Aubert {\it et al.}  (\babar\ Collaboration),  Phys.\ Rev.\ Lett.\  {\bf 98}, 211802 (2007).

\bibitem{Staric:2007dt}
  M.~Staric {\it et al.}  (Belle Collaboration), Phys.\ Rev.\ Lett.\  {\bf 98}, 211803 (2007).

\bibitem{:2007uc}
  T.~Aaltonen {\it et al.}  (CDF Collaboration),  Phys.\ Rev.\ Lett.\  {\bf 100}, 121802 (2008).

 \bibitem{Petrov:2006nc}
  A.~A.~Petrov,
  Int.\ J.\ Mod.\ Phys.\  A {\bf 21}, 5686 (2006).
  
\bibitem{Golowich:2007ka}
  E.~Golowich, J.~Hewett, S.~Pakvasa, and A.~A.~Petrov,
  Phys.\ Rev.\  D {\bf 76}, 095009 (2007).










\bibitem{Aubert:2007if}
  B.~Aubert {\it et al.}  (\babar\ Collaboration), Phys.\ Rev.\ Lett.\  {\bf 100}, 061803 (2008).
 
\bibitem{:2008rx}
M.~Staric {\it et al.}  (Belle Collaboration), Phys.\ Lett.\  B {\bf 670}, 190 (2008).

\bibitem{Acosta:2004ts}
D.~E.~Acosta {\it et al.}  (CDF Collaboration), Phys.\ Rev.\ Lett.\  {\bf 94}, 122001 (2005).



\bibitem{calorimetro}
L.~Balka   {\it et al.}, Nucl. Instrum. Methods A {\bf 267}, 272 (1988); S.~Bertolucci   {\it et al.}, Nucl. Instrum. Methods A {\bf 267}, 301 (1988); M.~Albrow   {\it et al.}, Nucl. Instrum. Methods A {\bf 480}, 524 (2002); and G.~Apollinari   {\it et al.}, Nucl. Instrum. Methods A {\bf  412}, 515 (1998).

\bibitem{muoni}
G.~Ascoli   {\it et al.}, Nucl. Instrum. Methods A {\bf 268}, 33 (1988).

\bibitem{COT} 
 T.~Affolder  {\it et al.},  Nucl. Instrum. Methods A {\bf 526}, 249 (2004).

\bibitem{SVX} 
A.~Sill  {\it et al.}, Nucl. Intrum. Methods A {\bf 447}, 1 (2000).

\bibitem{L00}
C.~S. Hill  {\it et al.}, Nucl. Instrum. Meth. A {\bf 530}, 1 (2004).  

\bibitem{ISL}
A. Affolder  {\it et al.},  Nucl. Instrum. Meth. A {\bf 453}, 84 (2000).

\bibitem{XFT}
E.~J.~Thomson {\it et al.}, IEEE Trans. Nucl. Sci. {\bf 49}, 1063 (2002);  R. Downing {\it et al.},  Nucl. Instrum. Methods,  A {\bf 570}, 36 (2007).

\bibitem{SVT} 
L.~Ristori and G.~Punzi, Annu. Rev. Nucl. Part. Sci. {\bf 60}, 595 (2010); W.~Ashmanskas  {\it et al.}, Nucl. Instrum. Methods, A {\bf  518}, 532 (2004).

\bibitem{tesi-angelo}
A.~Di Canto, Ph.D. Thesis, University of  Pisa, Fermilab Report No. FERMILAB-THESIS-2011-29 (2011).

\bibitem{johnson}
N.~L.~Johnson, Biometrika {\bf 36}, 149 (1949).
 
\bibitem{gronau}
S.~Bar-Shalom, G.~Eilam, M.~Gronau, and J.~L.~Rosner, Phys.\ Lett.\  B {\bf 694}, 374 (2011).

\bibitem{Bigi:1986dp}
I.~I.~Bigi and A.~I.~Sanda, Phys.\ Lett.\  B {\bf 171}, 320 (1986).

\bibitem{Golden:1989qx}
M.~Golden and B.~Grinstein, Phys.\ Lett.\  B {\bf 222}, 501 (1989).

\bibitem{Buccella:1994nf}
 F.~Buccella {\it et al.},  Phys.\ Rev.\  D {\bf 51}, 3478 (1995).
 
 \bibitem{Xing:1996pn}
Z.-Z.~Xing, Phys.\ Rev.\  D {\bf 55}, 196 (1997).

  \bibitem{Du:2006jc}
D.-S.~Du, Eur.\ Phys.\ J.\  {\bf 50}, 579 (2007).

\bibitem{Grossman:2006jg}
Y.~Grossman, A.~L.~Kagan, and Y.~Nir, Phys.\ Rev.\  D {\bf 75}, 036008 (2007).
 

\bibitem{Bigi:2011re}
I.~I.~Bigi, A.~Paul, and S.~Recksiegel, J. High Energy Phys.\ 06 (2011) 089.

\bibitem{Bigi:2011em}
I.~I.~Bigi and A.~Paul, arXiv:1110.2862.

 \bibitem{bhh-paper}
T.~Aaltonen {\it et al.} (CDF Collaboration),  Phys.\ Rev.\ Lett.\ {\bf 106},  181802 (2011).
 
\end{thebibliography}
\end{document}